\documentclass[11pt,oneside]{memoir}
\usepackage{amsmath,amssymb}
\usepackage{amsfonts}
\usepackage{graphicx}
\usepackage{cancel}
\usepackage{fullpage}
\usepackage[hidelinks]{hyperref}
\usepackage{color}
\usepackage[style=phys, sorting=none,backend=bibtex]{biblatex}

\newcommand{\Lagr}{\mathcal{L}}
\newcommand{\order}[1]{\mathcal{O}\!\left(#1\right)}

\newcommand{\nor}[1]{\colon\!\!#1\!\colon\!\!}

\newcommand{\derOrd}[2]{\frac{d#1}{d#2}}
\newcommand{\der}[2]{\frac{\partial#1}{\partial#2}}

\newcommand{\llangle}{\langle\!\langle}
\newcommand{\rrangle}{\rangle\!\rangle}

\newcommand{\qimplies}{\quad\implies\quad}

\newcommand{\non}{\nonumber\\}

\title{Classical and quantum aspects of non-linear sigma models with a squashed sphere target space}
\date{}

\author{Daniel Schubring}

\maxtocdepth{subsection}
\setsecnumdepth{subsection}
\begin{document}
	
	\maketitle
	
	Various aspects of non-linear sigma models with an $SU(N)\times U(1)$ symmetric target space are considered. In the case $N=2$, three-dimensional topological defects are discussed which are relevant for frustrated magnetic systems and which may offer a new perspective on the Skyrme model. An apparent discrepancy between the large $N$ expansion and the weak coupling expansion noted earlier in the literature is reviewed and clarified. A systematic approach to the operator product expansion at sub-leading order in large $N$ is developed and the spinon two-point function is expanded as a trans-series in which all ambiguities in the Borel plane are shown to cancel.

\tableofcontents

	\chapter{Introduction}
	Non-linear sigma models are ubiquitous in theoretical physics for several reasons. Perhaps most importantly, they can arise as effective field theories for real systems which have some underlying global symmetry which is inherited by the target space of the sigma model. An early example of this is the use of the classical Heisenberg model and its extension the $O(N)$ model \cite{stanley1968dependence,stanley1968spherical,polyakov1975interaction} as a model for spin systems. Another important example is the use of the principal chiral model (PCM) as an effective theory for strong interactions \cite{gell1960axial,skyrme1961non}, and this is the context where the term `sigma model' originated in the first place.
	
	Beyond the direct applicability as effective field theories, sigma models have also proved invaluable as relatively simple toy models for studying non-perturbative aspects of quantum field theory. The $CP^{N-1}$ model was developed for this purpose as a close cousin to four dimensional gauge theories \cite{eichenherr1978n,d19781n,witten1979instatons} and it later found more direct relevance as an effective field theory in condensed matter physics \cite{read1990spin}. Sigma models also play a central role in string theory and this has spurred on much of the study of higher loop beta functions for general target spaces (see e.g. \cite{callan1988sigma} and references therein). While there are severe restrictions placed on the target space in order for the string theory to be consistent, integrable asymptotically free sigma models which are more familiar in other areas of physics may arise as duals to spin chains in the AdS/CFT correspondence \cite{alday2007comments,bykov2010worldsheet,BYKOV2012100,basso2014bethe}.
	
	This thesis deals with a particular class sigma models which is relevant in all of these contexts. We will focus on a sigma models with a \emph{squashed sphere} target space \cite{CAMPOSTRINI1994680,azaria1995massive,basso2013integrability}. On the one hand these models may be understood as a deformation of the $O(2N)$ model which only preserves $SU(N)\times U(1)$ symmetry. From another perspective these sigma models may be understood as a Higgsed $CP^{N-1}$ model.
	
	To be more clear about this, the Lagrangian of the $O(2N)$ model may be written as
	\begin{align}
		\Lagr_{O(2N)}= \frac{1}{2g^2}\partial \bar{z}\cdot \partial z,\label{Lagr oN z}
	\end{align}
	where $z$ is an $N$ component complex unit vector field, $|z|^2=1$, and $g$ is the coupling constant. The target space of the sigma model is the unit $2N-1$ dimensional sphere $S^{2N-1}$. The action in this form has an obvious $U(N)$ global symmetry, but the grouping of real and imaginary parts in $z$ is arbitrary and in fact the full global symmetry group is $O(2N)$.
	
	The sphere $S^{2N-1}$ described in terms of $z$ may be mapped to complex projective space $CP^{N-1}$ by identifying all $z$ differing only by a phase factor. Likewise the $CP^{N-1}$ sigma model may be constructed by gauging the phase of the $z$ fields with an auxiliary $U(1)$ gauge field $A$.
	\begin{align}
		\Lagr_{CP^{N-1}}= \frac{1}{2g^2}\left(\partial + i A\right) \bar{z}\cdot \left(\partial - i A\right)z. \label{Lagr cpN A}
	\end{align}
The auxiliary field $A$ may be integrated out to get an action in terms of $z$ alone, but in some sense this Lagrangian better indicates the relevant physics of the model in two spacetime dimensions. The $z$ field is associated to $N$ particles (and their antiparticles) called \emph{spinons}. In the $O(2N)$ model these are the true asymptotic states in the spectrum, but in the $CP^{N-1}$ model the presence of the $A$ field may be seen as confining the spinons into bound state mesons \cite{witten1979instatons,HABER1980458,campostrini1992cp,chubukov1995confinement}.

The squashed sphere model arises by coupling this gauge field to a Stueckelberg field $\phi$, i.e. a scalar field transforming as $\phi\rightarrow \phi+\theta$ under gauge transformations $z\rightarrow e^{i\theta}z$,
\begin{align}
	\Lagr = \frac{1}{2g^2}\left(\partial + i A\right) \bar{z}\cdot \left(\partial - i A\right)z+\frac{N\gamma}{2}\left(\partial\phi-A\right)^2.\label{Lagr squashed sphere A psi}
\end{align}
We may fix the gauge by choosing $\phi=0$, in which case $A$ has a `mass term' with parameter $\gamma$ and a factor $N$ for later convenience. We may further integrate out the gauge field to get a more geometric description of the squashed sphere model, and this will be our starting point in chapter \ref{Sec 2}. But for now note that owing to the dualities in two spacetime dimensions ($d=2$)\footnote{Throughout the text the number of dimensions $d$ will refer to the number of continuous arguments of the fields rather than being tied to the physical interpretation. Depending on context these dimensions may be interpreted as entirely spatial, as will be the case for the time-independent fields of chapter \ref{Sec 3}, or they might have an interpretation as coordinates of Euclidean spacetime, which is more appropriate when discussing the quantum mechanical spectrum.} there are many alternate descriptions of this model. The $\phi$ field may be traded for one or more massless fermions through the bosonization map (see e.g. \cite{fujikawa2004anomalies}) It may also be T-dualized to get a description in terms of an axion field coupled to the instanton charge of the $CP^{N-1}$ model \cite{BALOG199616,BALOG2000367}.

As the parameter $\gamma$ goes from zero to infinity the $A$ field is suppressed and we interpolate between the $CP^{N-1}$ and $O(2N)$ models.\footnote{A comparison of the notation in this text with various other papers in the literature is found in Appendix \ref{appendix notation}.} As discussed above, both limits are widely studied as effective theories in condensed matter and toy models in high-energy, but in some respects they have very different properties. For $\gamma>0$ the confinement of the $CP^{N-1}$ model is broken and unbound spinons may appear as asymptotic states as they do in the $O(2N)$ model. However for $\gamma$ very small (in a sense to be made precise later), free spinons are much more massive than their bound states and the model is much like the $CP^{N-1}$ model together with an additional degree of freedom associated with the $\phi$ field. As $\gamma$ increases the mass of the spinons decreases with respect to the bound states, and eventually there is a threshold value where the bound states leave the spectrum entirely and only the free spinons remain. The model is actually integrable at this threshold value \cite{CAMPOSTRINI1994680,basso2013integrability} and this parameter restriction will play a special role in what follows.

As is generically true for sigma models in two spacetime dimensions with positive curvature target spaces, the squashed sphere sigma model is asymptotically free and thus it shares certain qualitative features with four-dimensional gauge theories like quantum chromodynamics (QCD). The flip side of asymptotic freedom is strong coupling in the infrared, and this leads to problems with naive perturbation theory in terms of $g$. Perturbative expansions of physical quantities like correlation functions must be supplemented by non-perturbative corrections scaling as powers of $\exp\left[-1/\left(Ng^2\right)\right]$. These non-perturbative corrections may be seen as arising from the vacuum expectation values (VEVs) of operators in an operator product expansion (OPE) \cite{shifman1979qcd}.

This situation holds both for QCD and for asymptotically free sigma models in 2D, but in the latter case we have more tools at our disposal, such as integrability and a simpler vector-like large $N$ limit. In Chapter \ref{Sec 5}, I will find the OPE for the spinon two-point function in the squashed sphere model, extending some work with Sheu and Shifman on the supersymmetric (SUSY) $O(N)$ model \cite{schubring2021treating}. As in QCD, the perturbative part of the OPE will not be Borel summable due to {renormalon} ambiguities in the Borel plane (see e.g. \cite{beneke1999renormalons}). However, following previous work on the bosonic $O(N)$ model \cite{dave1984237,beneke1998308}, I will show how these ambiguities cancel with ambiguities arising from the non-perturbative VEVs to all orders in the OPE.

From a modern point of view this OPE is an example of a trans-series, and in the resurgence program the non-perturbative corrections which cancel the renormalons are considered to arise from semi-classical objects in the path integral which appear after a twisted compactification of the spatial direction \cite{dunne:2012ae}. This scenario has been investigated for many of the simple homogeneous sigma models discussed here \cite{dunne:2012ae,chermanEtAl2013,dunne:2015ywa} including the squashed sphere model itself at $N=2$ \cite{demulder2016resurgence}. However usually the object under investigation has been the ground state energy rather than a full correlation function involving an expansion in terms of VEVs of operators, as we do here. Indeed a recent investigation of a single VEV in the SUSY $CP^{N-1}$ model suggested that there are shifts in the location of the renormalon ambiguity after compactification which are still poorly understood \cite{ishikawa2020infrared}. Part of the motivation for considering the squashed sphere model is that it shares many features with the $CP^{N-1}$ model, but there is an integrable limit in which it is more tractable to calculate the full OPE, and this provides a very strong consistency check on the calculation.

The calculation in Chapter \ref{Sec 5} is an example of using the squashed sphere model as a toy model to make incremental progress towards better understanding the operator product expansion and renormalons in QCD. But as was alluded to in the opening paragraph of this introduction, sigma models may also be directly applicable as effective field theories for spin systems in condensed matter, or even the low energy limit of QCD itself.\footnote{At least in the limit of a large number of colors, and with additional fields corresponding to a full tower of mesons \cite{Witten:1979kh} (see also \cite{sakai2005low,sutcliffe2010skyrmions}).} In Chapter \ref{Sec 3}, which is mostly reproduced from a paper with Naya, Shifman, Wang \cite{schubring2021skyrmeHopf} I consider topological defects in a model of a frustrated magnet which reduces to the squashed sphere model in the continuum limit.

Topological defects in the squashed sphere model may be interesting simply because the limiting cases of $S^{2N-1}$ and $CP^{N-1}$ are not homeomorphic and at first consideration seem very different in terms of allowable topological defects. In two dimensions we may define a topological charge for the $CP^{N-1}$ model associated to the non-trivial second homotopy group. In a condensed matter context this is associated with \emph{2D magnetic Skyrmions} which are a subject of intense experimental and theoretical investigation (see \cite{han2017skyrmions} for a review).

On the other hand, given a target space $S^{2N-1}$ the second homotopy group is trivial and there is no obvious topological charge in two dimensions,\footnote{There may still be soliton-like saddle points of the action which lack topological stability \cite{dunne:2015ywa}. Also we may lift the topological defects in $CP^{N-1}$ to $S^{2N-1}$ much as will be discussed for the 3D case, but this will lead to the field winding around a great circle of $S^{2N-1}$ at spatial infinity. } but for $N=2$ a topological charge may be defined in \emph{three dimensions}. In a high energy context this topological charge is essentially the baryon charge in the two flavor $SU(2)$ Skyrme model \cite{skyrme1961non}, but 3D magnetic Skyrmions are only just beginning to be investigated on the condensed matter side.

Chapter \ref{Sec 3} discusses two closely related models which extend the squashed sphere model by higher derivative terms which may stabilize topological defects in 3D. For the unsquashed $S^3$ case, the topological defects will be much like those in the Skyrme model for low energy QCD, even though the microscopic model will actually be a theory of a frustrated magnet. As the target space is squashed towards the $CP^1$ model, the topological defects will deform to knotted loops which may be identified as string-like extensions of the 2D magnetic Skyrmions.

On the one hand this model suggests a perspective on the topological defects in the ordinary Skyrme model as loops of string. This may end up not being a very useful picture, but at the very least it is a new idea which is still relatively unexplored. On the other hand, the topological defects studied in this chapter may potentially be realized in frustrated magnets in the real world.

\section{A brief review of literature on the squashed sphere model}\label{Sec review}

The case $N=2$ for which the squashed sphere has a topological charge in 3D is special for many reasons. $S^3$ is homeomorphic to the Lie group $SU(2)$, and so the $N=2$ squashed sphere may also be understood as a special case of another family of models which are deformations of the $SU(N)$ PCM. These deformations of general principal chiral models are known as Yang-Baxter models \cite{klimcik2003yang} and they are classically integrable \cite{Klimcik:2008eq}. Yang-Baxter models have attracted a lot of interest as a possible integrable deformation of the $AdS_5\times S^5$ superstring \cite{delduc2014integrable,delduc2014derivation}, in which case they are often referred to as \emph{$\eta$-deformations}. The $N=2$ squashed sphere is often studied as a paradigm example of such integrable deformations (see e.g. the reviews \cite{thompson2019introduction,hoare2022integrable}).

The context of integrability is where the $N=2$ squashed sphere model first explicitly appeared in the physics literature, with the work of Cherednik \cite{cherednik1981integrability}. Not long after, Wiegmann \cite{WIEGMANN1985209}, building on work with Polyakov \cite{POLYAKOV1983121}, considered the $N=2$ case as a means to prove the integrability of the $O(3)$ model. In that work Wiegmann proposed an exact S-matrix for the $N=2$ case, and also discussed the exact spectrum, which is rather subtle in the $CP^1$ limit.

Depending on your point of view, the squashed sphere for $N>2$ was discussed either earlier or later than these works on the $N=2$ case. It is well-known that $CP^{N-1}$ has an anomaly which spoils integrability in the quantum theory \cite{GOLDSCHMIDT1980392,abdalla1981anomaly}, but the anomaly may be canceled by coupling to fermions \cite{abdalla1986integrable}, and in \cite{koberle1987deconfinement} the exact S-matrix for the $CP^{N-1}$ model minimally coupled to massless fermions was proposed. In \cite{abdalla1990numerical}, as a side note to a numerical study, the authors made the point that upon integrating out the fermions the squashed sphere model is recovered. So the $CP^{N-1}$ model minimally coupled to fermions, which was first introduced not long after the $CP^{N-1}$ model itself \cite{DADDA1979125,witten1979instatons}, is in this sense an early appearance of the squashed sphere model in the literature.

In this respect, the SUSY $CP^1$ model \cite{witten1977supersymmetric} might actually be considered to be the earliest consideration of the squashed sphere model. The orthogonality constraint between the two fermions and two bosons in the SUSY $CP^1$ model may be solved in terms of one unconstrained fermion, and upon integrating out the fermion we recover a particular trajectory of the $N=2$ squashed sphere model (this is clarified further in \cite{basso2013integrability}). Furthermore, as was shown already in \cite{witten1977supersymmetric}, the unconstrained fermion may also be bosonized in a more conventional way, leading to the $CP^1$ model coupled to an axion field. 

The $CP^{N-1}$ model for general $N$ coupled to an axion field was studied much later in \cite{BALOG199616} and the point was made that it is T-dual to the squashed sphere model, but there is a non-trivial change of scheme between the two.\footnote{The discussion in this case will have some relevance to our discussion of scheme dependence in Chapter \ref{Sec 4}. } Some of the same authors later investigated the model for $N=2$ using the thermodynamic Bethe ansatz (TBA) \cite{BALOG2000367}, and found an expression for the mass gap in terms of the renormalization group invariant of the weak coupling expansion. A similar application of the TBA was done earlier in an integrable two-parameter model of Fateev \cite{FATEEV1996509}, and that model also reduces to the $N=2$ squashed sphere model in one limit of parameters.

Besides all these studies in integrability and dualities in the high-energy community, the squashed sphere model was also involved in some more down-to-earth applications in the condensed matter community. In \cite{dombre1989nonlinear}, Dombre and Read showed that a particular trajectory in the $N=2$ squashed sphere model is an effective theory for the antiferromagnetic Heisenberg model on a triangular lattice in two spatial dimensions. This effective theory was explored further in \cite{chubukov1994quantum} in a genuinely $2+1$ dimensional way in which both quantum and thermal fluctuations are taken into account. In \cite{azaria1990nonuniversality}, a $d=2+\epsilon$ expansion which is more closely related to the methods in this paper was considered and it was shown that the model flows to an $O(4)$ symmetric fixed point in the infrared. However some doubts over the validity of this analysis led to a follow-up work \cite{azaria1995massive} where the authors compare the model at two-loops in the weak coupling expansion to the large $N$ analysis considered earlier by Campostrini and Rossi \cite{CAMPOSTRINI1994680}.

To a certain extent I will use the comparison of the weak coupling expansion and large $N$ expansion in \cite{azaria1995massive} as a bridge in Chapter \ref{Sec 4} to connect the material in Chapter \ref{Sec 3}, which is closely connected to the study of frustrated antiferromagnets initiated by Dombre and Read, to the more technical considerations of renormalons in the large $N$ approach in Chapter \ref{Sec 5}. The careful consideration of renormalization group methods in this context will also present an opportunity to discuss some work on the squashed sphere model coupled to a Wess-Zumino term in Sec \ref{Sec 4.3}, which has many connections to the literature on integrable deformations discussed earlier in this section. See in particular \cite{kawaguchi2011yangian} for an earlier example of the squashed sphere model with a Wess-Zumino term, and also \cite{kawaguchi2011251} which shows that the $SU(2)_R$ which is explicitly broken in the squashed sphere model is restored via non-local currents as a $q$-deformed $SU(2)$ algebra.

To conclude this brief review let us note the squashed sphere model has an even more direct connection to the study of integrability in the context of AdS/CFT. It is perhaps somewhat well-known that the integrable $O(6)$ non-linear sigma model appears as a certain limit of the $AdS_5\times S^5$ superstring which is dual to certain high spin operators in $\mathcal{N}=4$ super Yang-Mills \cite{alday2007comments}. But also in the case of the $AdS_4\times CP^3$ superstring which is dual to the Aharony-Bergman-Jafferis-Maldacena (ABJM) model \cite{aharony2008} such a limit may be taken \cite{bykov2010worldsheet}, and the result is a massless fermion with a Thirring interaction coupled to the $CP^3$ sigma model. As emphasized by Basso and Rej \cite{basso2013integrability,basso2014bethe}, this is just a fermionic description of the squashed sphere model, with the running value of the Thirring coupling lying on an integrable trajectory.

\section{Outline of this dissertation}

The following is an outline of this dissertation which will also make clear which material is new and which material is present in my earlier papers on the subject.

Chapter \ref{Sec 2} introduces the various representations of the squashed sphere model which will be used in later chapters, and also discusses the geometry of the target space with the aim of calculating the renormalization group equations. Sections \ref{Sec 2.1}, \ref{Sec 2.2}, and \ref{Sec 2.3} and the accompanying appendices \ref{appendix coordinates} and \ref{appendix degenerate} largely cover the same material as \cite{schubring2020sigma} describing a fiber bundle approach to finding the Riemann tensor of a homogeneous space and using that to calculate the beta function. But the material is rewritten in light of my better understanding of the previous literature on the subject. In these sections the squashed sphere model is generalized slightly to a model which may be understood as a Higgsed Grassmannian model. This model is still relatively unstudied compared to other simple non-linear sigma models (although see \cite{bratchikov19881}).

In the later sections \ref{Sec 2.4} and \ref{Sec 2.5} the special features of the model for $N=2$ are discussed. In particular the topological charges are introduced. Much of the introductory material for \cite{schubring2021sigma} and \cite{schubring2021skyrmeHopf} is covered here.

Chapter \ref{Sec 3} discusses topological defects in a model of a frustrated magnet which is an extension of the squashed sphere model with additional higher derivative terms much like the Skyrme model. This chapter is a version of \cite{schubring2021skyrmeHopf} with much less material on the details of the numerical simulations and slightly more material on the construction of the models, the comparison to ansatzes, and the basic idea of the equivalence between the Hopfions and Skyrmions.

Section \ref{Sec 3.1} introduces the notion of a Hopfion and serves as an introduction for the rest of the chapter. Chapter \ref{Sec 3} has two main goals. One goal is to investigate topological defects in a simple toy model for frustrated magnets which is expected to demonstrate behavior characteristic of more realistic models. The toy model was abstracted from the more complicated consideration of a frustrated magnet on a pyrochlore lattice in \cite{batistaEtAl2018}, and the pyrochlore lattice model is considered further in  Appendix \ref{appendix pyrochlore}, which is new material. The toy model itself and its continuum limit are introduced in Section \ref{Sec 3 lattice}. Topological defects in a numerical simulation of the model are compared with the rational map ansatz \cite{houghton1998rational} in the continuum field theory, and are found to roughly agree for low charge.

The other goal of this chapter is to investigate how the topological defects deform as the squashing parameter of the squashed sphere model is adjusted. This is studied numerically in the frustrated magnet model in Fig \ref{fig Q10}. It is seen that topological defects deform to knotted string-like solutions.

To better consider these solutions theoretically, a slightly different model is introduced in section \ref{Sec 3 squashed skyrme}. This model is a squashed deformation of the Skyrme model which has been considered earlier in \cite{nasir2002effective,ward2004skyrmions}. An ansatz for a string solution in this model (a `baryon string') is presented in the later part of  \ref{Sec 3 squashed skyrme}, and this section contains some additional material on the strings in the non-trivial $\gamma>0$ case compared to the published version of \cite{schubring2021skyrmeHopf}.

Chapter \ref{Sec 4} serves as a bridge between the main results in chapter \ref{Sec 3} and chapter \ref{Sec 5} which are otherwise rather disconnected. The renormalization group (RG) flow for the squashed sphere model as an effective theory for 3D frustrated magnets was considered in \cite{azaria1995massive} both in terms of the weak coupling expansion and the large $N$ expansion. But these two methods appear to have a discrepancy, and it is implied in \cite{azaria1995massive} that the large $N$ expansion is only valid without further modification in the limit of $\gamma$ very small (near the $CP^{N-1}$ model). This issue is reconsidered and it is shown in Sec \ref{Sec 4.2} how the real source of the discrepancy is the well-known scheme dependence of a single parameter beta function at three loops and higher. All scheme independent terms agree between the two approaches.
	
	The large $N$ approach to the squashed sphere model is introduced in Sec \ref{Sec 4.0} and considered further in \ref{Sec 4.1} for the special case of $d=1$. Beginning with $d=1$ allows us to show that results from the large $N$ approach agree with exact results found from wave function methods which are discussed in Appendix \ref{appendix wavefunction}, so it gives strong evidence there is nothing intrinsically wrong with the large $N$ approach for the squashed sphere model. This section also serves as an example from my papers \cite{schubring2021lessons, schubring2021yukawa} on using ordinary quantum mechanical systems as a pedagogical tool to introduce quantum field theory methods.
	
	 On the other hand, the intrinsic validity of the two-loop weak coupling expansion is also shown in Sec \ref{Sec 4.3} by introducing a Wess-Zumino (WZ) term \cite{wess1971consequences,novikov1982hamiltonian} in the model. With the addition of this term the model has a Wess-Zumino-Witten (WZW) fixed point \cite{Witten:1983ar,knizhnik1984current} in the infrared, and it is shown that the weak coupling expansion agrees with exact results coming from CFT methods. This model of an $N=2$ squashed sphere with a WZ term is a well studied model in the context of integrable deformations \cite{kawaguchi2011yangian,kawaguchi2014deformation,demulder:2017zhz}, and this section is based on our paper \cite{schubring2021sigma}.
	 
	 Besides the agreement of the RG flow with the CFT, section \ref{Sec 4.3} also gives strong evidence for the existence of a parameter restriction of the model which is preserved by the RG flow to all orders, and also raises the possibility that there may be additional fixed points in the PCM with a WZ term besides the obvious WZW one.
	 
	 Chapter \ref{Sec 5} continues the investigation of the weak coupling expansion in the squashed sphere model by considering renormalons in the perturbative expansion of the spinon correlation function. On the one hand this chapter provides a new illustration of the operator product expansion approach championed by my advisor and his collaborators long ago \cite{shifman1979qcd,novikov1984two,Novikov:1984rf}. On the other hand it may have implications for the resurgence program popular in the last decade investigating renormalons in closely related non-linear sigma models, albeit compactified on a cylinder \cite{dunne:2012ae,chermanEtAl2013,dunne:2015ywa,demulder2016resurgence}. This chapter is new material, although it follows the same approach as my paper with Sheu and Shifman on the SUSY $O(N)$ model \cite{schubring2021treating}.
	 
	 In section \ref{Sec 5 back field} a background field method is introduced which allows for relatively easy calculation of the OPE. The background field method maintains manifest $SU(N)$ symmetry and I originally used it in \cite{schubring2020sigma} as an alternative method to calculate the RG flow. This method makes the anomalous dimension of the spinon field particularly transparent and as a side note this is shown to be related to the Stueckelberg field $\phi$ discussed earlier in the introduction. In the latter part of section \ref{Sec 5 Large N} the background field method will be considered in the large $N$ limit and it is shown how the lowest order propagator $(p^2+m^2)^{-1}$ arises from VEVs of the background field.
	 
	 In section \ref{Sec 5 subleading} the $\order{1/N}$ corrections to the spinon correlation function are considered in the background field approach. The part of the OPE involving subleading operator VEVs such as $\langle F^{\mu\nu}F_{\mu\nu} \rangle$ may be fully calculated in this approach, but the part involving subleading corrections coming from the perturbation series for the coefficient functions is a bit more subtle. The details of the calculation of the coefficient functions in this section are essentially the same as those presented in the first appendix of \cite{schubring2021treating}, although the interpretation is somewhat different.
	 
	 In section \ref{Sec.5 ambiguity operator} we begin to move away from the interpretation of the OPE as having a strict factorization scale, and start to consider renormalon ambiguities. As stressed by David \cite{dave1984237,david1986operator}, the operator VEVs in the OPE themselves have ambiguities which cancel with renormalons in the coefficient functions. In this section the ambiguities of the operator VEVs are calculated via two approaches: an analytic approach which is limited to the integrable parameter restriction of the squashed sphere model, and an approach involving a UV cutoff which works for any value of the squashing parameter. The UV cutoff approach is developed in more detail than has been done previously, and it suggests that previous applications of the method may need to be reconsidered \cite{ishikawa2020infrared}.
	 
	 Finally in section \ref{Sec 5 mellin} and its accompanying appendix \ref{appendix asymptotic} the ambiguities of the full OPE are shown to cancel to all orders following the method of \cite{beneke1998308} involving a Mellin transform. Compared to similar calculations in the earlier works \cite{beneke1998308} and \cite{schubring2021treating}, some additional subtleties of the Mellin transform in the context of the OPE are discussed and the operator VEVs of the expansion are identified much more clearly.
	 
	 A second summary of the results of this dissertation is given in the conclusion, chapter \ref{Sec 6}, which focuses more on the ideas this work raises for future research.
	 
	\chapter{Lagrangian and geometry of the target space} \label{Sec 2}
	In the form that we have so far introduced the squashed sphere model \eqref{Lagr squashed sphere A psi} the geometry of the target space of the sigma model is a little obscure. Gauge fixing $\phi$ to vanish and integrating out $A$ we get a Lagrangian in the form
	\begin{align}
			\Lagr = \frac{1}{2g^2}\left[\partial \bar{z}\cdot\partial z+\beta\left(\bar{z}\cdot \partial z\right)^2\right], \label{Lagr squashed sphere standard}
	\end{align}
	where $\beta\equiv\left(1+Ng^2\gamma\right)^{-1}$. If we express the constrained field $z$ in terms of $2N-1$ real unconstrained coordinate fields on the target space $\varphi$, this may be written in the schematic form
	\begin{align}
		\Lagr=\frac{1}{2g^2}\delta^{\mu\nu}G_{\alpha\beta}(\varphi)\partial_\mu\varphi^{\alpha}\partial_\nu\varphi^{\beta}., \label{Lagr generic sigma model}
	\end{align}
	This is a standard form for a generic sigma model \cite{meetz1969realization,ECKER1971481,friedan1980nonlinear}, and $G_{\alpha\beta}$ is the metric of the target space. Given a particular choice of coordinate system $\varphi$, we can do a straightforward calculation to rewrite \eqref{Lagr squashed sphere standard} in the generic form \eqref{Lagr generic sigma model} and simply read off the components of the metric in that coordinate system. The results for the metric and other geometric quantities for a coordinate system extending the standard coordinates on $CP^{N-1}$ are presented in Appendix \ref{appendix coordinates}.
	
	Part of the reason explicit knowledge of the metric is useful is that in $d=2$ the model is renormalizable, and under a change of renormalization group (RG) scale $\mu$ the metric $G$ itself will change. The beta function for $G$ in a general sigma model has been calculated in terms of quantities like the Riemann curvature tensor \cite{friedan1980nonlinear}, and we can use these existing results to find beta functions for the parameters $g$ and $\beta$. We have not yet shown that the squashed sphere model itself is stable under the RG flow, i.e. that it can be described by only the two parameters $g, \beta$ for all $\mu$. But as we proceed it will become clear why this must be the case, and the two-loop RG equations will also bear this out.
	
	\section{The squashed sphere as a homogeneous space}\label{Sec 2.1}
	
	It will be useful to put the squashed sphere sigma model \eqref{Lagr squashed sphere standard} in yet another form which emphasizes the fact that it is a homogeneous manifold $SU(N)/SU(N-1)$. Given a standard `vacuum' value $z_0\in S^{2N-1}$, which we will concretely take to be the unit vector with $1$ in the last component, then we may choose a $U\in SU(N)$ that satisfies $U z_0 = z$.
	
	Of course this condition does not uniquely specify $U$. We may multiply $U$ on the right by a spacetime dependent matrix $V\in H$, where $H=SU(N-1)$ refers to the stabilizer of $z_0$, \begin{align}
		UV z_0 = U z_0 =z.\label{Def U}
	\end{align}
	If we rewrite the Lagrangian in terms of $U$, we would see a natural $H$ gauge invariance, and more mathematically this condition may be thought of as describing a projection map of a fiber bundle $\pi: SU(N)\rightarrow S^{2N-1}$,
		\begin{align}
		\pi(U)\equiv U z_0. \label{Def proj map}
	\end{align}
	In the other direction, our particular choice of $U$ corresponding to $z$ may be described by a section $\sigma$, which will not be globally defined on all of $S^{2N-1}$,
	\begin{align}
		\sigma(z)=U, \qquad \pi\circ \sigma = I. \label{Def section}
	\end{align}

	We can use the maps $\pi$ and $\sigma$ to push-forward and pull-back objects between $SU(N)$ and the squashed sphere. In particular the metric on the squashed sphere may be pulled back to a degenerate metric $\pi^* G$ on $SU(N)$, and this may be easier to work with to find quantities such as the curvature tensor. We may think of $\sigma$ as mapping a patch of the squashed sphere to a submanifold in $SU(N)$ and then use a variant of the Gauss equation for submanifolds (see e.g. \cite{do1992riemannian}) to find the Riemann tensor on the squashed sphere itself. This was the approach taken in \cite{schubring2020sigma} and an abbreviated account is given in Appendix \ref{appendix degenerate}.
	
	For this approach to make sense, the degenerate metric $\bar{G}$ on $SU(N)$ must be such that the induced metric on the submanifold $\sigma^* \bar{G}$ is independent of the choice of section $\sigma$. This condition is automatically satisfied if we start with a metric $G$ and set $\bar{G}=\pi^* G$, but it is a quite restrictive condition on metrics $\bar{G}$ in general. This condition is another way to state the notion of \emph{gauge invariance} under right multiplication by $V$ mentioned above. Somewhat heuristically, this requirement of gauge invariance together with invariance under the $SU(N)$ left translation group is preserved by the metric $\bar{G}$ throughout the RG flow, and this is what ensures the stability of the squashed sphere.\footnote{The question of stability under the RG flow for general homogeneous sigma models was carefully considered in \cite{becchi1988renormalizability}.}
	
	To make this $SU(N)$ description of the squashed sphere more concrete note that the condition \eqref{Def U} fixes the last column of $U$ to be the vector $z$. The other column vectors $e_a$, may be freely chosen up to the constraint that $U$ be a unitary matrix.
\begin{align}
	U=	\left(\begin{array}{ccccc}
		e_1 & e_2 &\dots & e_{N-1} & z
	\end{array}\right).\label {Def U e}
\end{align}			
The columns of $U$ may be used as a basis to expand the derivatives $\partial z$. This allows us to write \eqref{Lagr squashed sphere standard} as
\begin{align}
	\Lagr&= \frac{1}{2g^2}\left[\sum_a |\bar{e}_{a}\cdot \partial z|^2+(1-\beta)|\bar{z}\cdot\partial z|^2\right].\label{Lagr squashed sphere e}
\end{align}	
These terms are related to the left-invariant Lie algebra element $J_\mu$,
\begin{align}
	J_\mu \equiv -i U^{-1}\partial_\mu U. \label{Def J}
\end{align}
The components of $J$ may be written using \eqref{Def U e},
\begin{align}
	J = -i\left(\begin{array}{cc}
		\bar{e}_a\cdot\partial e_b &	\bar{e}_a\cdot\partial z \\[2mm]
			\bar{z}\cdot\partial e_b & \bar{z}\cdot \partial z
	\end{array}\right).\label{Def J explicit}
\end{align}	
These components may be extracted by expanding $J=J^i \tau^i$ in a basis of the Lie algebra generators $\tau^i$, where we use the convention $\text{Tr}\left(\tau^i\tau^j\right)=2\delta^{ij}$. We can decompose the Lie algebra $\mathfrak{g}$ of $SU(N)$ as
\begin{align*}
	\mathfrak{g}=\mathfrak{h}+\mathfrak{a}+\mathfrak{b},
\end{align*}
	where $\mathfrak{h}$ is associated to the gauged $SU(N-1)$ space, $\mathfrak{b}$ is the set of generators that commute with all of $\mathfrak{h}$, and $\mathfrak{a}$ includes the remaining generators. In the simple cases we consider here $\mathfrak{b}$ only contains one element which we call the \emph{phase} generator, and we reserve the Lie algebra index $0$ for it in the general case.
	
	In this notation the Lagrangian \eqref{Lagr squashed sphere e} becomes
	\begin{align}
		\Lagr&= \frac{1}{2g^2}\left[\sum_{i\in \mathfrak{a}}\left(J^i\right)^2+(1-\beta)\frac{2(N-1)}{N}\left(J^0\right)^2\right].\label{Lagr squashed sphere J}
	\end{align}	
And given the general form of the sigma model \eqref{Lagr generic sigma model}, we can use this to determine the components of the metric in the non-coordinate basis on the squashed sphere given by pushing forward left-invariant vector fields on $SU(N)$ using $\pi_*$.
\begin{align}
	G_{ij}=\delta_{ij}C_i \qquad C_{i\in \mathfrak{a}}=1,\quad C_0 = (1-\beta)\frac{2(N-1)}{N}.\label{Eq.2 metric}
\end{align}
where no sum is implied and $C_i$ vanishes for $i\in \mathfrak{h}$.
	
\section{Grassmannian generalization}\label{Sec 2.2}

This whole construction	may be generalized to homogeneous spaces with gauge group $H=SU(L)\times SU(M)$ and $L+M=N$. The large $N$ limit of this class of sigma models was considered in \cite{bratchikov19881}, but this is still a relatively obscure model. In $d=2$ it is just equivalent to a massless fermion minimally coupled to the $U(1)$ gauge field of a Grassmannian sigma model, and in this sense it may be integrable for certain values of parameters \cite{abdalla1986integrable} much like the squashed sphere model.

The complex Grassmannian $Gr(M,N)$ is the space of all $M$-dimensional linear subspaces of $\mathcal{C}^{N}$. Since each linear subspace has an $N-M=L$-dimensional orthogonal complement this can equally well be described by the dual space $Gr(L,N)$. The duality between $L$ and $M$ will eventually be manifest, but for the moment consider a representation of the $M$-dimensional linear subspaces in terms an orthonormal basis of $M$ linearly independent $N$-dimensional column vectors, $z^i_a$. The index $i$  runs from $1$ to $N$ and the index $a$ runs from $1$ to $M$. This can be thought of as a rectangular matrix with $M$ column vectors,
$$z=\left(z_0, \cdots, z_M\right).$$
Since a change of basis does not change the linear subspace, there should be an equivalence relation under multiplying $z$ on the right by a unitary matrix $V_M\in U(M)$. This equivalence relation may be implemented by an auxiliary $U(M)$ gauge field in the Lagrangian, much as in the $M=1$ $CP^{N-1}$ case \eqref{Lagr cpN A}.

The Lagrangian is defined in terms of the auxiliary $U(M)$ gauge field $A$,
\begin{align}\Lagr_{Gr} = \frac{1}{2g^2}\text{Tr}_M\left[\left(\partial_\mu z^\dagger +i A_\mu z^\dagger \right)\left(\partial_\mu z -i z A_\mu\right)\right].\label{Lagr Grass z}\end{align}

As in the squashed sphere case, this may be lifted to $SU(N)$ by choosing a standard $z_0$ and $U\in SU(N)$ satisfying the condition,
$$U z_0 = z,\qquad z_0\equiv\left(\begin{array}{c}
	0_{L\times M}\\
	I_{M}
\end{array}\right).$$
Here $	I_{M}$ is the $M\times M$ identity matrix, and $0_{L\times M}$ are extra zeros to fill out the full $N\times M$ matrix.

This defining condition on $U$ fixes the last $M$ columns to be $z_a$, but there is still an $SU(L)$ gauge freedom in choosing the first $L$ columns $e_a$.
$$U=\left(e_1, \dots, e_L, z_1, \dots, z_M\right).$$

Combined with the natural equivalence relation on $z$ this leads to an $SU(L)\times SU(M)\times U(1)$ gauge group, although we will break the $U(1)$ part shortly. The Lie algebra may again be decomposed as $\mathfrak{h}+\mathfrak{a}+\mathfrak{b}$, with $\mathfrak{b}$ still containing only one diagonal phase generator that commutes with the block diagonal matrices in $\mathfrak{h}$. Repeating everything as in the previous section, the Grassmannian Lagrangian is just
\begin{align}
	\Lagr_{Gr}&=\frac{1}{2g^2}\sum_{i\in\mathfrak{a}}\left(J_\mu^{i}\right)^2.\label{Lagr Grass J}
\end{align}
Clearly we may reverse the steps that led to this point in order to recover the equivalent of \eqref{Lagr Grass z} in terms of $e$ rather than $z$, and that is how the duality between $L$ and $M$ is seen.

The Grassmannian model may now be deformed by adding a squashing term
	\begin{align}
	\Lagr&= \frac{1}{2g^2}\left[\sum_{i\in \mathfrak{a}}\left(J^i\right)^2+(1-\beta)\frac{2(\mathcal{N}-1)}{\mathcal{N}}\left(J^0\right)^2\right],\label{Lagr squashed Grass}
\end{align}	
where by definition,
\begin{align}
	\mathcal{N}&\equiv LM+1.
\end{align}
The unusual factor multiplying the squashing term is chosen for later convenience. Note that $\mathcal{N}=N$ when $M=1$ and $L=N-1$, and so this normalization agrees with \eqref{Lagr squashed sphere J}. Given this Lagrangian the metric is just the equivalent of \eqref{Eq.2 metric} with $N$ replaced by $\mathcal{N}$.

\section{Riemann tensor and 2-loop beta function}\label{Sec 2.3}
	  Given the metric in terms of the left-invariant vector field basis, it can be lifted to $SU(N)$ and the Riemann tensor may be calculated there. The mathematical validity of this approach is discussed in Appendix \ref{appendix degenerate}. The practical details were inspired by a calculation in \cite{milnor1976curvatures} for non-degenerate metrics.
	  
	  The key simplification that takes place upon lifting to the Lie group is that the Lie bracket of the left-invariant vector fields $\tau_i$ just involve the same structure coefficients $f_{ijk}$ as the Lie algebra
	  \begin{align}
	  	[\tau_i,\tau_j]&=2i\sum_k f_{ijk}\tau_k.\\
	  	[\tau_i,\tau_j]_L&\equiv \bar\nabla_{\tau_i}\tau_j-\bar\nabla_{\tau_j}\tau_i=-2\sum_k f_{ijk}\tau_k.\label{Def tauBracket}
	  \end{align}
	  In the top line $\tau_a$ are considered to be Hermitian matrices with the normalization $\text{Tr}(\tau_i\tau_j)=2\delta_{ij}$. In the bottom line they are considered as vector fields. We must be slightly careful to distinguish the two brackets since we are using the physicist's convention where $\tau_i$ are Hermitian rather than anti-Hermitian. %Following the notation in Appendix \ref{appendix degenerate} the covariant derivative is barred since it is properly understood as a non-unique metric compatible and torsion free connection associated to the degenerate metric $\bar{G}$.
	  
	  To begin recall that metric is diagonal in our choice of basis,
	  \begin{align}
	  	\bar{G}(\tau_i,\tau_j)=\frac{1}{g^2}C_i \delta_{ij}.\label{Eq.2 metricLI}
	  \end{align}
	  
	  Since the metric is constant in this basis, by metric compatibility
	  \begin{align*}
	  	\bar{G}(\bar{\nabla}_{\tau_i} \tau_j,\tau_k)+\bar{G}(\tau_j,\bar{\nabla}_{\tau_i} \tau_k)=0.
	  \end{align*}
	  Then by repeatedly using the vanishing torsion condition, $\bar{\nabla}_X Y =[X,Y]+\bar{\nabla}_Y X$, we can derive a relation in terms of Lie brackets, which we then can write in terms of structure coefficients \eqref{Def tauBracket} and metric components \eqref{Eq.2 metricLI},
	  \begin{align}
	  	\bar{G}(\bar{\nabla}_{\tau_i} \tau_j,\tau_k)&=\frac{1}{2}\left[\,\,\bar{G}([\tau_i,\tau_j]_L,\tau_k)-\bar{G}([\tau_j,\tau_k]_L,\tau_i)+\bar{G}([\tau_k,\tau_i]_L,\tau_j)\,\,\right]\non
	  	&=-\frac{1}{g^2}f_{ijk}\left(C_k-C_i+C_j\right).
	  \end{align}
	  Note that we used the fact that the structure coefficients are completely antisymmetric in our choice of Lie algebra basis. Also no summation convention over repeated Lie algebra indices is implied in this section.
	  
	  The connection must respect this equation and also the torsion-free condition, which implies
	  \begin{align}
	  	\bar{\nabla}_{\tau_i} \tau_j=-\sum_k f_{ijk}C_{ij;k}\tau_k + \eta_{ij}.\label{connectionExplicit}
	  \end{align}
  The coefficients are defined as
  \begin{align}
  	C_{ij;k}\equiv 1-\frac{C_i-C_j}{C_k}\chi_k,
  \end{align} 
with $\chi_k$ being an indicator function which is $0$ if $C_k=0$ and $1$ otherwise. The term $\eta_{ij}$ is an arbitrary set of vectors satisfying $\pi_* \eta_{ij}=0$ and which are symmetric under permutation of $i, j$. This explicitly shows the non-uniqueness of the connection for degenerate metrics mentioned in Appendix \ref{appendix degenerate}. By the lemma on the Riemann tensor \eqref{Eq.AppA riemannTensorEquality}, the choice of $\eta$ will not affect our calculation of the Riemann tensor on the base space, and so in the following we will simply take $\eta=0$.  
	  
	  Now the components of the Riemann tensor may be calculated
	  \begin{align}
	  R^l_{\,\,ijk}&=\hat\tau^l\left(\left[\bar\nabla_{[\tau_i,\tau_j]_L}-\bar\nabla_{\tau_i}\bar\nabla_{\tau_j}+\bar\nabla_{\tau_j}\bar\nabla_{\tau_i}\right]\tau_k\right)\non
	  &=\sum_m(2 f_{ijm}f_{klm}C_{mk;l} +f_{ikm}f_{ljm}C_{ik;m}C_{jm;l}-f_{jkm}f_{lim}C_{jk;m}C_{im;l}).\label{Eq.2 riemannAppendix}
	  \end{align}
  
  For our purposes the Riemann tensor will be useful since it is involved in the beta function for the target space metric $G_{ij}$. In a dimensional regularization scheme with $d=2+\epsilon$ the beta function to two loops is well-known \cite{friedan1980nonlinear}
  \begin{align}
  	\mu\der{}{\mu}G_{ij}=-\epsilon G_{ij}+\frac{1}{2\pi}R_{ij}+\frac{1}{8\pi^2}R_{i lmn}R_{j}^{\,\,\,lmn}.\label{Eq.2 general RG}
  \end{align}
  
   This may be calculated for the Grassmannian extension of the squashed sphere \eqref{Lagr squashed Grass} by finding the general form of the structure coefficients $f_{ijk}$ for the directions in $\mathfrak{a}+\mathfrak{b}$ and summing with the aid of some combinatorics. The result for the two independent parameters in the metric turns out to be
   \begin{gather}
   	\mu\der{}{\mu}\left(\frac{1}{g^2}\right)=-\frac{\epsilon}{g^2}+\frac{1}{\pi}(\mathcal{N}-1+\beta)\frac{N}{\mathcal{N}}+\frac{g^2}{2\pi^2}\left(4\mathcal{N}-6N\frac{N}{\mathcal{N}}(1-\beta)+(3\mathcal{N}-1)\left(\frac{N}{\mathcal{N}}\right)^2(1-\beta)^2\right),\label{Eq.2 RG 1}\\
   	\mu\der{}{\mu}\left(\frac{1-\beta}{g^2}\right)=-\frac{\epsilon(1-\beta)}{g^2}+\frac{1}{\pi}(\mathcal{N}-1)\frac{N}{\mathcal{N}}(1-\beta)^2+\frac{g^2}{2\pi^2}(\mathcal{N}-1)\left(\frac{N}{\mathcal{N}}\right)^2(1-\beta)^3.\label{Eq.2 RG 2}
   \end{gather}

	For $\beta=1$, these equations reduce to the 2-loop beta function for the Grassmannian model \cite{brezin1980generalized}. For $\mathcal{N}=N$, they reduce to the 2-loop beta function for the squashed sphere model \cite{azaria1995massive}.
	
	Note that one curious feature of the RG flow of these manifolds is that even though the restriction $\beta=0$ is preserved to one-loop (i.e. it is an Einstein manifold), the two-loop term causes $\beta$ to run. In the squashed sphere special case ($\mathcal{N}=N$) the limit $\beta=0$ is the sphere which has extended $O(2N)$ symmetry and this ensures stability under the RG flow for higher loops. But the general case has no such guarantee.
	
	We shall consider the RG flow further in Chapter \ref{Sec 4}, in particular see Fig. \ref{Fig 4.1}. For now we will consider some special features of the target space for $N=2$ which will be more useful in Chapter \ref{Sec 3}.
	
	\section{The special case $N=2$}\label{Sec 2.4}
	
	As discussed in the introduction, the case $N=2$ has many special features which are not seen in general. From the fiber bundle perspective, the stabilizer of $z_0$ is trivial and the squashed sphere is just diffeomorphic to $SU(2)$ itself. Using the Pauli matrices $\sigma^i$ as a basis for the Lie algebra, the diagonal $\sigma^3$ is the phase generator in this case,
	\begin{align}
	\Lagr_{N=2}&= \frac{1}{2g^2}\left[\left(J^i\right)^2-\beta\left(J^3\right)^2\right].\label{Lagr squashed sphere N=2}
\end{align}		
	
	Due to the isomorphism of Lie algebras, $J^i$ is also expressible in terms of $SO(3)$ matrices. We may define $R\in SO(3)$ such that
		\begin{align}
		U^{-1}\sigma^j U = R^{j}_{\,\,k}\sigma^k.\label{Def R}
	\end{align}
	Then we can express $J^i$ through
\begin{align}
	\left(R^{-1}\partial R\right)_{ij} = 2\epsilon_{ijk}J^k.\label{Eq.2 J R}
\end{align}	
The 2-component complex $z$ may also be projected to a 3-component real unit vector $\mathbf{S}$,
\begin{align}
	S^i\equiv -\bar{z}\sigma^iz, \qquad \left(S^i\right)^2=1\label{Def S}
\end{align}
Transformations of $\mathbf{S}$ by $R$ are compatible with transformations of $z$ by $U$.

The Lagrangian \eqref{Lagr squashed sphere standard} in the $CP^1$ limit may be expressed in terms of these real fields,
	\begin{align}
	\Lagr_{\beta=1} = \frac{1}{8g^2}\left(\partial \mathbf{S}\right)^2. \label{Lagr cp1 S}
\end{align}
If $g$ is absorbed into the metric, then it may be related to the radius of the target space. From the factor above it is seen that squashed sphere interpolates between the three-sphere with radius $g^{-1}$ and two-sphere with radius $(2g)^{-1}$.

\section{$U(1)$ gauge invariance and topological charge}\label{Sec 2.5}
So far in this chapter we have been discussing the fiber bundle mapping the Lie group $SU(N)$ to the squashed sphere and its Grassmannian generalization, and there is a non-Abelian gauge invariance associated to this structure. There is another fiber bundle which arises from projecting the squashed sphere to $CP^{N-1}$, and this is associated to a $U(1)$ gauge invariance which is made explicit through the use of the auxiliary $A$ field in \eqref{Lagr cpN A} and \eqref{Lagr squashed sphere A psi}. Here we will consider this $U(1)$ gauge invariance further for the special case of $N=2$, which will be important for the following chapter.

In terms of the field $U(x)\in SU(2)$, this gauge symmetry corresponds to right multiplication by the unitary matrix $V_R(x)=\text{diag}\left(e^{i\theta(x)}, e^{-i\theta(x)}\right)$, from which it is easily shown that the $J$ currents transform as,
	\begin{align}J^3\rightarrow J^3+\partial \theta,\qquad \left(\begin{array}{c}
			J^1 \\
			J^2
		\end{array}\right)\rightarrow\left(\begin{array}{cc}
			\cos 2\theta & -\sin 2\theta\\
			\sin 2\theta & \cos 2\theta
		\end{array}\right)\left(\begin{array}{c}
			J^1 \\
			J^2
		\end{array}\right).\label{Eq.2 J gauge transf}
	\end{align}
	
	Clearly the structure in the $CP^1$ action, $\left(J^1\right)^2+\left(J^2\right)^2$, is gauge invariant since there is no explicit $J^3$ dependence and the quadratic form is invariant under rotations of $J^1, J^2$. There are two other obvious gauge invariant structures that may be constructed. The two-form $J^1_\mu J^2_\nu-J^2_\mu J^1_\nu$ is also invariant under rotations of $J^1, J^2$. And given that $J^3$ transforms like a vector potential
	\begin{align}
		A_\mu\equiv J^3_\mu,\label{Def A}
	\end{align}
	the gauge invariant field strength tensor $F_{\mu\nu}\equiv\partial_\mu A_\nu-\partial_\nu A_\mu$ may also be constructed. In fact by expressing the $J$ currents in terms of the $z$ field it can be quickly shown that these two quantities are not independent\footnote{This may also be seen by considering $J_\mu$ as an $SU(2)$ non-Abelian gauge field, and using the fact that the non-Abelian field tensor vanishes since it is pure gauge.},
	\begin{align}
		F_{\mu\nu}= \partial_\mu J^3_\nu -\partial_\nu J^3_\mu= 2\left(J^1_\mu J^2_\nu -J^2_\mu J^1_\nu\right).\label{Def F tensor}
	\end{align}
	
	Just as the gauge invariant $CP^1$ action \eqref{Lagr cp1 S} may be written entirely in terms of the real unit vector field $S^i$, so may the gauge invariant $F_{\mu\nu}$ tensor.\footnote{One relatively quick way to verify this is to use the global symmetry to choose $S^i=(0,0,1)$ and $z^\alpha=(0,z_1)$ at a point.}
	\begin{align}
		F_{\mu\nu}=\frac{1}{2}\epsilon_{ijk}S^i\partial_\mu S^j \partial_\nu S^k.\label{Eq.2 F and S}
	\end{align}
	
	Apart from the factor of $1/2$ this is essentially a Jacobian $\left|\partial_\mu S \times  \partial_\nu S\right|$ and so this two-form may be integrated over any 2D surface in spacetime to find the area spanned on the $S^2$ target space. If the surface integrated over is itself homeomorphic to $S^2$, then the result (divided by $2\pi$) will be an integer measuring the number of windings around the target space. In $d=2$ this may be considered to be the instanton charge which is involved in the topological theta term for $CP^1$, and in a condensed matter context it is known as the \emph{magnetic Skyrmion} or \emph{baby Skyrmion} charge (see e.g. \cite{Nagaosa2013,han2017skyrmions,Back2020}). But note that it is perfectly valid to consider this 2-form in the more general squashed sphere case and in $d>2$. More will be said on this point in the next chapter.
	
	Note also that by the constraint the vectors $\partial_\mu \mathbf{S}$ must be perpendicular to $\mathbf{S}$, and so there can only be two linearly independent $\partial_\mu \mathbf{S}$. This implies that $F_{\mu\nu}$ is a closed form,
	\begin{align*}
		\epsilon^{\lambda\mu\nu}\partial_\lambda F_{\mu\nu}=0.
	\end{align*}
	If $\mathbf{S}$ is defined everywhere (i.e. there are no monopoles) and takes a unique value at the boundary then our spacetime is effectively $S^d$,  and by the triviality of the de Rham cohomology class for $d>2$ $F_{\mu\nu}$ must also be exact. So even if we were lacking the extra squashed sphere structure that gives us a natural definition of $A$ \eqref{Def A} we could globally construct a gauge field.
	
	The exact same considerations allowing us to define a two-form $F$ for a target space homeomorphic to $S^2$ allow us to define a 3-form measuring the windings around the squashed sphere target space homeomorphic to $S^3$. We can define a four component real unit vector $q^a$ in terms of either $U$ or $z$,
	\begin{align}
		U=q^0I-iq^i\sigma^i,\qquad z=\left(-q^2-iq^1,\,\,q^0+iq^3\right).\label{Def q}
	\end{align}
	Then we may define a topological charge 3-form $\rho$ in the exact same way,
	\begin{align}
		\rho_{\lambda\mu\nu}=\frac{1}{2\pi^2}\epsilon_{abcd}q^a\partial_\lambda q^b\partial_\mu q^c \partial_\nu q^d.\label{Eq.2 top charge q}
	\end{align}
This time we have normalized by the total volume of $S^3$ in order to ensure integer topological charge.

The topological charge may also be expressed in terms of the $J$ fields. At $\beta=0, g^2=1$, the fields $J^i_\mu$ are essentially vielbeins for the metric on the unit 3-sphere, and a normalized volume form may be written as
\begin{align}
	\rho_{\lambda\mu\nu}&=-\frac{1}{2\pi^2}J^1_{[\lambda} J^2_\mu J^3_{\nu]}=\frac{i}{12\pi^2}\text{Tr}\left(J_{[\lambda} J_\mu J_{\nu]}\right).\label{Def top charge}
\end{align}
In the squashed case $J^3$ must be normalized by a factor of $\sqrt{1-\beta}$ in order to be a proper vielbein, but this factor will be canceled after dividing out the total volume of the squashed sphere, so it is clear that $\rho$ takes the same form for any choice of the squashed sphere parameters. This expression may also be directly reduced to the expression for $\rho$ in terms of $q$ by using the identity,
\begin{align}
	J^i_\mu = q^i\partial_\mu q^0 - q^0\partial_\mu q^i+\epsilon_{ijk}q^j\partial_\mu q^k.
\end{align}

This topological charge $\rho$ will play an essential role in the following chapter on topological defects, and it will also appear in the definition of the Wess-Zumino term in $d=2$ in Sec \ref{Sec 4.3}.
	
	\chapter{Topological defects in 3D}\label{Sec 3}
	
\section{Hopfions vs 3D Skyrmions}\label{Sec 3.1}

In the previous section we have introduced a 2-form $F_{\mu\nu}$ in the $CP^1$ model which measures topological charge associated to objects known as magnetic or baby Skyrmions. The study of magnetic Skyrmions in 2D has attracted a huge amount of experimental and theoretical interest (e.g. \cite{Nagaosa2013,han2017skyrmions,Back2020}), and the study of corresponding objects in 3D has also increasingly been attracting attention (e.g. \cite{milde2013unwinding,seki2021direct,sutcliffe2017skyrmion,kent2021creation}).

In 3D, the magnetic Skyrmions are extended in the third dimension to form stringy topological defects, and the $F_{\mu\nu}$ tensor measures the flux of these strings across a 2D surface. In the absence of monopoles\footnote{However monopoles generically are expected to be relevant in chiral magnets \cite{milde2013unwinding}.} the flux across any closed surface vanishes\footnote{Networks of topological defects such as this have some close analogies to fluids coupled to an electromagnetic field \cite{schubring2015field,schubring2015dissipative}.} and it may seem that there is nothing that might prevent a loop of string from decaying to the vacuum.

However a $U(1)$ Chern-Simons 3-Form $A\wedge F$ may be constructed from the tensor $F$, and its corresponding gauge field $A$,
\begin{align}
	Q= -\frac{1}{8\pi^2}\int d^3x \,\epsilon^{\lambda\mu\nu}A_\lambda F_{\mu\nu}. \label{def Q Hopf}
\end{align}
This is just an integral for what is known as the Hopf charge $Q$~\cite{whitehead1947expression}, and topological defects with non-zero Hopf charge are referred to as \emph{Hopfions}. A Hopfion may be understood as magnetic Skyrmion string that is twisted or knotted in a topologically non-trivial way before being reconnected to form a loop,\footnote{This notion can be made precise through the definition of the quantities \emph{twist} and \emph{writhe}, see e.g. \cite{moffatt1995helicity,moffatt2014helicity} in the fluid context.} and recently a twisted Hopfion has been realized experimentally \cite{kent2021creation}. The Hopf charge is very closely analogous to the notion of \emph{helicity} in fluids, where $A$ plays the role of the fluid velocity and $F$ the vorticity. In this sense knotted `Hopfions' have been constructed experimentally in fluids as well \cite{kleckner2013creation}.

On the theoretical and numerical side by far the most studied model involving Hopfions is the model of Faddeev and Niemi \cite{faddeev1976some,faddeev1997stable,faddeev1976some,faddeev1997toroidal,battye1998knots,sutcliffe2007knots}, which adds an $F^2$ `kinetic term' for the gauge field in the $CP^1$ action in order to stabilize the Hopfions. The issue of energetic stability is important. A model that naively looks very similar to the Faddeev-Niemi model is the two-component superconductor model \cite{babaev2002hidden} where the gauge field is an independent degree of freedom that is not directly expressed in terms of the $\mathbf{S}$ fields. But although field configurations with non-vanishing Hopf charge can be constructed, there is nothing to prevent the solutions from shrinking to a size beyond the limits of applicability of the field theory and decaying \cite{jaykka2011supercurrent}.\footnote{With the inclusion of an additional current-current interaction this model may be stabilized \cite{rybakov2019stable}.}

In much work on magnetic Skyrmions the stabilization comes from a Dzyaloshinskii-Moriya interaction~\cite{Dzyaloshinski1960OnTM}, but more recently there has been a lot of interest in topological spin textures in inversion-symmetric magnets~\cite{okubo2012multiple,leonov2015multiply,lin2016ginzburg,Wang2020} where the magnetic Skyrmions are stabilized by competing magnetic interactions. The 3D lattice model we will consider in Sec \ref{Sec 3 lattice} is a frustrated spin system of this type, and in one limit is very close to an effective theory studied by Sutcliffe~\cite{sutcliffe2017skyrmion} as a medium for Hopfions.

The model presented here differs from the frustrated spin systems above in that the spins in the ground state are non-collinear, and the effective field theory of the system is described by a squashed sphere model where the squashing parameter $\beta$ is associated to the degree of collinearity in the underlying lattice model. This is very similar to how the squashed sphere model was originally introduced to the condensed matter community through the work of Dombre and Read \cite{dombre1989nonlinear} on triangular antiferromagnets with a non-collinear ground state. A more recent 3D example deals with frustrated spin systems on the pyrochlore lattice \cite{batistaEtAl2018}, and there it was suggested that the model should support topological defects much like the \emph{3D Skyrmions} in the Skyrme model \cite{skyrme1961non,WITTEN1983433,adkins1983static}, which have topological charge measured by the three-form in \eqref{Def top charge}.

In Appendix \eqref{appendix pyrochlore} I will extend the results on the pyrochlore lattice model in \cite{batistaEtAl2018} slightly by explicitly deriving the additional stabilizing terms in the effective field theory and showing that Skyrmions in the hedgehog ansatz are stable. The model will be modified somewhat from \cite{batistaEtAl2018} by imposing a hard constraint to focus only on the $SO(3)$ degree of freedom and also by allowing the degree of collinearity to be a tunable parameter so that the effective theory becomes a squashed sphere model rather than simply a PCM.\footnote{Strictly speaking the effective theory of the model in \cite{batistaEtAl2018} is not the PCM since there is always has some cubic anisotropy for any value of the parameters, but the anisotropy can be eliminated by fine tuning an additional higher-order neighbor interaction.} But this model can be abstracted even further, and the lattice model of Sec \ref{Sec 3 lattice} which was our main focus in \cite{schubring2021skyrmeHopf} arose as an attempt to construct a simple toy model which nevertheless demonstrates features which should be common to generic 3D spin systems with $SO(3)$ Goldstone modes.

\subsection{Relation between the Hopf and Skyrme charges}

Since we will be studying topological defects in the $N=2$ squashed sphere model, a natural question that arises is what happens to the topological defects in the limit $\beta\rightarrow 1$, in which the target space goes to $CP^1$ rather than something homeomorphic to $S^3$? Strictly at $\beta=1$ we expect that the model has Hopfions, and in fact in this limit the lattice model of Sec \ref{Sec 3 lattice} will be a minor variation of the model \cite{sutcliffe2017skyrmion} where the Hopfions were demonstrated numerically. But in the opposite limit $\beta=0$ the model looks much like the Skyrme model and this should have 3D Skyrmions.

The resolution is simply that the Hopfions and 3D Skyrmions are the exact same objects. This may be seen by simply substituting \eqref{Def A} for the $A$ field and \eqref{Def F tensor} for the $F$ field into \eqref{def Q Hopf} for the Hopf charge. The result is the same as \eqref{Def top charge} for the 3D Skyrme charge,
\begin{align}
	Q= -\frac{1}{2\pi^2}\int d^3x \,\epsilon^{\lambda\mu\nu}J^1_\lambda J^2_\mu J^3_\nu. \label{def Q Skyrme}
\end{align}

This is the main point that we wish to stress, \emph{a Hopfion may be considered to be a three-dimensional Skyrmion, and vice-versa}. The $z$ field description of a 3D Skyrmion may be directly mapped to the $\mathbf{S}$ field using \eqref{Def S}, and the result will have Hopf charge equal to its original 3D Skyrme charge. On the other hand, a description of a Hopfion involves a map to a $S^2$ target space which may always be globally lifted to a map to the $S^3$ target space. This is maybe not so obvious that a lift to $S^3$ is always possible for an arbitrary Hopfion which may be twisted and knotted in complicated ways, but it follows from the homotopy lifting property for the Hopf fibration (see e.g. proposition 4.48 in \cite{Hatcher:478079}).\footnote{Another approach to understanding this is through the fact that a gauge field $A$ may be globally defined given the closed two-form $F$ on spatial $S^3$, and if $A$ is identified with $J^3$ it may be integrated to construct $U$.} Of course this lift is not unique, which is a reflection of the $U(1)$ gauge symmetry, but every such lift will have a 3D Skyrme charge which is equal to the original Hopf charge.

So the picture I wish to present is that just as Hopfions are usually considered to be loops of stringy topological defects, the ordinary 3D Skyrmions in the Skyrme model may also be considered in this respect. The difference between the topological defects at $\beta<1$ and $\beta=1$ is not one of topology, rather it is a question of whether the $A=J^3$ field contributes to the energy or not, and the difference is very much like difference between the global strings of the XY model versus the local strings of the Abelian Higgs model. To discuss this more fully, in Sec \ref{Sec 3 squashed skyrme} I will first focus on a model \cite{nasir2002effective,ward2004skyrmions} that has slightly nicer higher derivative stabilizing terms than the effective field theory for the lattice model in Sec \ref{Sec 3 lattice}.

\subsection{Additional notions of topological charge}

To conclude this section, let us briefly discuss some other notions of topological charge appearing in the Skyrme model, and their relation to the ideas discussed above. First of all, note that the equivalence between the 3D Skyrme charge and Hopf charge itself is not a new idea, it is clearly discussed in~\cite{radu2008stationary} and~\cite{han2017skyrmions} for instance, and it also underlies Ward's treatment \cite{ward2004skyrmions} of the squashed sphere model which will be introduced in the next section.

However this idea is distinct from a completely different notion of Hopfions in the Skyrme model~\cite{meissner1985toroidal,cho2001monopoles,Cho2008}, where a field configuration $U$ is restricted to only take values in a subspace $CP^1\subset SU(2)$. In that case since the $U$ field does not cover $SU(2)$ the Skyrme charge vanishes, but a different notion of Hopf charge\footnote{The $F$ tensor for this second notion of Hopf charge is defined in terms of the unit vector $n$ which is considered in the context of rational maps in Sec.~\ref{Sec 3 lattice} rather than $S$.} may still be defined in terms of the $CP^1$ subset. This construction is rather similar to the idea in $d=2$ of embedding $CP^1$ instantons into the $O(4)$ model as a means to study resurgence \cite{dunne:2015ywa}.

Finally, note that there is a third way in which the charge $Q$ may be understood which is more familiar from Yang-Mills theory. The current $J_\mu$ in \eqref{Def J} may also be understood as a non-Abelian gauge field associated to gauge symmetry under right multiplication by $SU(2)$ matrices. It is pure gauge and the non-Abelian field strength tensor vanishes. If we consider the Chern-Simons three-form associated to this \emph{non-Abelian} gauge symmetry (as opposed to the Abelian gauge symmetry involved in the definition of the Hopf charge), we have the charge
$$\mathcal{K}=-\frac{1}{16\pi^2}\int d^3 x \, \epsilon^{\lambda\mu\nu}\left(J^k_\lambda \partial_\mu J^k_\nu + \frac{1}{3}\epsilon_{ijk}J_\lambda^iJ_\mu^jJ_\nu^k\right).$$
Rewriting $\partial_{[\mu}J^k_{\nu]}$ as a product of two gauge fields in a manner similar to \eqref{Def F tensor} we see that this just reduces to the expression for the baryon charge \eqref{def Q Skyrme}, so $\mathcal{K}=Q$. This non-Abelian Chern-Simons charge $\mathcal{K}$ is interesting in 4D Yang-Mills theory because 4D instantons can be understood as interpolating between 3D vacua with different values of $\mathcal{K}$~\cite{jackiw1976vacuum,CALLAN1976334,shifman1994instantons}.

This perspective on the charge is well-illustrated by the construction of Atiyah and Manton, where Skyrmion configurations with non-zero $Q$ are generated from an initial trivial configuration by integrating over $SU(2)$ instantons~\cite{atiyah1989skyrmions,atiyah1993geometry}. These configurations are reasonably good approximations to the minimal energy configurations in the Skyrme model. In fact the Skyrme model may be extended to a BPS model involving a whole tower of vector mesons by expanding the Yang-Mills action in $d=4$ holographically to produce an energy functional in $d=3$ \cite{sutcliffe2010skyrmions}. Even the truncation of the BPS model to the lowest vector meson already improves the description of nuclei compared to the Skyrme model proper~\cite{naya2018skyrmions}.

\section{The squashed Skyrme model and baryon strings}\label{Sec 3 squashed skyrme}

		So far we have have been discussing the squashed sphere sigma model which involves a Lagrangian quadratic in derivatives, but due to Derrick's theorem~\cite{derrickstheorem} higher order terms are necessary to stabilize the topological defects. The Lagrangian of the full Skyrme model~\cite{skyrme1961non} is
\begin{align}
	\Lagr_{Skyrme}=\frac{f_\pi^2}{2}\left[\left(J^k_\lambda\right)^2 + \frac{1}{2M^2}\epsilon_{ijk}\epsilon_{klm}J_\mu^iJ_\nu^jJ_\mu^lJ_\nu^m\right],
\end{align}
where $f_\pi$ has dimensions of energy and is equivalent to $g^{-1}$ in our earlier notation, and $M$ is some new dimensionful parameter often written as $ef_\pi$. Using \eqref{def Q Skyrme}, the energy may be written as,
\begin{gather*}
	E_{Skyrme}=\int d^3x\,\Lagr_{Skyrme}=E_{BPS}|Q|+	\frac{f_\pi^2}{2}\int d^3x\,\left(J^k_\lambda \pm \frac{1}{2M}\epsilon_{ijk}\epsilon^{\lambda\mu\nu}J_\mu^iJ_\nu^j\right)^2,\non\qquad E_{BPS}\equiv 6\pi^2 \frac{f_\pi^2}{M}.
\end{gather*}
This form of the Skyrme energy functional clearly shows the BPS bound $E\geq  |Q|E_{BPS}$. This expression for the energy functional may easily be generalized to the squashed sphere case,
\begin{align*}
	E_{BPS}|Q|+	\frac{f_\pi^2}{2}\int d^3x\left[\left(J^a_\lambda \pm \frac{1}{2M}\epsilon_{ija}\epsilon^{\lambda\mu\nu}J_\mu^iJ_\nu^j\right)^2+(1-\beta)\left(J^3_\lambda \pm \frac{1}{2M(1-\beta)}\epsilon_{ij3}\epsilon^{\lambda\mu\nu}J_\mu^iJ_\nu^j\right)^2\right],
\end{align*}
where Latin indices from the beginning of the alphabet like $a$ are only summed over $1,2$. Note that the $\epsilon_{ij3}J^iJ^j$ expression in the $\beta$ dependent term is proportional to the $F$ tensor \eqref{Def F tensor} defined above. Expanding the squares leads to the Lagrangian,
\begin{align}
	\Lagr=\frac{f_\pi^2}{2}\left[\left(J^k_\lambda\right)^2 -\beta\left(J^3_\lambda\right)^2+ \frac{1}{2M^2}\epsilon_{ijk}\epsilon_{klm}J_\mu^iJ_\nu^jJ_\mu^lJ_\nu^m+\frac{\beta}{8M^2(1-\beta)}\left(F_{\mu\nu}\right)^2\right],\label{lagr Ward}
\end{align}
which also satisfies the BPS bound
\begin{align}
	E\geq 12\pi^2\frac{f^2_\pi}{2M}|Q|\equiv E_{BPS}|Q|. \label{eq BPS bound}
\end{align}
The new term quartic in derivatives is exactly that of the Faddeev-Niemi model, so squashing the target space of the Skyrme model while maintaining the BPS bound naturally leads to an interpolation between the Skyrme and Faddeev-Niemi models which we are calling the \emph{squashed Skyrme model}. This model was considered earlier by Nasir and Niemi~\cite{nasir2002effective} and Ward and Silva Lobo~\cite{ward2004skyrmions,lobo2011generalized,silva2011lattices}.

It may seem that there is a difficulty in extending to the limit $\beta=1$ due to the prefactor $(1-\beta)^{-1}$ of the Faddeev term. If $f_\pi$ and $M$ are taken fixed as $\beta$ is varied this is indeed the case. This parametrization will be referred to as the \emph{fixed bound parametrization} since the energy satisfies the BPS inequality with an energy $E_{BPS}$ that is constant with $\beta$.

But if $M^2$ is allowed to vary with $\beta$, then there is no problem taking the $\beta=1$ limit. In particular, the \emph{Ward parametrization}~\cite{ward2004skyrmions},
\begin{align*}
	\frac{f_\pi^2}{2}=\frac{1}{4\pi^2\left(3-\beta\right)},\qquad \frac{f_\pi^2}{2M^2}=\frac{1-\beta}{4\pi^2\left(3-2\beta\right)},
\end{align*}
is based on requiring that the identity map from a base space with spherical $S^3$ geometry to the $S^3$ target space has unit energy for all $\beta$, and it leads to a fairly constant dependence on $\beta$ of the energy of a $Q=1$ Skyrmion in flat space as well. No matter which parametrization for $f_\pi$ and $M$ is chosen, the results for any other parametrization may be recovered by adjusting the energy and length scales. %Table \ref{table SqSkyrme} involves a simulation in the fixed bound parametrization, but the rescaled results agree with Ward up to an error of $\sim0.1\%$ from finite size effects.

%\begin{table}
%	\begin{center}
%		\begin{tabular}{c|c|c|c|c}
%			$\beta$ & $E/E_{BPS}$ & $E_{h}/E_{BPS}$ & $E_{W}$ &  $\frac{\langle (J_\mu^1)^2 \rangle}{\langle (J_\mu^3)^2 \rangle}$ \\
%			\hline \hline
%			0.0 &1.2323 &1.2331 & 1.2323& 1.0 \\
%			0.1 &1.2339 &1.2348& 1.2324& 0.9873 \\
%			0.2 &1.2392 &1.2403 &1.2324 & 0.9737 \\
%			0.3 &1.2497 &1.2513 & 1.2322& 0.9592 \\
%			0.4 &1.2679 &1.2702 & 1.2319& 0.9439 \\
%			0.5 &1.2981 &1.3015 & 1.2315& 0.9279 \\
%			0.6 &1.3486 &1.3535 & 1.2311& 0.9103 \\
%			0.7 &1.4370 &1.4442 &1.2309 & 0.8912 \\
%			0.8 &1.6111 &1.6224 & 1.2316& 0.8695 \\
%			0.9 &2.0530 &2.0650 & 1.2269& 0.8519\\
%			
%		\end{tabular}
%		\caption{A simulation of a $Q=1$ soliton in the squashed Skyrme model. $E$ is the energy in the fixed bound parametrization. The simulation was carried out on a cubic lattice with $100^3$ sites (except for $\beta=0.9$ where the length was doubled to $200^3$) and lattice spacing $a=0.2$ in units where $M=1$. An arrested Newton flow method was used for the minimization as described in~\cite{battye2002skyrmions}, with the time evolution implemented by a fourth-order Runge-Kutta method with time step $\Delta t = 0.1$. $E_h$ is the optimal energy in the spherically symmetric hedgehog ansatz for this same parametrization. $E_W$ is the energy in the Ward parametrization which was found by rescaling $E$. To better indicate the departure from the hedgehog ansatz, the values of $\left(J^1\right)^2$ and $\left(J^3\right)^2$ are averaged over the domain of the simulation and compared.}
%		\label{table SqSkyrme}
%	\end{center}
%\end{table}

Any parametrization which allows for a well-defined $\beta=1$ limit will involve the energy $E_{BPS}$ in the BPS bound \eqref{eq BPS bound} tending to zero. This makes sense since in the Faddeev-Niemi model the minimal energy solutions obey a weaker $E\geq K Q^{3/4}$ inequality for some value of $K$~\cite{vakulenko1979stability,ward1999hopf} and moreover the minimal energy Hopfions found numerically~\cite{battye1998knots,battye1999solitons,sutcliffe2007knots} appear to come close to saturating this bound.

For $\beta$ close to but less than $1$, the energies of solitons with small values of $Q$ may be very close to the energies in the Faddeev-Niemi model, and this is not disallowed by \eqref{eq BPS bound} since the value of $E_{BPS}$ may be very small. But no matter how small $E_{BPS}$ may be, eventually for large enough $Q$, $E_{BPS}Q> K Q^{3/4}$. So for $\beta<1$ the energies of the large $Q$ solitons can not scale asymptotically as $Q^{3/4}$, and thus if the Faddeev-Niemi model indeed has this asymptotic behavior there must be a dramatic difference for large $Q$ solitons if $\beta$ is even slightly below $1$.

\subsection{Position curves}\label{sec Position curves and strings}

Intuitively a Hopfion is often described as a loop of string whereas a single Skyrmion in the Skyrme model is spherically symmetric and multiple Skyrmions form polyhedral clusters. While we have shown that the baryon charge and Hopf charge are identical, let us comment a bit more on how these two pictures are resolved.

\begin{figure}[t]
	\centering
	\includegraphics[width=\textwidth]{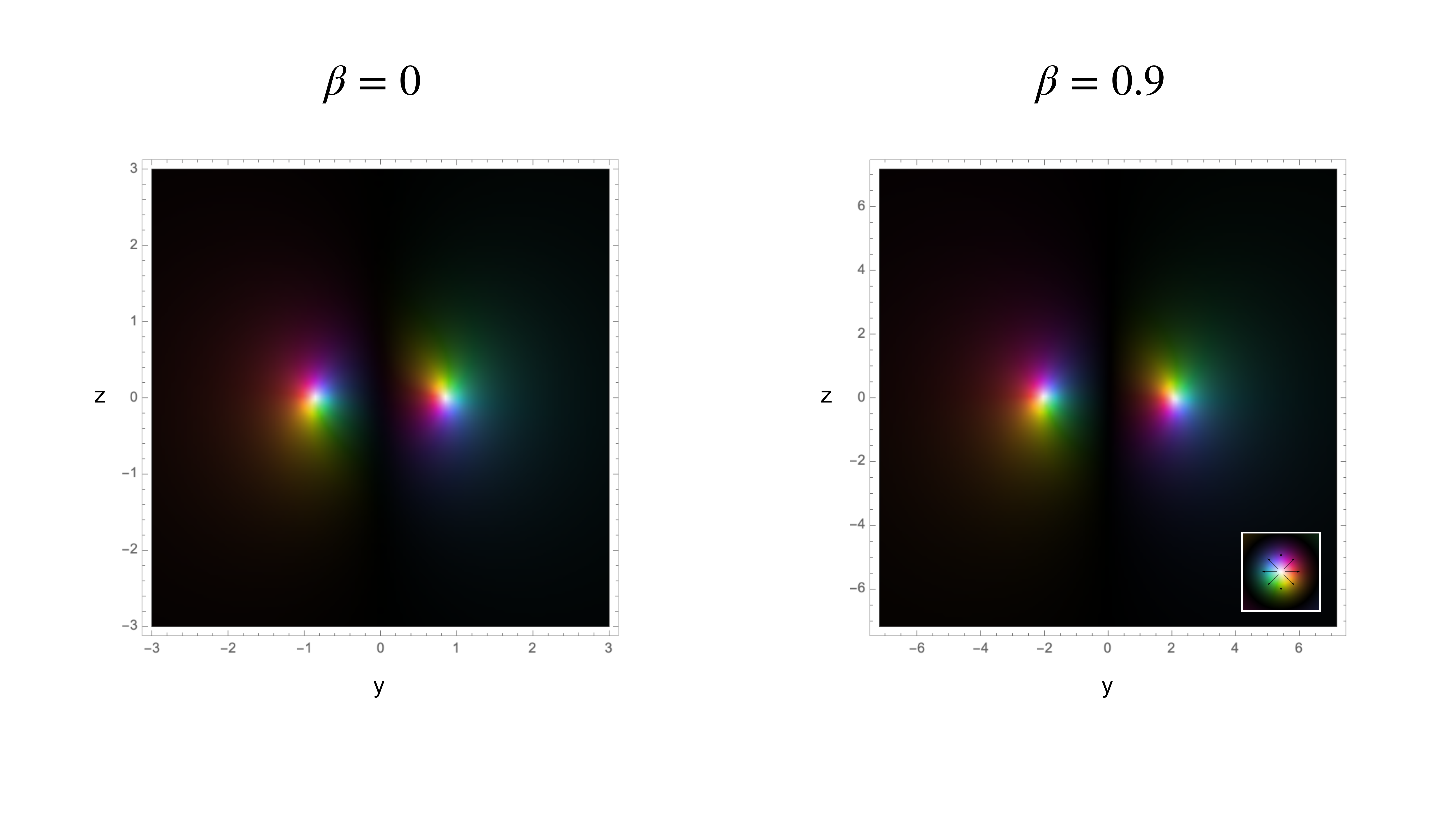}
	\caption{On the left is a cross-section in the $yz$-plane of the hedgehog Skyrmion in the Skyrme model proper, mapped to the unit vector $S$ field, as discussed in the text. The hue of a color denotes the azimuthal angle of $S$, and the brightness denotes the polar angle. On the right is the minimal energy $Q=1$ soliton in the squashed Skyrme model at $\beta=0.9$ in the fixed bound parametrization. Both plots may be compared to similar plots for true Hopfions with gauge invariance such as Fig. 1b of~\cite{sutcliffe2017skyrmion}, and  Fig. 1b and Fig. 3 in~\cite{kent2021creation}.}
	\label{fig Hedgehog cross section}
\end{figure}

The $Q=1$ Skyrmion in the Skyrme model satisfies the \emph{hedgehog ansatz},
\begin{align}
	U(x^\mu) = \cos f(r)\,I+ i \sin f(r)\frac{x^i}{r}\sigma^i,\label{def Hedgehog}
\end{align}
for some radial profile function $f(r)$ which equals $\pi$ at $r=0$ and vanishes at infinity. Considering \eqref{Def U} and \eqref{Def S}, the third component of the unit vector $S$ field in the hedgehog ansatz is,
$$S^3=\cos^2 f -\frac{x^2+y^2-z^2}{r^2}\sin^2 f.$$
The boundary condition on the $\mathbf{S}$ field at infinity $S^3=+1$ is also satisfied along the $z$ axis, and the furthest departure from the boundary condition $S^3=-1$ is satisfied in a loop in the $xy$-plane with radius $r_0$ such that $f(r_0)=\pi/2$. A curve such as this where $S^3=-1$ is referred to as the \emph{position curve}, and it may be thought of as the core of a stringy topological defect.

A cross-section in the $yz$-plane of the Skyrmion at both $\beta=0$ and $\beta=0.9$ is plotted in Fig. \ref{fig Hedgehog cross section}. Even though the squashing parameter is very different, by eye the topological defects are almost indistinguishable.\footnote{The scale of both figures is set to agree in the Ward parametrization.} The two intersections of the position curve loop with the plane are clearly seen, and it may be seen from the colors representing the orientation of $S$ how the 2D magnetic Skyrmion charge in the $yz$-plane (quantified by $F_{23}$ \eqref{Eq.2 F and S}) is concentrated around the position curve.

Note that due to the dependence on the $J^3$ field the energy density and topological charge density of the hedgehog Skyrmion are actually spherically symmetric and not concentrated near the position curve. However even the $Q=1$ Hopfion in the Faddeev-Niemi model at $\beta=1$ is approximately spherically symmetric in this sense as well \cite{ward2004skyrmions} so there is very little change in the unit charge topological defects throughout the entire range of the squashing parameter. But for higher charge solutions the situation is more interesting, and a simulation for $Q=7$ in the squashed Skyrme model is presented in \cite{schubring2021skyrmeHopf}, and results for the lattice model will be shown here in Sec \ref{Sec 3 lattice}.

Note that at $\beta=0$ the global symmetry of system under right multiplication is increased from $U(1)$ to $SU(2)$, and this has some consequences for the notion of considering the position curve as a physical core of a string. Any soliton configuration $U$ may be transformed to a new field $U \rightarrow V^{-1}U V$ with the same boundary conditions at infinity and the same energy. And for a general $V\in SU(2)$ this transformation will not leave the position curve invariant. For the hedgehog configuration this just ends up being equivalent to the degeneracy of the solution under spatial rotations, but for higher charge configurations this means that the shape of the position curve itself may deformed in certain ways without changing the energy. However for $\beta>0$ the symmetry is reduced to a {$U(1)\rtimes Z_2$} subgroup\footnote{ {The extra discrete $Z_2$ global symmetry arises from those internal $SU(2)_R$ transformations which flip the third axis in isospin space. Concretely the $Z_2$ subgroup may be chosen as $\{ I,\sigma^1\}\subset SU(2)_R$.}} which leaves the locus of the position curve unchanged, and only translates the field along the position curve.

\subsection{Baryon string ansatz}

Some insight may be gained by considering the structure of the $U$ field around the position curve in the hedgehog ansatz in \eqref{def Hedgehog} and Fig. \ref{fig Hedgehog cross section}, and abstracting this to a new ansatz of a cyllindrically symmetric straight string with a position curve aligned with the $z$-axis,
\begin{align}
	U(\rho,z,\phi)= \cos g(\rho) e^{-i\phi\sigma^3}+i\sin g(\rho)\left(\cos\left(\frac{2\pi z}{L}\right)\sigma^1+\sin\left(\frac{2\pi z}{L}\right)\sigma^2\right).
\end{align}
Here $g(\rho)$ is some new profile function depending on $\rho=\sqrt{x^2+y^2}$, and $L$ is some parameter describing the rate of twisting along the string. The profile function vanishes at infinity and $g(0)=\pi/2$. The baryon charge \eqref{def Q Skyrme} integrated over a length $\Delta z$ is found to be,
\begin{align*}
	Q= -\frac{1}{L}\int dz d\rho \,g^\prime(\rho)\sin \left(2g(\rho)\right) =\frac{\Delta z}{L},
\end{align*}
so every segment of length $L$ has baryon charge $1$.

Outside the core of the string, where $g\approx 0$, $U$ is restricted to a $U(1)$ subgroup, and the principal chiral model effectively reduces to a 3D XY model. Unless the $U(1)$ subgroup is gauged (as it is at $\beta=1$) the energy per length of an isolated straight string will be logarithmically divergent.

To better understand the energy per length due to the core of the string, the structures involved in the energy density \eqref{lagr Ward} may be expressed in terms of the ansatz,
\begin{gather}
	(J^1)^2+(J^2)^2=g^{\prime2}+\left[\left(\frac{2\pi}{L}\right)^2+\frac{1}{\rho^2}\right]\cos^2g \sin^2 g \non
	(J^3)^2=\left(\frac{2\pi}{L}\right)^2 \sin^4g + \frac{1}{\rho^2}\cos^4g	\non
	\frac{1}{8}\left(F_{\mu\nu}\right)^2=g^{\prime 2}\left[\left(\frac{2\pi}{L}\right)^2+\frac{1}{\rho^2}\right]\cos^2g \sin^2 g\non
	-\frac{1}{4}\text{Tr}[J_\mu,J_\nu]^2=\left[\sin^2g \left(\frac{2\pi}{L}\right)^2 +\cos^2g \frac{1}{\rho^2}\right]g^{\prime 2}+\left(\frac{2\pi}{L}\right)^2\frac{1}{\rho^2}\cos^2g \sin^2g.\label{eq Energy string ansatz}
\end{gather}
The $\rho^{-2}$ term in the $\left(J^3\right)^2$ structure is what causes the infrared divergence of an isolated string. Of course this is typical for global strings or vortices and does not preclude either a network of oppositely oriented long strings or strings forming closed loops with radius of curvature much larger than the string thickness, in which cases this ansatz may still be useful. In the latter case, due to the reduction of the model to a global $U(1)$ theory outside the string core, and due to the expression \eqref{Def F tensor} relating the curl of $J^3$ to the dual of $F$ tensor which has a constant $2\pi$ flux across the string core, the $J^3$ field may be calculated outside the string core using the Biot-Savart formula, much as is done calculating the fluid velocity outside knotted loops of vorticity~\cite{kleckner2013creation}.

The energy inside the core may be calculated by optimizing the above expressions for the energy \eqref{lagr Ward} and \eqref{eq Energy string ansatz} over profile functions $g$ and length per baryon charge $L$. To a get rough sense of the influence of the various terms in the Lagrangian note that since $g$ is dimensionless, it must be possible to write it as a function $g( \rho/R)$ where $R$ is a length scale which is on the order of the string core radius. Then we can write the energy of a $Q=1$ segment of the string schematically in terms of $L, R, M$,
\begin{align}
	E= f_\pi^2\left[C_1 L + C_2 \frac{R^2}{L}+C_3 \frac{1}{LM^2}+C_4 \frac{L}{R^2M^2}\right],\label{Eq.3 energy schematic}
\end{align}
where the $C$ coefficients may all be calculated from dimensionless integrals involving $g$. From this expression we see that the energy is minimized when
\begin{align}
	L= \left(\frac{C_3}{C_1}\right)^{1/2}M^{-1},\quad R=  \left(\frac{C_3 C_4}{C_1 C_2}\right)^{1/4}M^{-1},\label{Eq.3 baryon string schematic solution}
\end{align}
This is of limited use on its own since $C_1$ will diverge for $\beta<1$, but let us first discuss the local string limit of $\beta=1$, and then we will return to the global string case of $\beta<1$.

It is straightforward to numerically optimize the profile function $g$ for $\beta=1$ using the Euler-Lagrange equations. Then the optimum energy of the straight string ansatz is found to be $E \approx 396$ per length $L\approx 3.95$, where to compare with previous results we have used the parametrization in Battye and Sutcliffe~\cite{battye1999solitons} where $f^2_\pi/2=4$ and  $M\sqrt{1-\beta}\rightarrow 1/2$.  The energy per charge already agrees reasonably well with the unstable toroidal solutions of Battye and Sutcliffe in Table 1 and Figure 9 of~\cite{battye1999solitons}, and it may be expected to become closer to the straight Skyrmion string ansatz as the charge increases.

For $\beta<1$ the $J^3$ field outside these large loop solutions will lead to a contribution to the energy \eqref{Eq.3 energy schematic} which is essentially equivalent to the calculation of the self-inductance of a loop of wire in classical electromagnetism.\footnote{Equation (5.158) in Jackson \cite{jackson1999classical}.} The far-field dipole field will contribute to the $C_1$ coefficient in \eqref{Eq.3 energy schematic} and the logarithmic divergence of the near-field is cut off by the scale set by the radius of curvature of the loop, leading to a contribution like $\sim f_\pi^2L(1-\beta)\log\left(\frac{QL}{R}\right).$

We can also say something about how the asymptotic large $\rho$ behavior of the profile function $g$ changes for $\beta<1$. The linearized equations of motion for $g$ may be found from \eqref{eq Energy string ansatz},
\begin{align*}
	\left[\rho^2g^{\prime\prime}+ \rho g^\prime -\left(\left(\frac{2\pi}{L}\right)^2\rho^2+1\right)g\right]+2(1-\beta)g+(1-\beta)M^{-2}\left[g^{\prime\prime}-\frac{1}{\rho}g^{\prime}-\left(\frac{2\pi}{L}\right)^2g\right]=0.
\end{align*}
The first terms in brackets are just the modified Bessel equation and they give the dominant behavior as $\rho\rightarrow\infty$. The $\beta$ dependent terms give the subleading behavior,
\begin{align}
g(\rho)=\frac{c}{\sqrt{\rho}}e^{-\frac{2\pi}{L}\rho}\left(1-\frac{3}{8\rho}+\frac{1-\beta}{\rho}+\order{\rho^{-2}}\right).
\end{align}
The prefactor $c$ is not so easy to determine, but the $\rho$ dependence follows from standard techniques \cite{bender1999advanced}. Up to this order only the squashed sphere model terms have affected things (there is no dependence on the scale $M$ introduced by the stabilizing terms yet). So it is quite possible that the effective theory for the magnetic model in the next section has the same asymptotic dependence at this order.

\section{A frustrated magnetic model and the rational map ansatz}\label{Sec 3 lattice}

Now we will introduce a simple spin system which at lowest order in the continuum approximation reduces to the same squashed sphere non-linear sigma model discussed in the previous section. The higher order derivative terms which may stabilize topological defects will be different from the rotationally symmetric Skyrme and Faddeev-Niemi terms, but are in many respects qualitatively similar.

\subsection{A description of the lattice model} \label{sec Lattice model intro}

The system is defined on an ordinary cubic lattice with lattice spacing $a$, and each site $x$ has three real unit vector spins $S^i_r(x)$, where the $i$ index refers to the three components of the unit vector, and the $r$ index labels the distinct spins at the site. The dot product between any two spins at a given site is constrained to be equal to a parameter $\kappa$ which is fixed for the entire system, i.e. ${\mathbf{S}}_r(x)\cdot \mathbf{S}_s(x)=\kappa$ for $r\neq s$ and all $x$. So the three spins at each site act like a rigid body with an orientation which may be described by a matrix $R(x)\in SO(3)$.
The spins may be written in terms of this $R(x)$ and a basis $\mathbf{e}_r$ which does not depend on $x$,
\begin{gather}
	{S}^i_r(x)=R(x)^i_{\,j}{e}^{j}_r,\qquad
	e^j_{r}\equiv
	\left(\begin{array}{ccc}
		\frac{\sqrt{3}}{2}\sin\theta& 0 &-\frac{\sqrt{3}}{2}\sin\theta\\
		-\frac{1}{2}\sin\theta &\sin\theta&-\frac{1}{2}\sin\theta\\
		\cos\theta & 	\cos\theta & 	\cos\theta
	\end{array}\right),\label{def R e}
\end{gather}
where the three vectors $\mathbf{e}_r$ are represented as a matrix with $r$ referring to different columns. The fixed parameter $\theta$ in this basis is directly related to the parameter $\kappa$,
\begin{align}
	\mathbf{S}_r(x)\cdot \mathbf{S}_s(x)=\kappa \equiv \frac{3}{2}\cos^2\theta-\frac{1}{2}.\label{def Kappa theta}
\end{align}

The spins interact as a frustrated classical Heisenberg model, with a ferromagnetic coupling $K_1<0$ between nearest neighbors at a distance of one lattice spacing $a$, and an antiferromagnetic coupling $K_2>0$ between sites at a distance of $2a$, which are indicated by doubled angled brackets in a slight abuse of notation,\footnote{For simplicity in this toy model, sites at the nearer distances of $\sqrt{2}a$ and $\sqrt{3}a$ are not taken to interact.}
\begin{align}
	H&=K_1\sum_{r,\,\langle x,y\rangle}\mathbf{S}_r(x) \cdot \mathbf{S}_r(y)+K_2\sum_{r,\,\llangle x,y\rrangle}\mathbf{S}_r(x) \cdot \mathbf{S}_r(y).\label{ham S}
\end{align}
Note that a given spin $\mathbf{S}_r(x)$ only interacts with spins $\mathbf{S}_r(y)$ with the same `species' $r$.
\begin{align}
	\mathbf{S}_r(x)\cdot \mathbf{S}_s(x)=\kappa \equiv \frac{3}{2}\cos^2\theta-\frac{1}{2}.
\end{align}

This model {was originally inspired} by the treatment of spins interacting on a pyrochlore lattice in~\cite{batistaEtAl2018} which is discussed further in Appendix \ref{appendix pyrochlore}. In the pyrochlore case $r$ takes four values corresponding to the four sites of the tetrahedral cells of the pyrochlore lattice. The dot product between spins $\kappa$ in that case is fixed so that the spins are in an `all-in-all-out' configuration which is preferred {in the presence of biquadratic spin interactions}. If we formally allow $\kappa$  to be a tunable parameter and restrict the interaction to third-nearest-neighbor sites so only spins with the same value of $r$ interact we obtain a very similar model to that considered here.

%This model can be regarded as the low-energy description for the broad class of realistic 3D noncollinear magnets, where SO(3) rotation is the only low-energy mode while other modes are gapped out. For instance, the Heisenberg model on pyrochlore lattice was demonstrated to have gapless SO(3) mode, where other modes are gapped in the presence of biquadratic spin interactions [34]. The parameter κ is expected to vary for different 3D lattices and different magnetic anisotropies. As a result, the model study in this paper is expected to provide a unified qualitative description for a series of realistic materials that are described by different values of κ.

{The motivation for making these abstractions was to create a simple lattice model that still captures the main qualitative features of a broad class of realistic 3D noncollinear magnets which involve $SO(3)$ Goldstone modes. The `squashing' parameter $\kappa$ that measures the degree of collinearity} is expected to vary for different 3D lattices and different magnetic anisotropies. So the model studied here is expected to provide a unified qualitative description for a series of realistic materials that are described by different values of $\kappa$.

Also note that at the limiting value $\kappa=1$ where the spins $\mathbf{S}_r$ {are perfectly collinear}, the lattice model reduces to a 3D version of inversion-symmetric frustrated magnets
which have been previously considered in 2D as a host to magnetic Skyrmions~\cite{leonov2015multiply,lin2016ginzburg}. A 3D extension of these frustrated magnets has already been considered {in the collinear case} \cite{sutcliffe2017skyrmion}, and Hopfions were investigated and the analogy to the Faddeev-Niemi model was pointed out.

But in the opposite limit of $\kappa=0$ this model will instead be shown to be closely analogous to the Skyrme model, so this toy model bears the same relationship to the effective theory of frustrated magnets in~\cite{sutcliffe2017skyrmion} as the squashed Skyrme model~\cite{nasir2002effective,ward2004skyrmions} discussed in the previous section bears to the Faddeev-Niemi model.

\subsection{Effective theory in the continuum limit}\label{sec Lattice model effective theory}

To show that this analogy is valid, let us now turn to the effective continuum description of the model. Following a similar procedure to Dombre and Read's continuum description of the triangular antiferromagnet~\cite{dombre1989nonlinear}, the Hamiltonian can be described up to fourth order in derivatives in terms of continuous fields $S^i_r(x)$,
\begin{align}
	H=-\frac{1}{2a}\left(K_1+4K_2\right)\sum_\mu \int d^3 x \left(\partial_\mu \mathbf{S}_r\right)^2 +\frac{a}{24}\left(K_1+16 K_2\right)\sum_\mu \int d^3 x \left(\partial^2_\mu \mathbf{S}_r\right)^2. \label{ham Continuum}
\end{align}
Now since rotational symmetry is broken by the fourth order terms, any sums over the spatial index $\mu$ will always be indicated explicitly, although sums over internal indices like $r$ or $i$ are still implied by the summation convention or context. The lack of rotational symmetry in the fourth order terms is the main difference between the effective description of this toy model and that considered by Lin and Hayami~\cite{lin2016ginzburg}. Here the interaction between the neighbors at distances $\sqrt{2}a$ and $\sqrt{3}a$ was set to zero whereas in~\cite{lin2016ginzburg,sutcliffe2017skyrmion} it was implicitly tuned to maintain rotational symmetry in the fourth order terms. Note that in the absence of any tuning such cubic anisotropies would generically be expected to be present, {and the presence of isotropy in these higher derivative terms is not essential for the stabilization of topological defects.}

For this Hamiltonian to have stable topological defects it is easily shown by an argument along the lines of Derrick's theorem~\cite{derrickstheorem} that the coefficients of both the second and fourth order terms must be positive,
$$-K_1 > 4 K_2 > -\frac{1}{4}K_1.$$

Moreover, for the Skyrmion size to be much larger than the lattice spacing and this continuum description to be valid we must be close to the Lifshitz transition $K_2= -\frac{1}{4}K_1$ where the sign of the quadratic term changes from positive to negative. Suppose that a Skyrmion field configuration has some length scale $L$ representing the radius, and the parameters are displaced from the Lifshitz transition by some small positive quantity $\epsilon$,
\begin{align}
	K_2= -\left(\frac{1}{4}-\epsilon\right)K_1.\label{def epsilon}
\end{align}
Then it can be shown using an argument similar to that leading to \eqref{Eq.3 baryon string schematic solution} that the radius of the Skyrmion is on the order
$L\sim \epsilon^{-1/2}a,$ where the exact coefficient depends on dimensionless integrals over the field configuration.

Now to proceed and better illustrate the connection to the squashed sphere sigma model, the spins $\mathbf{S}_r$ may be written in terms of the rotation matrix field $R(x)$ using \eqref{def R e}. The quadratic terms become
$$\sum_\mu \int d^3 x \left(\partial_\mu \mathbf{S}_r\right)^2 = \sum_\mu \int d^3 x \text{Tr}\left[\partial_\mu R^{-1}\partial_\mu R \,\mathbf{e}_r\otimes \mathbf{e}_r\right],$$
where
\begin{align}
	\mathbf{e}_r\otimes \mathbf{e}_r = \text{diag}\left(1-\kappa,\,1-\kappa,\, 1+2\kappa\right).
\end{align}
For $\kappa=0$ this is clearly equivalent to the principal chiral model, except that it is expressed in terms of $R\in SO(3)$ rather than $U\in SU(2)$. For $\kappa \neq 0$, the components of the diagonal matrix $\mathbf{e}_r\otimes \mathbf{e}_r$ will take different values and this will become a squashed sphere model. This can be seen by expressing the model in terms of the currents $J$ using \eqref{Eq.2 J R},
\begin{align}
	\sum_\mu \int d^3 x \left(\partial_\mu \mathbf{S}_r\right)^2 = 4(\kappa+2)\sum_\mu \int d^3 x \left[\left(J^i_\mu\right)^2-\frac{\kappa}{\frac{1}{3}\left(\kappa+2\right)}\left(J^3_\mu\right)^2\right].\label{lagr Quadr}
\end{align}
This is precisely the squashed sphere model in \eqref{Lagr squashed sphere J}, with the parameter $\beta$ expressed in terms of $\kappa$. The overall dimensionful parameter $f_\pi$ in the squashed sphere model depends on the prefactor of the quadratic terms given in the full Hamiltonian \eqref{ham Continuum}, and it is seen to be on the order $f_\pi \sim \left(\epsilon|K_1|a^{-1} \right)^{1/2}$.

Exactly the same chain of steps may now be followed to express the quartic terms of the Hamiltonian in terms of the $J$ fields and the parameter $\kappa$. After some calculation,
\begin{align}
	\sum_\mu \left(\partial^2_\mu \mathbf{S}_r\right)^2 = &8(1-\kappa)\sum_\mu \left[\left(\partial_\mu J_\mu^i\right)^2+4\left(J_\mu^i J_\mu^i\right)^2\right]+ 12\kappa \sum_\mu \left[\left(\mathcal{D}_\mu J_\mu^a\right)^2+4\left(J_\mu^a J_\mu^a\right)^2\right],\label{lagr Quartic}
\end{align}
where as discussed previously, $i$ runs over all components $1, 2, 3$, and $a$ is only taken over $1, 2$. The covariant derivative with respect to the gauge symmetry defined in \eqref{Eq.2 J gauge transf} is
$$\mathcal{D} J^a\equiv \partial J^a + 2\epsilon_{ab3}J^3\,J^b.$$
Note that the continuum model is completely gauge symmetric at $\kappa=1$, which must be the case considering that in the lattice model all three spins at each site are pointing in the same direction, so the rotation field $R(x)$ is only fixed up to rotations about the spin axis.

\subsection{Comparison to the rational map ansatz}

\begin{figure}[t]
	\centering
	\includegraphics[width=.8\textwidth]{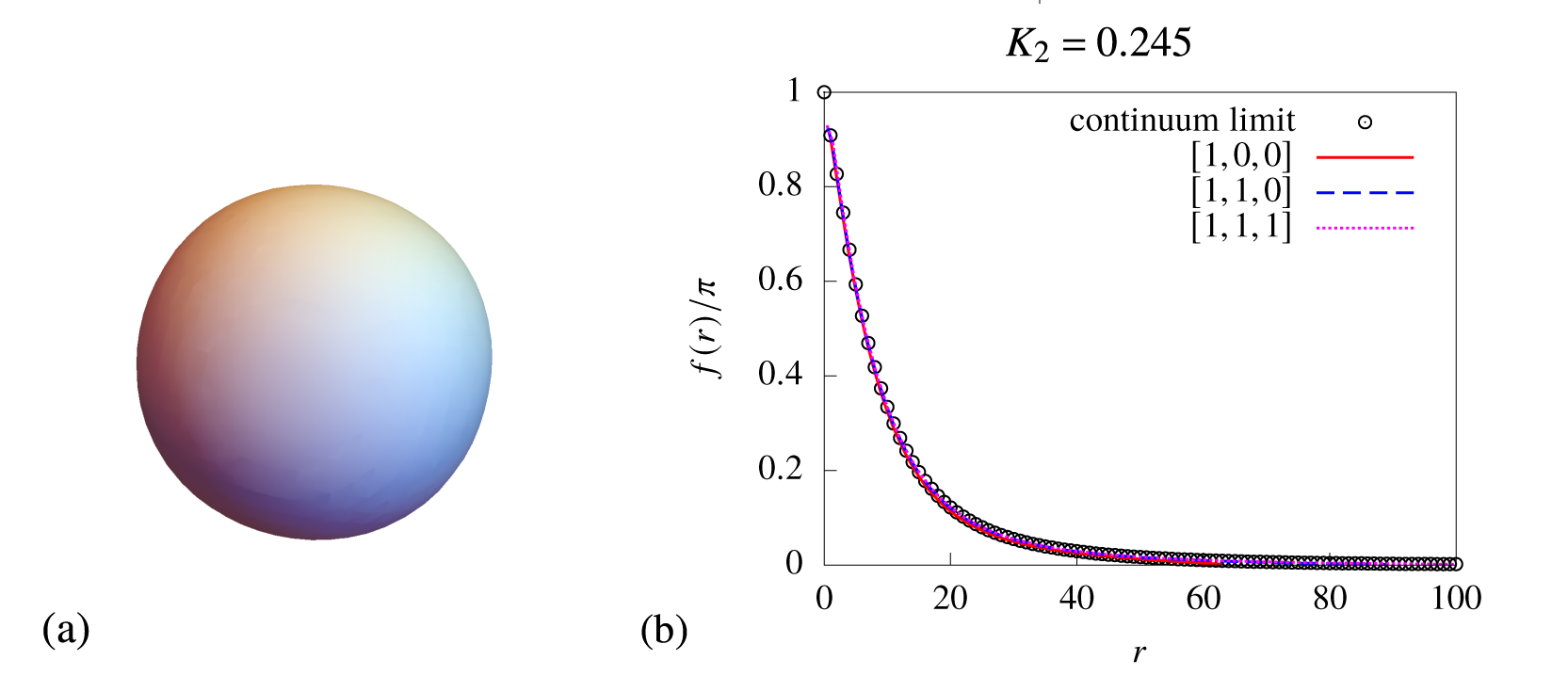}
	\caption{The unit charge Skyrmion solution on an $128^3$ lattice with periodic boundary conditions and $K_1=-1$, $\kappa=0$. (a) The topological charge density isosurface with $K_2=0.248$. (b) The profile functions for $K_2=0.245$ along different high-symmetry directions, along with the solution of the continuum theory Eq.~\eqref{ham Continuum}.}
	\label{fig profile lattice charge1}
\end{figure}

The numerical simulation of the lattice model is discussed in more detail in \cite{schubring2021skyrmeHopf}. Here I will focus on the comparison of ansatzes in the effective theory to the numerical results. 

In Fig \ref{fig profile lattice charge1}(a) an isosurface of the topological charge density \eqref{Eq.2 top charge q} is plotted and it is seen that despite the cubic anisotropies in the stabilizing terms the unit Skyrmion at $\kappa=0$ was found to be very spherically symmetric. In the continuum theory the spherically symmetric hedgehog ansatz \eqref{def Hedgehog} may be optimized, and in Fig \ref{fig profile lattice charge1}(b) it is shown that the profile function found in the continuum theory agrees very
closely with the profile function found in the lattice simulation by extracting the $SO(3)$ rotation angles along various directions on the lattice.\footnote{In accordance with the discussion below \eqref{def epsilon}, as $K_2$ is increased towards the Lifschitz point $K_2=1/4$ the size of the solution increases and finite size effects spoil some of the rotational invariance.} Much as in the squashed Skyrme model, as $\kappa$ is increased towards $1$ the unit charge solution remains approximately spherically symmetric.\footnote{For both models the lack of internal $SU(2)_R$ symmetry as $\kappa\rightarrow 1$ is most easily seen by comparing the average value of $J^3$ to $J^{1,2}$ \cite{schubring2021skyrmeHopf}.}

\begin{figure}[t]
	\centering
	\includegraphics[width=0.8\textwidth]{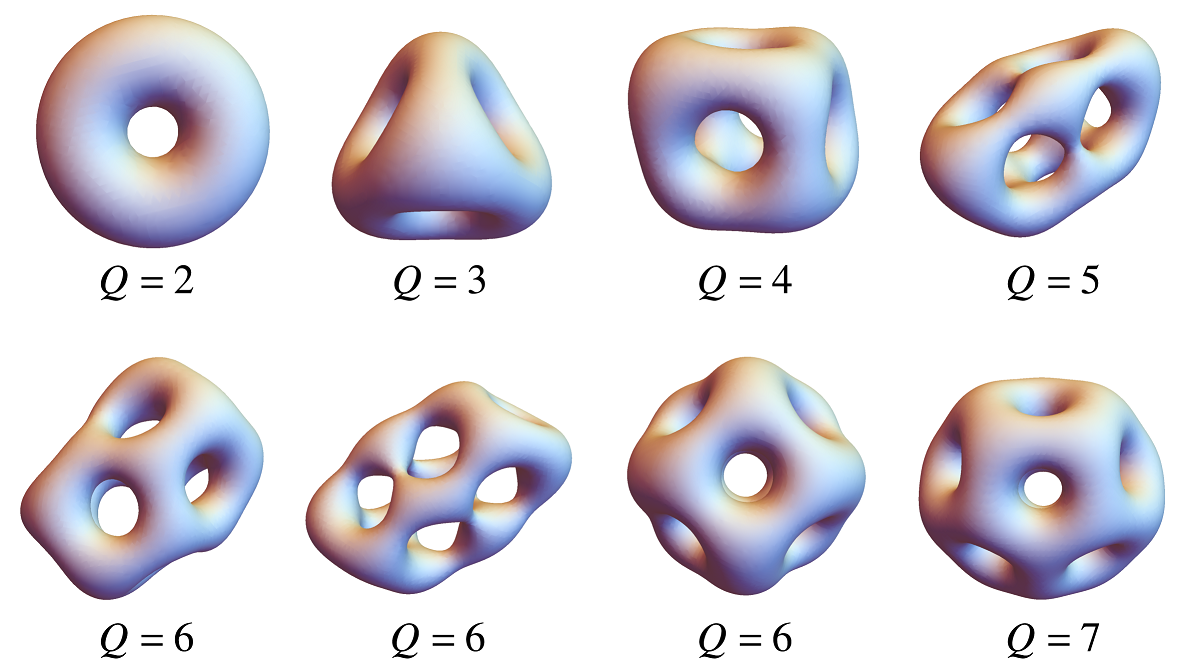}
	\caption{Charge density isosurfaces of the relaxed higher charge Skyrmion solutions on an $128^3$ lattice with $K_1=-1$, $K_2=0.248$, and $\kappa=0$.}
	\label{fig high charge}
\end{figure}

The charge density isosurfaces of the higher charge solutions are plotted in Fig \ref{fig high charge}. The solutions were found both by relaxing from initial configurations of lower charge solutions (in multiple different ways), and relaxing from an initial rational map ansatz configuration, as described below. For $Q=6$ three distinct locally stable solutions were found and they are ordered in terms of increasing energy going from left to right. The solutions were found to be robust under changing the type of boundary conditions, except for the highest energy $Q=6$ solution, which was seen for periodic boundary conditions but not boundary conditions fixed to the vacuum configuration.

One striking thing about Fig.~\ref{fig high charge} is the extent to which the solutions for $Q\leq 4$ (as well as the $Q=7$ and highest energy $Q=6$ solution) are qualitatively similar to the corresponding Skyrmions in the Skyrme model~\cite{BRAATEN1990147} as well BPS monopoles in $SU(2)$ Yang-Mills~\cite{hitchin1995symmetric}, both of which may be approximated by the rational map ansatz~\cite{houghton1998rational,battye1997symmetric,battye2002skyrmions},
\begin{align}
	U(r,w) = \cos f(r)\,I+ i \sin f(r){n}(w)\cdot{\sigma}.\label{def Rational map ansatz}
\end{align}
Here $n^i$ is a unit vector, which is a generalization of $n^i=x^i/r$ in the hedgehog ansatz \eqref{def Hedgehog}. The spatial coordinates $x^i=(x,y,z)$ are expressed in a spherical coordinate system $r,w$, where $w$ is a complex coordinate which is a function of the angles $\theta,\phi$,
\begin{align}
	w\equiv\quad\tan \frac{\theta}{2}e^{i\phi}=\quad\frac{x+iy}{r+z}. \label{def W}
\end{align}
The dependence of ${n}$ on $w$ may be expressed in terms of an analytic function $R(w)$,
\begin{align}
	{n}=\frac{1}{1+|R|^2}\left(2\,\text{Re} R, \,2\,\text{Im} R,\, 1-|R|^2\right).\label{eq n definition}
\end{align}
This function $R$ is the rational map from which the ansatz gets its name. It is a rational function $R=p/q$ where $p,q$ are polynomials with no common roots. The degree of $R$ is defined as the maximum degree of $p$ or $q$, and it turns out that the degree is simply equal to the baryon charge $Q$ of the ansatz $U$.

To determine how well the lower charge solitons found in the direct lattice simulation fit the rational map ansatz, the continuum description of the Hamiltonian \eqref{ham Continuum} was used to optimize the profile function $f$ given some rational map $R$. The quartic terms \eqref{lagr Quartic} in this case are a bit more complicated than the Skyrme model, where all integrals over angle either lead to an expression for the charge
\begin{align}
	Q=\frac{r^2}{4\pi}\int d\Omega\,\frac{1}{2}\sum_\mu \left(\partial_\mu {n}\right)^2,\label{eq Q n integral}
\end{align}
or a single non-trivial integral $\mathcal{I}_0$,
\begin{align*}
	\mathcal{I}_0\equiv \frac{r^4}{4\pi}\int  d\Omega\left[\frac{1}{2}\sum_\mu \left(\partial_\mu {n}\right)^2\right]^2.
\end{align*}
In the Skyrme model $\mathcal{I}_0$ may be easily expressed in terms of $R$ and minimized independently of the profile function~\cite{houghton1998rational}.

However the present model leads to four distinct angular integrals \eqref{def I} which are all coupled to the profile function and rather complicated if expressed in terms of $R$. In practice we simply took $R(w)$ to have the same discrete symmetry as it does in the Skyrme model, and for $Q\leq 4$ that completely fixes $R(w)$ so no minimization is necessary~\cite{houghton1998rational}.

\begin{table}
	\begin{center}
		\begin{tabular}{ c||c|c|c|c|c||c|c|c| }
			$Q$ & $R(w)$ & $\mathcal{I}_1$ & $\mathcal{I}_2$ & $\mathcal{I}_3$ & $\mathcal{I}_4$&  $E_{ansatz}/Q$ & $E/Q$ & $E/Q_{num}$ \\
			\hline
			$1$ & $w$ & $0.4$ & $1.6$ & $3.2$ & $-0.8$ & $59.81$ & $59.530$ & $61.308$\\
			$2$ & $w^2$ & $0.74926$ & $9.64307$ & $16.09$ & $-1.62537$& $59.92$ & $56.274$ & $57.687$	\\
			$3$ & $\frac{\pm i \sqrt{3} w^2-1}{w \left(w^2\mp i \sqrt{3}\right)}$ & $1.24335$ & $20.7566$ & $33.5133$ & $-2.37832$&  $57.92$ & $53.883$ & $55.189$	\\
			$4$ & $\frac{w^4+2 i \sqrt{3} w^2+1}{w^4-2 i \sqrt{3} w^2+1}$ & $1.97218$ & $30.2953$ & $48.4568$ & $-3.01391$&  $55.25$ & $52.608$ & $53.791$	\\
			%	$5$ & $\frac{w \left(w^4\pm 4 i w^2-3\right)}{-3 w^4\pm 4 i w^2+1}$ & $1.69048$ & $60.512$ & $94.9228$ & $-4.15476$	\\
			\hline
		\end{tabular}
		\caption{A comparison of the energy $E_{ansatz}$ of the rational map ansatz to the energy $E$ of the solitons found in the lattice simulation {with PBC}. All values are taken at $K_1=-1, K_2= 0.248$, {$\kappa=0$}, and the profile function was minimized in a finite volume of $r\leq 120$. $E$ is divided by both the exact charge $Q$ and a numerical charge $Q_{num}$ found from a finite difference approximation to \eqref{Eq.2 top charge q}.}\label{table I integrals}
	\end{center}
\end{table}

The structures of the Hamiltonian \eqref{ham Continuum} at $\kappa=0$ expressed in terms of the ansatz and averaged over solid angle are
\begin{gather}
	\frac{1}{4\pi}\sum_\mu\int d\Omega 	\left(J^i_\mu\right)^2 =\left(f'\right)^2+\frac{2Q\sin^2 f}{r^2},\label{lagr Quad f}\\
	\frac{1}{4\pi}\sum_\mu\int d\Omega \left(	 \left(J^i_\mu\right)^2\right)^2=\frac{3}{5}\left(f^\prime\right)^4+\frac{2\sin^2 f}{r^2}\left(f^\prime\right)^2\mathcal{I}_1+\frac{\sin^4 f}{r^4}\mathcal{I}_2\label{lagr Quart J4 f}
\end{gather}
\begin{align}
	\frac{1}{4\pi}\sum_{\mu}\int d\Omega  \left(\partial_\mu J^i_\mu\right)^2&= \frac{3}{5}\left(f^{\prime\prime}\right)^2+\frac{4}{5r}f^\prime f^{\prime\prime}+\frac{8}{5r^2}\left(f^{\prime}\right)^2\non
	&\qquad +4\frac{\cos^2 f}{r^2}\left(f^{\prime}\right)^2\mathcal{I}_1+\frac{\sin^2 f}{r^4}\mathcal{I}_3-\frac{\sin^4 f}{r^4}\mathcal{I}_2\non
	&\qquad -2\frac{\cos f\, \sin f}{r^2}f^{\prime\prime}\mathcal{I}_1+2\frac{\cos f\, \sin f}{r^3}f^{\prime}\left(-2Q+\mathcal{I}_1+2\mathcal{I}_4\right),\label{lagr Quart dJdJ f}
\end{align}
with the integrals defined as,
\begin{align}
	\mathcal{I}_1&\equiv \frac{r^2}{4\pi}\sum_\mu\int d\Omega  \left(\left(x^\mu\right)^2 \left(\partial_\mu n^i\right)^2\right) \non%&\rightarrow \left(1-q\right)\non
	\mathcal{I}_2&\equiv \frac{r^4}{4\pi}\sum_\mu\int d\Omega\left(	 \left(\partial_\mu n^i\right)^2\right)^2 \non
	\mathcal{I}_3&\equiv \frac{r^4}{4\pi}\sum_\mu\int d\Omega \left(\partial^2_\mu n^i\right)^2\non
	\mathcal{I}_4&\equiv \frac{r^3}{4\pi}\sum_\mu\int d\Omega \left(x^\mu\,\partial_\mu n^i\partial^2_\mu n^i\right).\label{def I}
\end{align}
The values of the $\mathcal{I}$ integrals are given in Table \ref{table I integrals}, including the hedgehog special case $R(w)=w$, which was used in calculating the profile function for the unit charge Skyrmion in Fig \ref{fig profile lattice charge1}. It is seen that the energy of optimal rational map ansatz in the continuum comes fairly close to the energy of the true minimum energy solitons in the lattice model.

\subsection{From nuclei to knots}

We have just seen that in the $\kappa=0$ limit the solutions of the frustrated magnetic model are quite similar to those in the Skyrme model. Even the solutions for $Q=5,6$ which are not quantitatively well described by the rational map ansatz still have charge density isosufaces which look like distorted polyhedral clusters. But in the $\beta=\kappa=1$ limit in which the squashed Skyrme model and the lattice model approach the Faddeev-Niemi model \cite{battye1998knots} and the frustrated magnetic model of Sutcliffe \cite{sutcliffe2017skyrmion}, respectively, we expect the appearance of the solutions to drastically change. In particular in \cite{sutcliffe2017skyrmion} it was shown that there is a solution at $Q=10$ that has a position curve forming a trefoil knot, so here we focus on the $Q=10$ solution.

\begin{figure}[tbp]
	\centering
	\includegraphics[width=\textwidth]{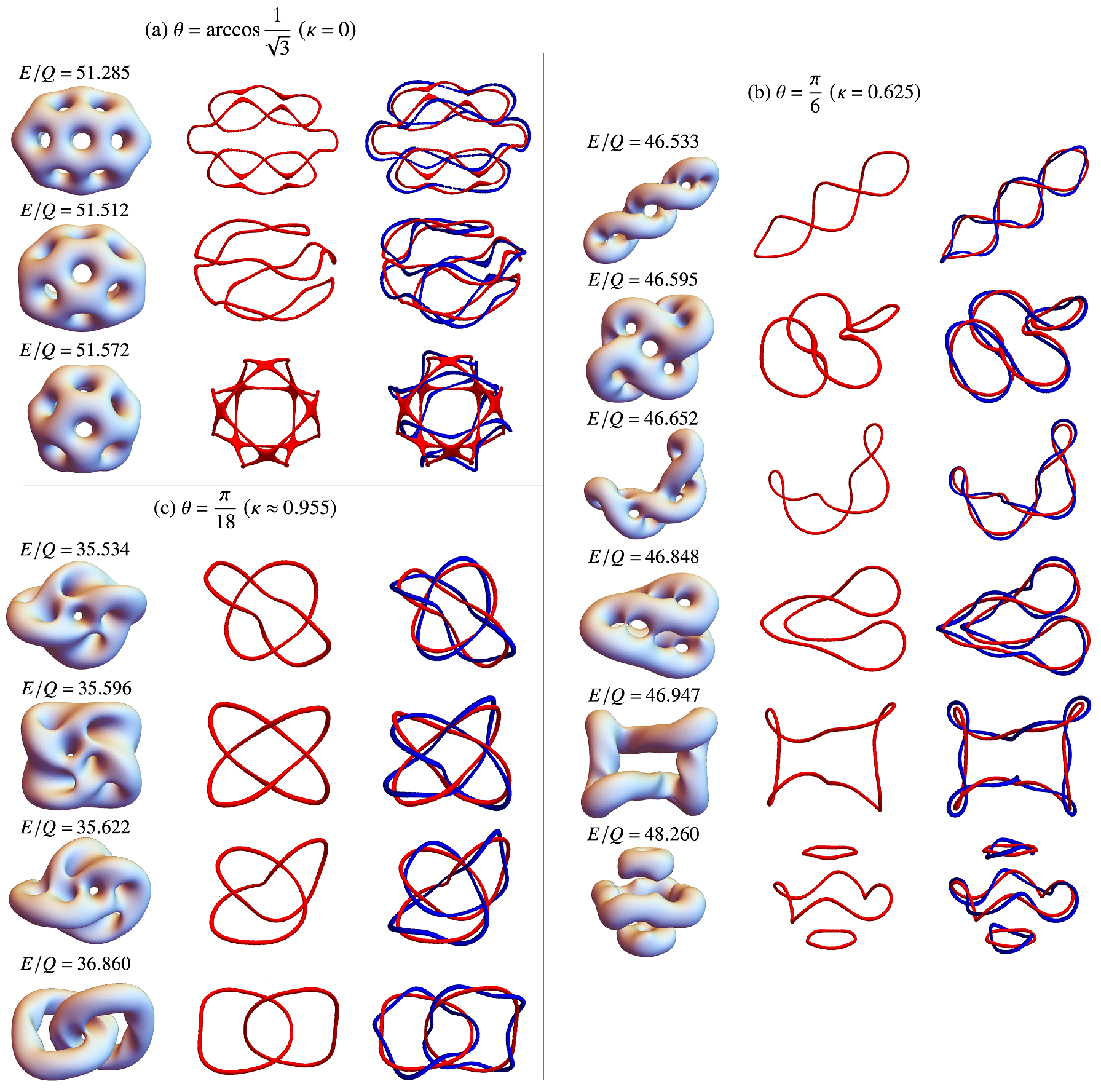}
	\caption{The charge density isosurfaces and the position curves (red) and linking curves (blue), for fully relaxed charge $Q=10$ Skyrmion solutions on an $128^3$ cubic lattice with $K_1=-1$, $K_2=0.248$, $\kappa=\{ 0,\, 0.625,\,0.955 \}$, and fixed boundary conditions.}
	\label{fig Q10}
\end{figure}

In Fig \ref{fig Q10}, multiple locally stable $Q=10$ solutions are shown for various values of $\kappa$. In addition to the charge density isosurfaces, the position curves are plotted in red. To better visualize the Hopf charge, the locus where $\mathbf{S}=\left(0,-1/\sqrt{2},-1/\sqrt{2}\right)$ is also plotted in blue, and this is referred to as the \emph{linking curve}. The number of times the position curve crosses over itself determines the ``writhe" and the number of times the linking curve twists around the position curve determines the ``twist." Individually the twist and writhe are not topological invariants but their sum in each case is equal to the Hopf charge $Q=10$.

As discussed in section \ref{Sec 3 squashed skyrme} exactly at $\kappa=0$ the shape of the position curve may change under action of the global $SU(2)_R$ symmetry. The position curves shown Fig \ref{fig Q10}.(a) are those that are locally stable for arbitrarily small nonzero values of $\kappa$. Note that in the rational map-like solutions at $\kappa=0$, the position curve may self-intersect somewhat like the $\chi$ solutions found in the Faddeev-Niemi model~\cite{sutcliffe2007knots}. 

As $\kappa$ increases the solutions become more `string-like,' and for $\kappa\approx 0.955$ there are topologically non-trivial position curves such as doubly-linked loops and a trefoil knot. Since at $\kappa=0$ these solutions are much like those in Skyrme model, which can be considered at some level to be a description of nuclei, this deformation is quite reminiscent of the old idea of Lord Kelvin suggesting that atoms might have something to do with knotted vortices \cite{LordKelvin}. 

%In papers on Hopfions and related topological defects, it is very common to cite the old idea of Lord Kelvin suggesting that atoms might have something to do with knotted vortices \cite{LordKelvin}. Here we have a model in which in one limit we have solutions much like those in the Skyrme model which at least to some extent may be considered models of the nuclei of an atom. As the squashing parameter is tuned to $\kappa\rightarrow 1$ the position curves of the `nuclei' deform to nontrivial structures including linked rings and knots, and so in this sense the model is much like a modern incarnation of Lord Kelvin's idea.\ds{maybe move this to intro}

	\chapter{Large N methods versus weak coupling}\label{Sec 4}
	In the last chapter we have been considering a squashed sphere model in three spatial dimensions purely classically, and have focused on topological defects involving non-renormalizable terms in an essential way. We now wish to move closer to the focus of much of the literature discussed in the introduction by considering the effects of thermal fluctuations in a partition function. From a classical statistical field theory perspective the temperature dependence can be absorbed into the definition of our parameter $g^2$. As the temperature (or equivalently $g^2$) increases there should eventually be a phase transition to a disordered phase. This phase transition may be studied through renormalization group methods which ultimately stem from studying a perturbative expansion in $g$, and hence are referred to as \emph{weak coupling} methods.
	
\begin{figure}
		\centering
		\includegraphics[width=0.45\textwidth]{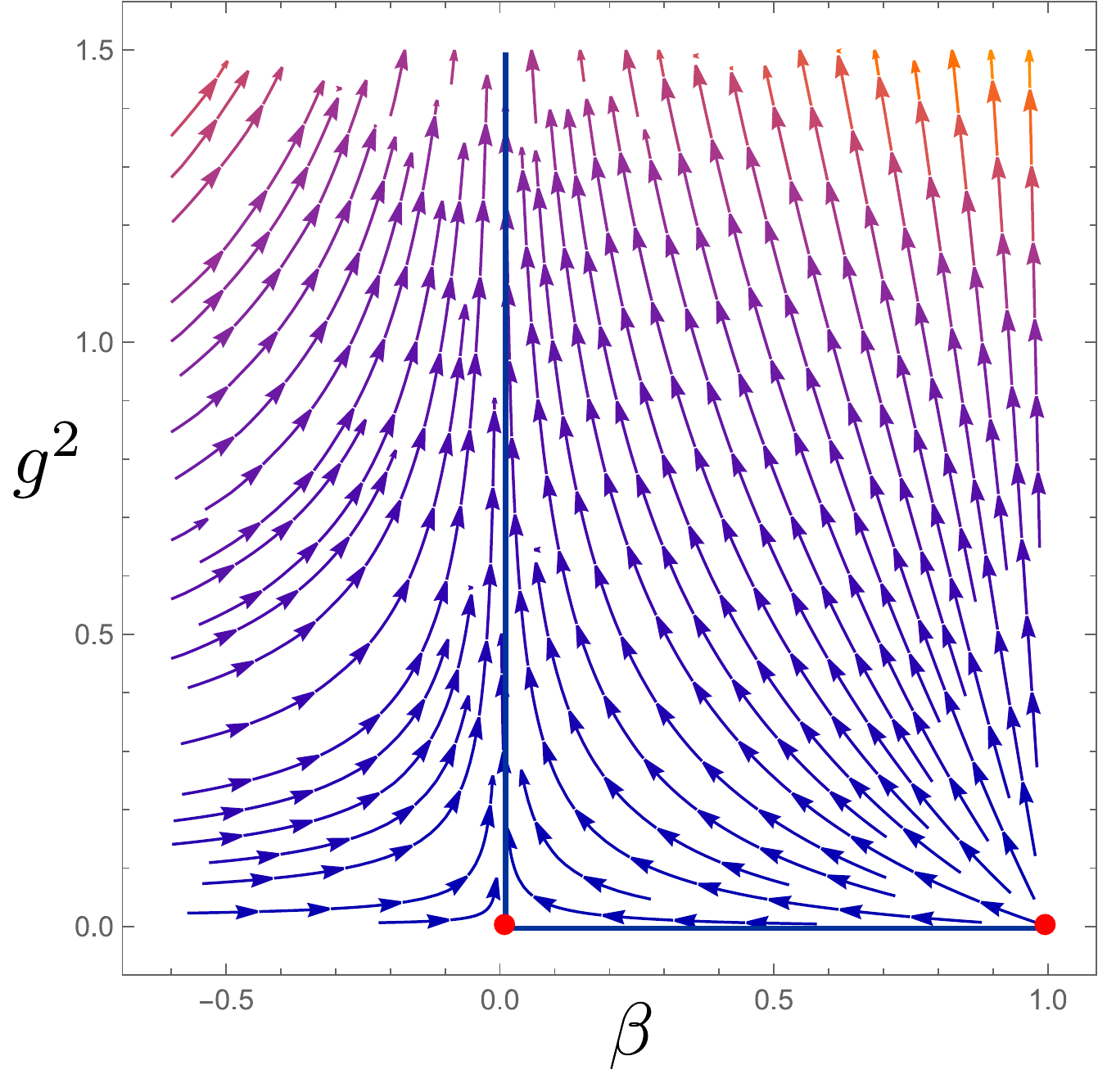}\qquad \includegraphics[width=0.45\textwidth]{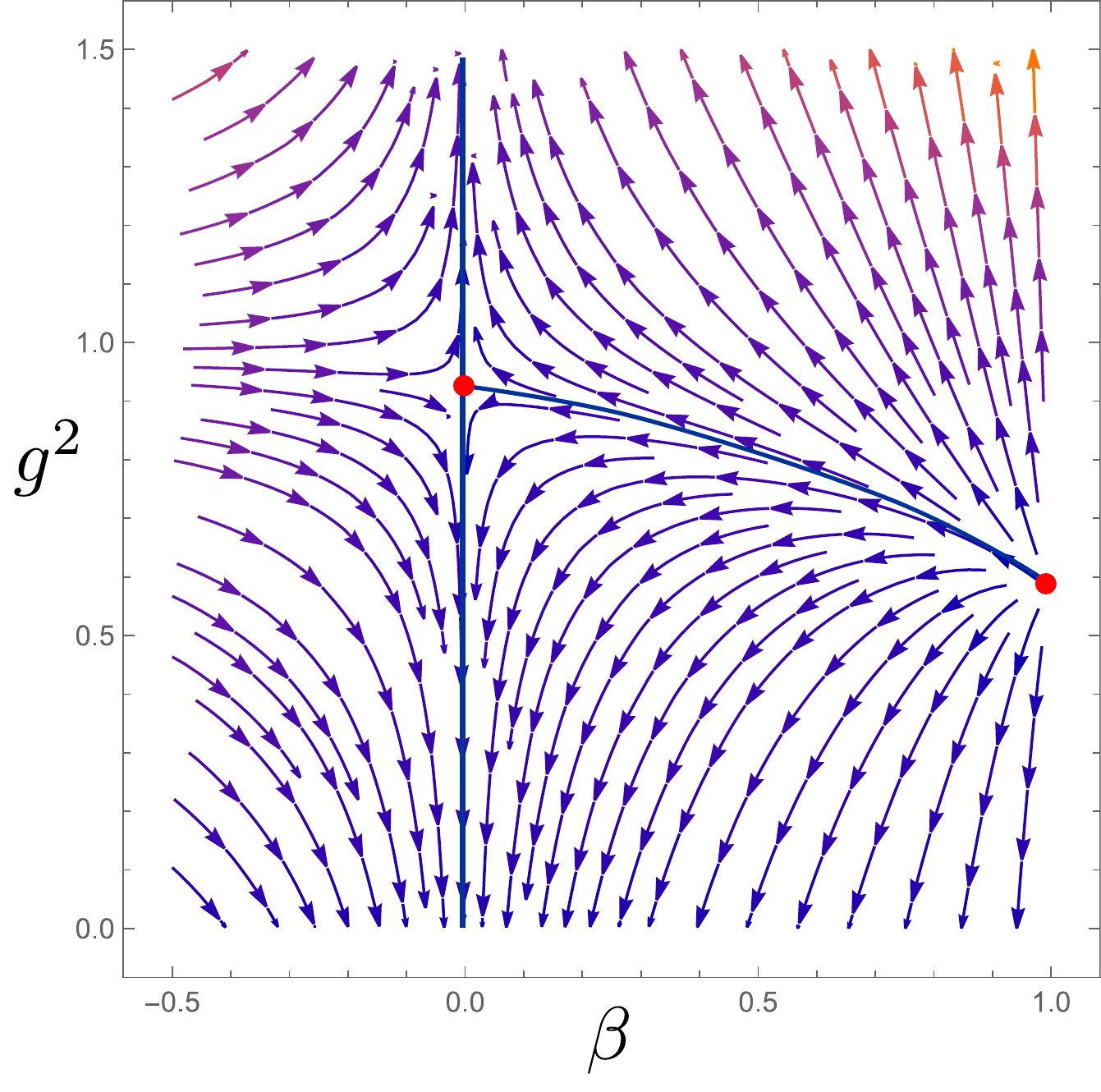}
		\caption{The RG flow for $d=2$ (left) and $d=2+\epsilon$ with $\epsilon=1$ (right). In both cases $N=4$. The arrows point towards the IR. The fixed points at $\beta=0$ and $\beta=1$ are indicated by a red dot.}\label{Fig 4.1}
	\end{figure}

	We have already performed such an analysis in Sec. \ref{Sec 2.3} for a slight generalization of the model. The 2-loop RG equations for the squashed sphere model itself originally appear in \cite{azaria1995massive}. The RG flow for the parameters $g^2, \beta$ is plotted in Fig \ref{Fig 4.1}. As follows from general considerations,  in $d=2$ there is no ordered phase. As is characteristic for asymptotically free theories the coupling $g^2$ increases in the infrared and diverges at a finite value $\Lambda$ of the RG scale $\mu$. Meanwhile the squashing parameter $\beta$ (not to be confused with the inverse temperature) tends to flow towards the more symmetric $\beta=0$ limit.

	For $d>2$, an ordered phase appears in which the coupling $g^2$ flows to zero in the infrared. Of course we were implicitly assuming the existence of such a phase with spontaneously broken symmetry in the previous chapter. All trajectories for $\beta>0$ have the same non-trivial UV fixed point at $\beta=1$ and there is a critical trajectory that flows to the $O(2N)$ symmetric Wilson-Fisher fixed point at $\beta=0$. Such a scenario was first studied for the most physically relevant $N=2$ case \cite{azaria1990nonuniversality}, but departures of experimental and numerical results from the expected $O(4)$ Wilson-Fisher critical exponents led to lingering doubts about the validity of this weak coupling analysis.
	
	To address these concerns, in \cite{azaria1995massive} the RG flow was additionally considered from the point of view of large $N$ methods. However it was stressed in that paper that the large $N$ results ``partially contradict those of the weak coupling analysis.'' To see the problem let us rewrite the RG equations \eqref{Eq.2 RG 1} and \eqref{Eq.2 RG 2} in terms of the 't Hooft coupling $\lambda\equiv Ng^2$, and also set $N=\mathcal{N},\,\epsilon =0$ for simplicity,
	 \begin{gather}
		\mu\der{}{\mu}\lambda=-\frac{\lambda^2}{\pi}\left[1-\frac{1-\beta}{ N}+\frac{\lambda}{2\pi N}\left(4-6(1-\beta)+\left(3-N^{-1}\right)(1-\beta)^2\right)\right]\label{Eq.4 RG weak coupling}\\
		\mu\der{}{\mu}\left(1-\beta\right)=-\frac{\lambda}{\pi}\beta(1-\beta)\left[1+\frac{\lambda}{N\pi}\left(1+\beta\right)\right].
	\end{gather}
	On the other hand, the beta function for $\lambda$ has also been calculated by large $N$ methods up to order $N^{-1}$ \cite{CAMPOSTRINI1994680},
	\begin{align}
\mu\der{}{\mu}\lambda=-\frac{\lambda^2}{\pi}\left[1+\frac{\lambda}{2\pi N}\left(1+\frac{3-2 \pi\gamma}{1+\lambda\left(\gamma - \frac{1}{\pi}\right)}\right)\right].\qquad \text{(Large $N$)}\label{Eq.4 RG large N}
	\end{align}
	The parameter $\gamma$, which was defined back in \eqref{Lagr squashed sphere A psi}, is an RG invariant to the lowest order of the large $N$ expansion,
	\begin{align}
		\gamma=\frac{1-\beta}{\lambda \beta}.\label{Def gamma}
	\end{align}
	
	 We will review the calculation of this beta function in Sec. \ref{Sec 4.2}. For now simply note that the large $N$ expansion would be expected to be exact in $\lambda, \beta, \gamma$ for each order in $N$. However if we substitute in the definition of $\gamma$ into \eqref{Eq.4 RG large N} and expand to first order in $\lambda$ we see a discrepancy,
	 \begin{align}
	\mu\der{}{\mu}\lambda=-\frac{\lambda^2}{\pi}\left[1-\frac{1-\beta}{ N}+\frac{\lambda}{2\pi N}\left(4-5(1-\beta)+2(1-\beta)^2\right)+ \order{\lambda^2}\right].\qquad\text{(Large $N$)}\label{Eq.4 RG large N 2}
	 \end{align}
	This is frustratingly close to the weak coupling result \eqref{Eq.4 RG weak coupling}, but it differs in the $\beta$ dependent two-loop terms. In \cite{azaria1995massive}, Azaria et al suggest that the large $N$ analysis is only valid near the fixed point at $\beta=1$. In order to study the behavior near $\beta=0$ they consider a double expansion in terms of both $1/N$ and $\beta$ about the $O(2N)$ fixed point.\footnote{This double expansion was first considered in \cite{chubukov1994quantum}, which studied the system as a quantum system in two spatial dimensions, considering the effect of both thermal and quantum fluctuations much as in the style of the textbook \cite{sachdev2011quantum}. See also \cite{chubukov1996crossover}.}

	Azaria et al in \cite{azaria1995massive} mention a few possibilities for the source of the discrepancy between the large N and weak coupling approaches. One point made is that the weak coupling expansion neglects non-perturbative corrections on the order of $\exp\left(-1/\lambda\right)$. Another point made is that Campostrini and Rossi in \cite{CAMPOSTRINI1994680} neglected the running of the squashing parameter $\gamma$ (this criticism is repeated in \cite{basso2013integrability}). The main culprit is suggested to be that there is a problem with introducing the auxiliary gauge field $A$ in \eqref{Lagr squashed sphere A psi} when the parameter $\gamma$ is large (this is echoed in \cite{chubukov1996crossover}).
	
	In this chapter I will instead argue that both the large $N$ and weak coupling approaches are valid, and that the discrepancy is instead simply due to a difference in renormalization scheme. Sigma models with target spaces that depend on more than one parameter may become scheme dependent already at two-loops rather than three (see e.g. \cite{metsaev1987order}), and in Sec. \ref{Sec 4.2} I will show exactly where this discrepancy arises.
	
	These issues of scheme dependence as well as the intrinsic lack of non-perturbative corrections may lead us to mistrust the weak coupling expansion. But in Sec. \ref{Sec 4.3} I will consider adding a Wess-Zumino (WZ) term to the squashed sphere sigma model, and show that the weak coupling beta function at two-loops does agree with exact results arising from the WZW conformal field theory.
	
	But first I will introduce large $N$ methods in the following section and then move on to considering the squashed sphere model in $d=1$ in Sec \ref{Sec 4.1}. The calculation is taken from a larger work \cite{schubring2021lessons} in which I consider a number of simple sigma models in $d=1$ both to introduce quantum field theory methods and to consider some subtleties of the large $N$ expansion. By beginning with this simpler case (which nevertheless requires regularization) we may confirm that the large $N$ expansion of the squashed sphere model actually leads to exact results for all values of $\gamma$.

	 %[Such an analysis was done to two-loops in \cite{azaria1995massive}\footnote{See also \cite{chubukov1994quantum} which studies the system as a quantum system in two spatial dimensions, and investigates the effect of both thermal and quantum fluctuations much as in the style of the textbook \cite{sachdev2011quantum}.}. The result was as we calculated in \ds{put in equation}, and the RG flow is shown in Fig \cite{}. In particular for $d>2$ there is a non-trivial IR fixed point with $O(2N)$ symmetry separating the ordered and disordered phases.  

\section{Introduction to large $N$ methods}\label{Sec 4.0}
Before moving to features specific to the one-dimensional case, let us introduce the large $N$ approach in an arbitrary dimension $d$. It is convenient to rescale the fields $z$ so that they have a conventional propagator, and the constraint becomes $|z|^2=\frac{N}{2\lambda}$.
\begin{align}
	\Lagr = \left(\partial+iA\right)\bar{z}\cdot\left(\partial-iA\right){z}+m^2|z|^2+\frac{N\gamma}{2}A^2+\alpha\left(|z|^2-\frac{N}{2\lambda}\right). \label{Lagr large N}
\end{align}
 The constraint is enforced with a Lagrange multiplier field $\alpha(x)$. Due to the constraint the addition of a term $m^2|z|^2$ with an arbitrary parameter $m^2$ will not affect any correlation functions of $z$ or $A$, it is just equivalent to adding a constant to action. But such a term will affect expectation values involving $\alpha$, and $m^2$ will be chosen so that the VEV of $\alpha$ vanishes to lowest order in the large $N$ expansion.\footnote{We can not straightforwardly apply the constraint in expectation values involving $\alpha$. Some of the consequences of this are discussed in \cite{schubring2021lessons}.}
 
 From here one may proceed by integrating out the $z$ fields which only appear to quadratic order. Due to the $N$ distinct $z$ fields there is an overall factor of $N$ multiplying all of the new terms generated. Given our choice of parameters $\lambda$ and $\gamma$ there is also a factor of $N$ on the terms in the action which did not involve $z$ fields. So the action in terms of $\alpha$ and $A$ is multiplied by an overall potentially large factor $N$, and thus correlation functions are well approximated by their saddle point values with each additional correction suppressed by a power of $1/N$.
 
 This is a conceptually clean way to phrase the large $N$ expansion, but in practice it is more convenient to keep the $z$ fields in the action and only explicitly integrate out $z$ for the terms linear and quadratic in the auxiliary fields $(\alpha, A)$ in order to find both the saddle point condition on the parameter $m$ and the propagators of the auxiliary fields. For any vertices involving higher powers of the auxiliary fields, or for external $z$ propagators coming from correlation functions, we instead make use of the bare $z$ propagator and the $\alpha |z|^2$ and $A\bar{z}\partial z$ vertices in the original action.
 
 Since the terms linear in $\alpha$ in the original action involve the factor $(|z|^2-\frac{N}{2\lambda})$, we immediately see that for the linear terms to cancel after $z$ is integrated out we must have the saddle point condition
 \begin{align}
 	\int\frac{d^dk}{(2\pi)^d}\frac{1}{k^2+m^2}=\frac{1}{2\lambda}.\label{Eq.4 saddle point}
 \end{align} 
For $d= 2$ this \emph{tadpole} integral is logarithmically divergent, and enforcing this condition with fixed $m$ immediately implies the lowest order RG flow of $\lambda$. This condition will also be useful in dealing formally with some integrals arising in large $N$ perturbation theory since we can exchange a divergent integral for the parameter $\lambda$ and vice versa.

However, in $d=1$ the saddle point integral converges to $\frac{1}{2m}$, and we have the very simple condition $m=\lambda$.

To find the propagator of $\alpha$ we integrate out two $\alpha |z|^2$ vertices, leading to the quadratic term in the action
$$-\frac{N}{2}\int \frac{d^d p}{\left(2\pi\right)^d}\,J(p;m,d)\alpha(-p)\alpha(p)$$
where $J(p; m,d)$ is shorthand for the \emph{bubble} integral
\begin{align}
	J(p;m,d)&\equiv\int\frac{d^dk}{(2\pi)^d}\frac{1}{(k^2+m^2)\left((p-k)^2+m^2\right)}\label{Def Jintegral}\\&=\frac{\Gamma\left(2-\frac{d}{2}\right)}{(4\pi)^{d/2}m^{4-d}}\,\,{}_2 F_1\left(1,2-\frac{d}{2};\frac{3}{2};-\frac{p^2}{4m^2}\right).\label{Eq.4 hypergeometric}
\end{align}
So the propagator of $\alpha$ is just
\begin{align}
	-\frac{1}{NJ(p;m,d)}.\qquad \text{($\alpha$ propagator)}\label{Propagator alpha}
\end{align}
In $d=1$ it simplifies considerably,
\begin{align}
	J(p;m,1)=\frac{1}{m}\frac{1}{p^2+4m^2}.\label{Eq.4 J d1}
\end{align}
Usually we will suppress the mass and dimension argument of $J(p)$ when it is clear by context.

	\begin{figure}
	\centering
	\includegraphics[width=0.8\textwidth]{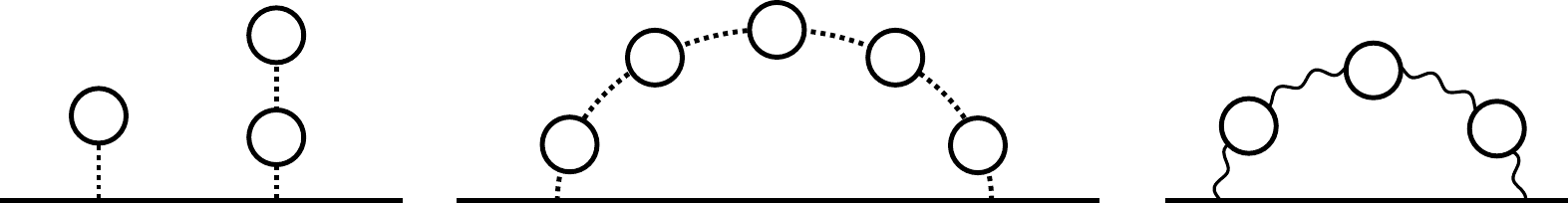}
	\caption{Examples of excluded tadpole and bubble diagrams in large $N$ perturbation theory. The $z$ field is represented by a solid line, the $\alpha$ field by a dashed line, and the $A$ field by a wavy line. }\label{Fig 4 excluded}
\end{figure}

Similarly we may find the $A$ propagator by forming a bubble diagram with two $A\bar{z}\cdot\partial z$ vertices. This leads to quadratic terms
\begin{align}
\frac{N}{2}\int \frac{d^d p}{\left(2\pi\right)^d}\,A_{\mu}(-p)\left[\left(\gamma+\lambda^{-1}\right)\delta_{\mu\nu}-\int \frac{d^d q}{(2\pi)^d}\frac{\left(p_\mu+2q_\mu\right)\left(p_\nu+2q_\nu\right)}{\left(q^2+m^2\right)\left(\left(q+p\right)^2+m^2\right)}\right]A_
\nu(p).\label{Eq.4 A quadratic terms}
\end{align}
This may be separated into longitudinal and transverse parts parallel and perpendicular to $p$, respectively. Of course in $d=1$ there is only one direction so we will defer calculating the transverse part until Sec \ref{Sec 4.2}. The longitudinal part may be found by contracting with the projector $p_\mu p_\nu/p^2$ and some formal manipulations of the integral. It is the same for all $d$,
\begin{align}
	\frac{1}{N\gamma}\frac{p_\mu p_\nu}{p^2}.\qquad \text{($A$ propagator, longitudinal part)}\label{Eq.4 Propagator A longitudinal}
\end{align}
This vanishes in the $O(2N)$ limit, and diverges in the gauge invariant $CP^{N-1}$ limit, as we might expect.

Now that we have summed over all tadpole and bubble integrals to fix $m^2$ and the auxiliary field propagators, these sub-diagrams must not further be included in large $N$ diagrams, see Fig \ref{Fig 4 excluded}. The exclusion of these sub-diagrams is actually closely linked to the constraint defining the sigma model, see Chapter 4 of \cite{schubring2021lessons}.

\begin{figure}[t]
	\centering
	\includegraphics[width=0.6\textwidth]{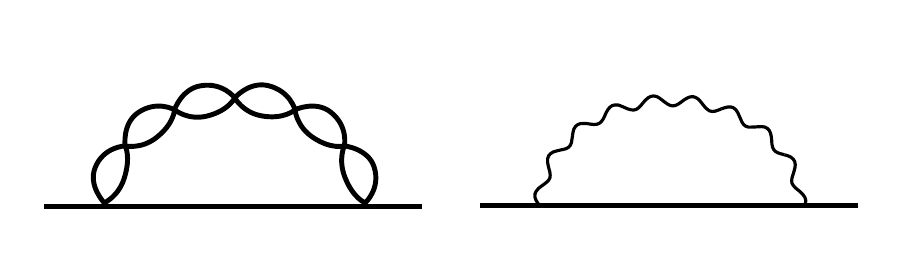}
	\caption{The summation over all bubble chains of arbitrary length constructed from the 4-point vertex $g^2\beta(\bar{z}\partial z)^2$ (as on the the left) is completely equivalent to the large $N$ propagator of the $A$ field (right).}
	\label{Fig.4 bubble chains}
\end{figure}

To conclude this section I will briefly remark on the interpretation in Azaria et al \cite{azaria1995massive} that the use of the $A$ field rather than the quartic vertex involves some approximation that is not justified in certain limits. Actually whenever we introduce a Hubbard-Stratonovich field like the $A$ field to decouple a four-point interaction term, the resulting diagrams are completely equivalent to leaving in the original four-point interaction, so this does not involve any additional approximation. The original $A$ propagator is like a delta function that glues together two $A\bar{z}\partial z$ vertices to form a $g^2\beta(\bar{z}\partial z)^2$ vertex, where the correct vertex factor $g^2\beta$ is coming from the bare propagator
$$\frac{1}{N(\gamma+ \lambda^{-1})}=g^2\beta.$$
The procedure of integrating out the $z$ bubble diagrams which leads to the additional integral term in \eqref{Eq.4 A quadratic terms} just corresponds to summing up chains of four-point vertices of arbitrary length as in Fig \ref{Fig.4 bubble chains}. These chains are all at the same order in the large $N$ expansion since each vertex gives a factor $g^2\beta\sim N^{-1}$, and each bubble provides a compensating factor of $N$.

Actually even the $\alpha$ propagator may be thought of from this perspective. If instead of a rigid constraint, we consider the sigma model to arise from a limit of a scalar field theory with a quartic interaction, then the $\alpha$ field is just the Hubbard-Stratonovich field that decouples the quartic interaction, and the $\alpha$ propagator is also seen to arise a limiting case of bubble chain diagrams. This is shown in chapter 6 of \cite{schubring2021lessons}.

\section{Squashed sphere model in $d=1$}\label{Sec 4.1}

\begin{figure}
	\centering
	\includegraphics[width=0.4\textwidth]{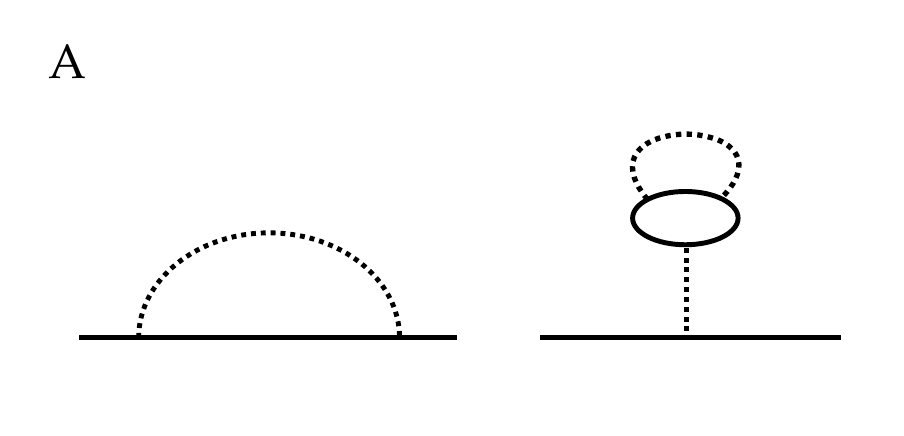}\qquad\includegraphics[width=0.4\textwidth]{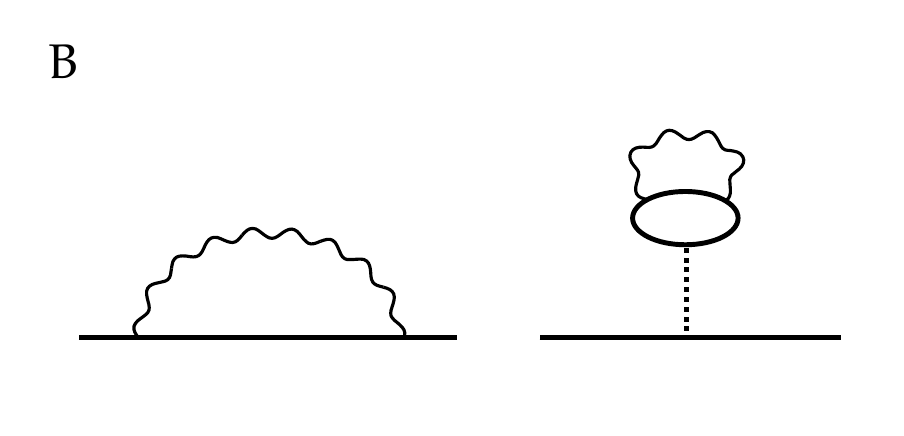}
	\caption{Corrections to the spinon propagator. The diagrams in A are due only to the $\alpha$ field and are also present in the $O(2N)$ model. The diagrams in B involve the gauge field. }\label{Fig 4 self energy}
\end{figure}

In any dimension the spinon propagator in the large $N$ limit is simply
\begin{align*}
\left\langle \bar{z}^i(p)z^j(-p)\right\rangle^{(0)}=\frac{\delta^{ij}}{p^2+m^2},
\end{align*}
where a superscript with parenthesis refers to the order in the large $N$ expansion. The spinon mass $m_z$ is given by the pole of the two-point function,
\begin{align*}
	m_z^2 = m^2 + \order{N^{-1}},
\end{align*}
and the $\order{N^{-1}}$ corrections are given by the self-energy corrections in Fig \ref{Fig 4 self energy}. Now specializing to the $d=1$ case, let us first focus on the corrections due to the $\alpha$ field in Fig \ref{Fig 4 self energy}(a), which are also present in the $O(2N)$ model.

\subsection{Corrections in the $O(2N)$ model}

Given the form of $J(k;m,1)$ \eqref{Eq.4 J d1}, the $\alpha$ arc diagram leads to
\begin{align*}
	\Pi_{\alpha, arc}(p)=\frac{m}{N}\int \frac{dk}{2\pi}
	\frac{k^2+4m^2}{\left((k+p)^2+m^2\right)}.
\end{align*}
This integral is clearly divergent so we need to consider some form of regularization. We will use the following simple cutoff scheme:
\begin{align*}
	\Pi_{\alpha, arc}(p)=	\frac{m}{N}\int \frac{dk}{2\pi}\left(1+\frac{\bcancel{-2pk }+p^2+ 3m^2}{k^2+m^2}\right)\quad= \frac{m}{N}\left(M+\frac{p^2+3m^2}{2m}\right).
\end{align*}
First the power-law divergence is subtracted off and labeled as $M$
\begin{align*}
	M \equiv \int \frac{dk}{2\pi}1.%\label{3.regularization}
\end{align*}
Then the logarithmically divergent part linear in $k$ in the numerator is taken to vanish due to symmetric integration, and only a finite part remains.

This may seem a rather naive prescription, but in fact it is closely related to the rules of dimensional regularization. The main practical difference from a dimensional regularization treatment of the integral is that in this case the power law divergences are kept track of through $M$ instead of being taken to vanish. This is useful since all power law divergences either cancel between diagrams, or have a real interpretation as contact terms.

Using the same scheme, the $\alpha$ tadpole diagram leads to
\begin{align*}
	\Pi_{\alpha, tadpole}&=-\frac{4m^4}{N}\int \frac{dk}{2\pi}\frac{dq}{2\pi}
\frac{k^2+4m^2}{\left((k+q)^2+m^2\right)\left(q^2+m^2\right)^2}\quad =-\frac{m}{N}\left(M+2m^2\right),
\end{align*}
so the total self-energy due to the $\alpha$ field is
\begin{align}
	\Pi_\alpha(p)\equiv \Pi_{\alpha, arc}(p)+\Pi_{\alpha, tadpole}= \frac{p^2-m^2}{2N}.
\end{align}

To find the $\order{N^{-1}}$ shift in the pole of the two-point function $$\frac{\delta^{ij}}{p^2+m^2+\Pi_\alpha(p)},$$ we may simply substitute in the value of the lowest order pole $p^2=-m^2$ into the self-energy
\begin{align*}
m_z^2=m^2+\Pi_\alpha(-m^2)=m^2\left(1-\frac{1}{N}\right).
\end{align*}
So in the $O(2N)$ model, where there is no $A$ correction, the spinon mass is corrected to
\begin{align}
	m_z=m\left(1-\frac{1}{2N}\right).\label{Eq.4 spinon alpha correction}
\end{align}

This is actually already the exact answer, which may be verified with the wave function methods of Appendix \ref{appendix wavefunction}.

Instead of just the $z$ fields we may consider two-point functions of the composite operators
\begin{align}
	O_{r,s}\equiv \bar{z}_{a_1}\bar{z}_{a_2}\dots \bar{z}_{a_r}z_{b_1}\dots z_{b_s},\label{Def O}
\end{align}
where the particular values of the $a$ and $b$ indices are not so important, but to avoid mixing of states we require that $a_i\neq b_j$ for any $i,j$. The two-point functions of the $O_{r,s}$ operators have poles at the bound state masses $m_j$, where $j=r+s$,
\begin{align}
	m_j=\frac{g^2}{2}\left(2Nj+j(j-2)\right).\label{Eq.4 d1 spectrum}
\end{align}

Again it turns out that this is an exact expression, and the $\order{N^{-2}}$ corrections may explicitly be shown to vanish in large $N$ perturbation theory \cite{schubring2021lessons}.

It may be illuminating to rewrite this in terms of the number of real fields $N_R\equiv 2N$ which may be extended to odd values. For $N_R=3$, we just have the well-known expression for the spectrum of a quantum rotor, $m_j=\frac{g^2}{2}j(j+1)$.

\subsection{Gauge theories in $d=1$}
To introduce the peculiarities of gauge theories in one dimension, consider the $CP^1$ model with an auxiliary gauge field $A$ and normalization $|z|^2=1$,
	\begin{align*}
	\Lagr_{CP^1}= \frac{1}{2g^2}\left(\partial + i A\right) \bar{z}\cdot \left(\partial - i A\right)z.
\end{align*}
Gauge theories in $d=1$ are very simple. Since $A$ has only one component it can always be set to zero by a field redefinition. Simply take $z(\tau)\rightarrow e^{i\phi(\tau)}z(\tau)$ with
$$\phi(\tau)=-\int_{\tau_0}^\tau d\sigma A(\sigma).$$
Then the Lagrangian becomes $$\frac{1}{2g^2}\partial \bar{z}\cdot \partial z,$$ which is seemingly identical to the $O(4)$ model. 

But recall that the $CP^1$ model may also be written in terms of three real fields \eqref{Lagr cp1 S},
$$\frac{1}{8g^2}\left(\partial {S}\right)^2,$$
which is identical to the $O(3)$ model, which has a different spectrum and correlation functions. How is the apparent discrepancy resolved?

Consider the even energy levels $j=2l$ in the $O(4)$ model using \eqref{Eq.4 d1 spectrum},
$$m_{2l}=2g^2 l(l+1).$$
These are exactly the energy levels in the $O(3)$ model after the difference $g^2\rightarrow 4g^2$ in the normalization is taken into account. It makes sense that only the even energy levels of the $O(4)$ model are involved in the $O(3)$ model since the ${S}$ fields are bilinears of the $z$ fields. But also there are only 3 independent single particle states in the $O(3)$ model, whereas there are 9 independent two particle states in the $O(4)$ model. So not only are the odd energy levels of $O(4)$ entirely absent in $O(3)$, the multiplicity of the even levels is drastically reduced too.

In some sense what is happening is that the inclusion of the gauge field $A$ is filtering out everything but the gauge invariant sector of the theory, but it is leaving that sector completely unchanged. This may be understood from various points of view. In particular, how this occurs diagrammatically due to the additional $A$ fields in large $N$ diagrams is rather obscure, and that will be our main focus.

The main problem is that in $d=1$ the transverse propagator for the $A$ field vanishes. So even though if we integrate out the $z$ fields in the $CP^{N-1}$ action there may be higher order vertices coupling the $A$ field and the $\alpha$ field, it is difficult to see how to use these in a perturbative expansion since there is no quadratic term for the $A$ fields. But of course in our case there is a clear solution since we are modifying $CP^{N-1}$ to be the squashed sphere model with $\gamma$ not necessarily zero \eqref{Eq.4 Propagator A longitudinal}, and this has a constant longitudinal propagator $(N\gamma)^{-1}$ which we may work with.

The self-energy correction from the gauge field arc diagram is
\begin{align*}
	\Pi_{\parallel,arc}(p)=-\frac{1}{\gamma N}\int \frac{dk}{2\pi}\frac{\left(2p+k\right)^2}{(k+p)^2+m^2}\quad= -\frac{1}{\gamma N}\left(M+\frac{p^2-m^2}{2m}\right).
\end{align*}
The corresponding tadpole diagram only serves to cancel the divergence $M$, and so the total correction to the spinon mass due to the gauge field is $$\delta m_z^2=\Pi_\parallel (-m^2)= +\frac{m}{N\gamma},$$
or after taking the square root,
\begin{align}
	\delta m_z = + \frac{1}{2N\gamma}.\label{Eq.4 deltam1}
\end{align}
In the gauge invariant limit $\gamma\rightarrow 0$ the energy of the single spinon state diverges, and so the non-gauge invariant fundamental representation does indeed leave the spectrum.

\subsection{Full spectrum of the squashed sphere sigma model}

Now let us consider how the gauge field affects the spectrum in general. We may consider the correlation functions $\langle O^\dagger_{r,s}(\tau)O_{r,s}(0)\rangle$ of the composite operators defined in \eqref{Def O}. If we again perform a field redefinition in the path integral $z(\tau) \rightarrow e^{i\int^\tau_{0} A(\sigma)d\sigma}z(\tau)$, the $A$ fields and $z$ fields completely decouple in the Lagrangian so the correlation function factorizes,
\begin{align*}
	\left\langle O^\dagger_{r,s}(\tau)O_{r,s}(0)\right\rangle=\left\langle e^{i(r-s)\int^\tau_0 A(\sigma)d\sigma}\right\rangle_{A \text{ action}}\left\langle O^\dagger_{r,s}(\tau)O_{r,s}(0)\right\rangle_{O(2N)\text{ model} }.
\end{align*}
We see that if $r=s$ so that the operators $O_{r,s}$ are gauge invariant, then there is no change in the correlation function compared to the ordinary $O(2N)$ model. The formula \eqref{Eq.4 d1 spectrum} for the spectrum which was calculated in \cite{schubring2021lessons} is still valid for gauge invariant operators in the squashed sphere model.

If $r\neq s$ there is an additional factor which only depends on the quadratic Lagrangian $\frac{N\gamma}{2}A^2$. Using the propagator $$\left\langle A(\tau)A(0)\right\rangle = (N\gamma)^{-1}\delta(\tau),$$
and Wick's theorem we can show
\begin{align}
	\left\langle O^\dagger_{r,s}(\tau)O_{r,s}(0)\right\rangle=e^{-\frac{(r-s)^2}{2N\gamma }|\tau|}\left\langle O^\dagger_{r,s}(\tau)O_{r,s}(0)\right\rangle_{O(2N)\text{ model} }.
\end{align}
So states in the squashed sphere model with $r\neq s$ get a correction to their mass $\delta m_{r,s}$ which diverges as $\gamma \rightarrow 0$,
\begin{align}
	\delta m_{r,s}= +\frac{(r-s)^2}{2N\gamma }.\label{7.exact correction}
\end{align}
In particular this shows that the $\order{N^{-1}}$ correction $\delta m_z \equiv \delta m_{0,1}$ that we calculated via large $N$ perturbation theory was actually exact. And as we wished to show, gauge-invariant operators have the exact same correlation functions as $O(2N)$, and non-gauge invariant states leave the spectrum as we go to the gauge-invariant $CP^{N-1}$ limit.

This formula for the spectrum of the squashed sphere model will be derived from another perspective by considering the spectrum of the Laplace-Beltrami operator on the squashed sphere in Appendix \ref{appendix wavefunction}. In the remainder of this section we will sketch briefly a third method to find it solely by considering large $N$ diagrams and some combinatorics.

\begin{figure}[tbp]
	\centering
	\includegraphics[width=0.7\textwidth]{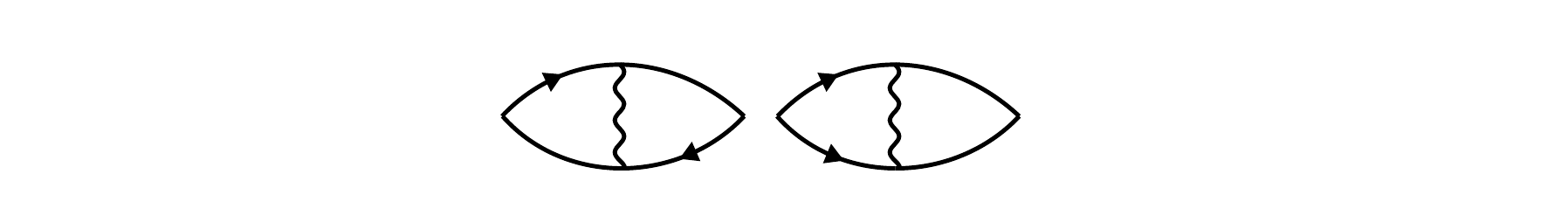}
	\caption{Corrections to the two-point functions for $O_{1,1}$ (opposite charges) and $O_{2,0}$ (like charges), respectively. The orientation of a $z$ line is indicated by an arrow.}
	\label{Fig.4 2body}
\end{figure}

The corrections to the two-point functions $\left\langle O^\dagger_{r,s}(\tau)O_{r,s}(0)\right\rangle^{(1)}$ come from two sources. There are corrections to the individual $r+s$ single particle lines, and also two-body interactions between two distinct lines as in Fig \ref*{Fig.4 2body}. Higher $n$-body interactions don't appear at this order of the large $N$ expansion.

The mass of a two-body bound state at lowest order in large $N$ is clearly just $2m$. Following the discussion of non-perturbative consequences of the constraint in \cite{schubring2021lessons}, it can be shown that if the two-body mass is corrected to $2m+\delta m_2$ then the exact 2-point function must be
\begin{align*}
	\frac{1}{m}\frac{1+\frac{\delta m_2}{2m}}{p^2+(2m+\delta m_2)^2}=\frac{1}{m}\frac{1}{p^2+4m^2}+\frac{p^2-4m^2}{2m^2(p^2+4m^2)^2}\delta m_2+\order{\delta m_2^2}.
\end{align*}
Evaluating the diagrams in Fig \ref*{Fig.4 2body} leads to a correction precisely of this form, so the two-body interaction energy can be identified as $$\pm 2\frac{1}{2N\gamma}$$
where like charges repel and opposite charges attract. So for the gauge invariant operators $O_{1,1}=\bar{z}_a z_b$ the large positive correction to the two single particle masses is precisely canceled by the large binding energy of the system, and overall the gauge field has no effect.

A simple combinatorial argument allows us to find the correction $\delta m_{r,s}$ for arbitrary states $O_{r,s}|0\rangle$.
\begin{itemize}
	\item There are $r+s$ single particle lines, each of which has a mass correction \eqref{Eq.4 deltam1}:
	$$\frac{r+s}{2N\gamma}.$$
	\item There are $rs$ diagrams where two particles with opposite charge interact:
	$$-\frac{2rs}{2N\gamma}.$$
	\item Finally, counting the number of combinations where particles with the same charge interact:
	$$+2\left(\frac{r(r-1)}{2}+\frac{s(s-1)}{2}\right)\frac{1}{2N\gamma}.$$
\end{itemize}
So in total, there is a correction of $\delta m_{r,s}=(r-s)^2\frac{1}{2N\gamma}$ which is exactly what was found above and in Appendix \ref{appendix wavefunction} using more abstract arguments.

%\footnote{In \cite{schubring2021lessons}, I left a four-point vertex $A^2|z|^2$ in the Lagrangian rather than using the constraint to eliminate the $|z|^2$ factor, and the cancellation of $M$ was said to occur through a diagram involving this vertex. It is easy to show using the methods of chapter 4 in \cite{schubring2021lessons} that choosing to include these diagrams or not merely changes the value correlation functions involving $\alpha$, and not the physical correlation functions.}
\section{Large N vs weak coupling in $d=2$}\label{Sec 4.2}
So we have just seen that the large $N$ expansion is actually quite innocuous and can give exact results for quantities such as the spectrum in $d=1$. Now we will consider the situation in $d=2$, which is our main focus. As far as the large $N$ expansion is concerned, this only differs from the $d=1$ case in terms of the saddle point expression for $m^2$ \eqref{Eq.4 saddle point} and the exact form of the bubble integral $J(p)$.

\subsection{Propagators for $d=2$}

The function $J(p)$ \eqref{Def Jintegral} in $d=2$ is
\begin{align}
J(p;m,2)= \frac{1}{2\pi p^2\xi}\log \left(\frac{\xi+1}{\xi-1}\right),\qquad \xi\equiv \sqrt{1+4 \frac{m^2}{p^2}}.\label{Eq.4 Jintegral 2d}
\end{align}
Two limits of this will be useful,
\begin{gather}
	J(0)=\frac{1}{4\pi m^2},\\
	J(p^2\gg m^2)=\frac{1}{2\pi p^2}\log\frac{p^2}{m^2}+\order{\frac{m^2}{p^4}\log\frac{p^2}{m^2}}.\label{Eq.4 J asymptotic limit}
\end{gather}
If needed, the higher order corrections in the $p^2\ll m^2$ limit follow simply from the series expansion of the hypergeometric function in \eqref{Eq.4 hypergeometric}, and a closed form for the $ p^2\gg m^2$ expansion is derived in \eqref{Eq.App bubble}.

The function $J(p)$ determines the $\alpha$ propagator \eqref{Propagator alpha}, and the longitudinal part of the $A$ propagator \eqref{Eq.4 Propagator A longitudinal} is the same as the $d=1$ case, but now we must calculate the transverse part as well. This is schematically like
\begin{align}
	\frac{1}{N}\Delta(p)\left(\delta_{\mu\nu}-\frac{p_\mu p_\nu}{p^2}\right).\qquad \text{($A$ propagator, transverse part)}
\end{align}
The function $\Delta(p)$ may be found by acting on the bracketed expression in \eqref{Eq.4 A quadratic terms} with a transverse projector $(d-1)^{-1}\left(\delta_{\mu\nu}-p_\mu p_\nu/p^2\right)$. We obtain after some manipulation
\begin{align*}
	\Delta^{-1}=\gamma+\frac{1}{d-1}\left[2\int \frac{d^d q}{(2\pi)^d}\frac{d-2}{q^2+m^2}+\left(p^2+4m^2\right)J(p)\right],
\end{align*}
The divergent integral came about in part by expressing $\lambda^{-1}$ in terms of the saddle point condition \eqref{Eq.4 saddle point}. Combined with the factor of $d-2$ this has a finite limit as $d\rightarrow 2$, so this can be reduced to
\begin{align}
	\Delta(p)=\left[\left(p^2+4m^2\right)J(p)+\gamma-\frac{1}{\pi}\right]^{-1}.
\end{align}
Notice that at $p^2=0$, $\left(p^2+4m^2\right)J(p)$ cancels exactly with the $-1/\pi$ term, which is a term that depended on the subtleties of dimensional regularization. A careful use of Paulli-Villars regularization also works \cite{d19781n}, but a naive cutoff regularization in which the term proportional to $d-2$ would simply vanish does not.\footnote{I stress this point because Campostrini and Rossi \cite{CAMPOSTRINI1994680} take this propagator as given and then go on to calculate in a cutoff regularization scheme. Apparently the results are consistent, but the discrepancy with the weak coupling results will be shown to arise from details of the regularization, so this is worth keeping in mind.}

For $0<\gamma<\frac{1}{\pi}$ the function $\Delta$ has a pole at $p^2=-m_A^2$ \cite{CAMPOSTRINI1994680}, which corresponds to the mass $m_A$ of a singlet bound state of spinons. Since $m_A$ is a physical mass to lowest order in large $N$, and only depends on $m$ and $\gamma$ (and not $\lambda$ or the RG scale $\mu$ explicitly), this implies that $\gamma$ is an RG invariant to lowest order in large $N$.

As a consequence of this argument and the definition of $\gamma$ in terms of $\beta$ \eqref{Def gamma}, the lowest order RG flow of $\lambda$ arising from the saddle point condition \eqref{Eq.4 saddle point} also determines the RG flow of $\beta$,
\begin{align*}
		\mu\der{}{\mu}\lambda&=-\frac{\lambda^2}{\pi}+\order{N^{-1}}\\
\mu\der{}{\mu}\left(1-\beta\right)&=-\frac{\lambda}{\pi}\beta(1-\beta)+\order{N^{-1}}
\end{align*}
This is already the lowest order truncation of the weak coupling RG equations \eqref{Eq.4 RG weak coupling}, and conversely if we begin from those more exact equations we can confirm that $\gamma$ is an RG invariant to lowest order in large $N$. So despite some criticisms in later works, Campostrini and Rossi were correct in taking $\gamma$ to be an invariant to the order that they were working in. Even if $\gamma$ is invariant, the squashing parameter $\beta$ does flow towards the $O(2N)$ symmetric limit in a manner that agrees to this order with the weak coupling approach.\footnote{See Appendix \ref{appendix notation} for a comparison of notation across papers.}

\subsection{Spectrum and integrability}
Now is perhaps a good time to discuss what is known about the spectrum. The bound state associated with the pole of $\Delta$ becomes marginally stable at $\gamma=1/\pi$ at which point its mass is $m_A=2m$.  For $\gamma>1/\pi$ the spinons are the only stable particles in the spectrum.

Extending beyond the lowest order of the $N$ limit, $\gamma$ does run, but it converges to a finite limit in the UV, and this finite limit $\gamma_0$ can be used as a true RG invariant (by construction) in place of $\gamma$ \cite{BALOG2000367,basso2013integrability}. At $\gamma_0=1/\pi$ the model is actually integrable, which was suggested by Campostrini and Rossi \cite{CAMPOSTRINI1994680} based on the fact that the squashed sphere model is equivalent to fermions minimally coupled to the $CP^{N-1}$ model \cite{abdalla1990numerical} which has a known factorizable S-matrix \cite{koberle1987deconfinement}. This was considered in more detail by Basso and Rej \cite{basso2013integrability} who stressed the running of $\gamma$ and the necessity of a Thirring term in the fermionic description.

For $N=2$ the model is integrable for all values of $\gamma_0$, and so the spectrum is exactly known, although the $\gamma_0\rightarrow 0$ limit is rather subtle \cite{WIEGMANN1985209}. As $\gamma_0$ goes to zero the mass of the spinons diverges while the adjoint two-body bound states remain finite, as was seen in the much simpler $d=1$ case last section. Besides the lowest adjoint multiplet there is a singlet multiplet of the same mass which may associated with the additional degree of freedom coupled to the $CP^{1}$ model. It decouples from the adjoint particles which comprise the $CP^{1}$ model in the $\gamma_0\rightarrow 0$ limit. Just as in the sine-Gordon model\footnote{The 2-body S-matrix involves a sine-Gordon factor, so this is an exact statement.\cite{WIEGMANN1985209}}, as the mass of the spinons increases more and more excited bound states appear in the spectrum. However in the limit $\gamma_0\rightarrow 0$ all of these excited states become marginally stable and decompose into the lowest adjoint multiplet.

For $N>2$, there is not an exact S-matrix for the region $0<\gamma_0<1/\pi$ so the situation is somewhat less certain, although some insight can be gained by considering bound states in the spinon-spinon potential via ordinary quantum mechanics \cite{CAMPOSTRINI1994680}, and the picture involving excited bound states seems to be much the same. Certainly the divergence of the mass of the spinons as $\gamma_0\rightarrow 0 $ is revealed by the next order correction in large $N$, and that will be our next focus.

Since there is integrability at $\gamma_0=1/\pi$ for all $N$, the thermodynamic Bethe ansatz (TBA) may be used to relate the weak coupling RG invariant $\Lambda$ and the exact spinon mass $m_z$ \cite{CAMPOSTRINI1994680,basso2013integrability},
\begin{align}
m_z = \frac{(2/e)^{1/N}}{\Gamma\left(1+\frac{1}{N}\right)}\Lambda.\label{Eq.4 TBA relation}
\end{align}
This agrees with the TBA relation for $N=2$ \cite{BALOG2000367} which is valid for all $\gamma_0$. Even in the absence of an exactly known relation between the RG invariants $m_z, \Lambda$ it makes more sense to compare these RG invariants rather than the two-parameter beta functions directly. So our approach will be to first calculate the $\order{1/N}$ correction to $m_z$, which will determine the next order RG flow of $\lambda$ \eqref{Eq.4 RG large N} stated above. Then we will compare this to the RG invariant $\Lambda$ arising from the two-loop beta functions \eqref{Eq.4 RG weak coupling} in the weak coupling approach.

\subsection{Spinon mass correction and large $N$ RG flow}

The expression for the physical spinon mass $m_z^2$ as a function of $M, \lambda, \gamma$ determines the beta functions $\beta_\lambda, \beta_\gamma$ (not to be confused with the squashing parameter $\beta$) through a Callan-Symanzik equation
\begin{align*}
M^2\derOrd{}{M^2}m_z^2(M,\,\lambda(M),\,\gamma(M))=0.
\end{align*}
This can be put in a useful form by writing $m_z^2=m^2+\Pi^{(1)}+\order{N^{-2}}$ where $\Pi^{(1)}$ is the correction to $m^2_z$ at lowest non-trivial order, which is obtained by setting $p^2=-m^2$ in the self-energy. Using this expansion and the lowest order equation
\begin{align*}
M^2\der{}{M^2}m^2(M,\lambda)+\frac{1}{2}\beta^{(0)}_\lambda\der{}{\lambda}m^2(M,\lambda),
\end{align*}
the mass correction and the correction to the beta function may be related
\begin{align}
	\beta_\lambda^{(1)}=-\frac{\lambda^2}{\pi}\frac{M^2}{m^2}\der{}{M^2}\Pi^{(1)}(M^2, m^2, \gamma).\label{Eq.4 callan symanzik}
\end{align}
In this expression the self-energy is expressed in terms of $m^2$ rather than $\lambda$, and the derivative only acts on explicit appearances of $M^2$. This expression depended on $\gamma$ being an RG invariant to lowest order, $\beta_\gamma^{(0)}=0$.

%\footnote{It is worth noting that since the free energy density is an RG invariant and is proportional to $m^2$ with no $\gamma_0$ dependence at lowest order of the large $N$ limit then it must also be proportional to $m_z^2$ at higher order. This is an alternative check on the large $N$ calculations which is considered in \cite{CAMPOSTRINI1994680} and related models \cite{biscari1990quantitative,campostrini1992cp}.}. 

The $\order{N^{-1}}$ correction to the mass $m_z$ depends on the self-energy diagrams in Fig \ref{Fig 4 self energy}, and it was first calculated in \cite{CAMPOSTRINI1994680}. The self-energy correction due to the $\alpha$ arc is
\begin{align}
	\Pi_{\alpha, arc}(p)=\frac{1}{N}\int \frac{d^2k}{(2\pi)^2}
	\frac{1}{J(k)\left((k+p)^2+m^2\right)},\label{Eq.4 alpha arc}
\end{align}
and the correction due to the corresponding tadpole is
\begin{align}
	\Pi_{\alpha, tadpole}&=-\frac{1}{N J(0)}\int \frac{d^2k}{(2\pi)^2}\frac{d^2q}{(2\pi)^2}
	\frac{1}{J(q)\left((k+q)^2+m^2\right)\left(q^2+m^2\right)^2}\non
	&=-\frac{1}{N}\int \frac{d^2k}{(2\pi)^2}
	\frac{1}{J(k)\left(k^2+4m^2\right)}-\frac{1}{NJ(0)}\int \frac{d^2k}{(2\pi)^2}
	\frac{1}{k^2+4m^2}.\label{Eq.4 alpha tadpole}
\end{align}
The first term in the tadpole diagram cancels most of the UV divergence of the arc diagram, but a $\log \log$ divergence remains as well as the $\log$ divergence due to the second term in the tadpole. Regulating with a UV cutoff $M$, the correction to $m_z^2$ arising from the $\alpha$ diagrams alone is
\begin{align}
	\Pi_\alpha(p^2=-m^2)=\frac{m^2}{N}\left(-\log \frac{M^2}{4m^2}+\log\log \frac{M^2}{m^2}+\text{finite}\right).\label{Eq.4 alpha corrections}
\end{align}
The finite part is not so important for our purposes, but it may be calculated to be the Euler gamma constant \cite{biscari1990quantitative}. This expression already determines the $\order{1/N}$ correction to the beta function for the $O(N)$ model.

%$$d\left(\frac{1-\beta}{\lambda\beta}\right)=\frac{1-\beta}{\beta}d\left(1/\lambda\right)-\frac{1}{\lambda\beta}d\beta-\frac{1-\beta}{\lambda\beta^2}d\beta=0\qimplies d\beta = \lambda\beta(1-\beta)d(1/\lambda)$$
%Considering the picture of the $d=2$ squashed sphere model as a massless fermion coupled to the $CP^{N-1}$ model, the particle associated to the $A$ field is like a bound state of the fermions, as in the massless Schwinger model. In the limit of $\gamma\rightarrow 0$ the mass of the particle goes to zero and this corresponds to the massless field that was coupled to the $CP^{N-1}$ model that decouples in the $\gamma\rightarrow 0$ limit.
The gauge field diagrams may be split into transverse and longitudinal components. The transverse arc leads to a correction,
\begin{align}
	\Pi_{\perp,arc}(p)=-\frac{4}{N}\int \frac{d^2k}{(2\pi)^2}\Delta(k)\frac{k^2p^2-(k\cdot p)^2}{k^2\left((k+p)^2+m^2\right)},\label{Eq.4 Pi perp arc}
\end{align}
and the transverse tadpole,
\begin{align}
	\Pi_{\perp,tadpole}&=\frac{1}{N J(0)}\int \frac{d^2k}{(2\pi)^2}\Delta(k)J(k)\non
	&=-\frac{1}{N J(0)}\int \frac{d^2k}{(2\pi)^2}\Delta(k)\frac{\gamma-\frac{1}{\pi}}{k^2+4m^2}+\frac{1}{N J(0)}\int \frac{d^2k}{(2\pi)^2}\frac{1}{k^2+4m^2}.\label{Eq.4 Pi perp tadpole}
\end{align}
The second term cancels with the second term of the $\alpha$ tadpole \eqref{Eq.4 alpha tadpole}.

The self-energy $\Pi_\parallel$ due to the longitudinal propagator of the $A$ field vanishes upon setting $p^2=-m^2$ and thus will not lead to a correction to $m_z^2$,%\footnote{This might seem to contradict our results for the $d=1$ case, where we did calculate a mass correction. The resolution is that when we generalize the $A$ propagator in $d=1$ to a propagator with indices it should be understood as being proportional to a Kronecker delta $\delta_{\mu\nu}$ rather than a longitudinal projector $p_\mu p_\nu/p^2$.  So the mass correction may still be understood as coming from the limit $d\rightarrow 1$ of the transverse self energy $\Pi_\perp$.}
\begin{align}
	\Pi_\parallel(p)=\frac{1}{\gamma N}\int \frac{d^dk}{(2\pi)^d}\frac{p^2+m^2}{k^2}\left(1-\frac{p^2+m^2}{(k+p)^2+m^2}\right).\label{Eq.4 Pi longitudinal}
\end{align} 
Given our results in $d=1$ in the last section, a brief remark is in order. The expression for $\Pi_\parallel$ actually did not depend on the number of dimensions $d$, and so one might wonder how it could possibly vanish given that we calculated a correction to the mass in $d=1$ last section. To clarify this, it may be rewritten in a form closer to the steps taken in that context,
\begin{align*}
\Pi_\parallel(p)=	-	\frac{1}{\gamma N}\int \frac{d^dk}{(2\pi)^d}\frac{p^2-m^2}{k^2+m^2}-\frac{4}{\gamma N}\left(p^2\delta_{\mu\nu}-p_\mu p_\nu\right)\int\frac{d^dk}{(2\pi)^d}\frac{k_\mu k_\nu}{k^2\left((p+k)^2+m^2\right)}.
\end{align*}
The first term gives the mass correction \eqref{Eq.4 deltam1} and the second term would seem to vanish as $d\rightarrow 1$ given it involves an orthogonal projector acting on a finite integral. But in fact if this integral depending on indices $\mu, \nu$ is decomposed into scalar integrals in the standard way it can be shown that this term does not vanish, and it cancels with the first term for $p^2=-m^2$. The resolution is that the index structure of the auxiliary $A$ field propagator in $d=1$ should be understood as being proportional to a Kronecker delta $\delta_{\mu\nu}$ rather than a longitudinal projector $p_\mu p_\nu/p^2$.  So the transverse self energy $\Pi_\perp$ may still give a correction to the mass in the limit $d\rightarrow 1$.

Now let us calculate the correction arising from $\Pi_\perp$ for $d=2$ which is our main focus in this section. This may be done by integrating over the angular part of \eqref{Eq.4 Pi perp arc} and setting $p^2=-m^2$, %from $\int \frac{d\theta}{2\pi}\frac{\sin^2\theta}{p^2+k^2+m^2+2pk\cos\theta}=\frac{(k^2+p^2+m^2)-\sqrt{(k^2+p^2+m^2)^2-4p^2k^2}}{4p^2q^2}$
\begin{align}
	\Pi_{\perp,arc}(p^2=-m^2)&=-\frac{1}{4\pi N}\int_0^{M^2} dk^2 \Delta(k)\left[1-\sqrt{1+4\frac{m^2}{k^2}}\right]\non
	&=\frac{m^2}{N^2}\int_0^{M^2}  \frac{dk^2}{k^2\left(\log \frac{k^2}{m^2}+2\pi\gamma -2\right)}+\text{finite}.\label{Eq.4 Pi perp arc 2}
\end{align}
Combining this with the tadpole \eqref{Eq.4 Pi perp tadpole} and the $\alpha$ field corrections \eqref{Eq.4 alpha corrections},
\begin{align}
	\Pi(p^2=-m^2)=\frac{m^2}{N}\left(\log\log \frac{M^2}{m^2}+\left(3-2\pi\gamma\right)\log\left(\log \frac{M^2}{m^2}+2\pi\gamma - 2\right)+\text{finite}\right)+\order{N^{-2}}.\label{Eq.4 large N mass correction}
\end{align}
Here the finite part will depend on $\gamma$, and it is calculable both numerically and analytically for the integrable $\gamma=1/\pi$ special case \cite{CAMPOSTRINI1994680}. Using the Callan-Symanzik equation \eqref{Eq.4 callan symanzik} this expression for $\Pi$ determines the large $N$ beta function \eqref{Eq.4 RG large N} introduced in the beginning of this chapter.

\subsection{Weak coupling invariant $\Lambda$ and scheme dependence}

Now we wish to compare this expression for the mass correction \eqref{Eq.4 large N mass correction} with the RG invariant $\Lambda$ found from the weak coupling RG flow. Any difference between $\Lambda$ and the physical spinon mass arising from an $N$ dependent factor (as in the expression \eqref{Eq.4 TBA relation} for $\gamma=1/\pi$) would only lead to a difference in the finite terms.

To find $\Lambda$ let us first consider the weak coupling RG equations \eqref{Eq.4 RG weak coupling}, which we reproduce here to order $1/N$ together with certain unknown $\beta$ dependent three-loop terms $C_\lambda$ and $C_\beta$,
\begin{gather*}
	\mu\der{}{\mu}\lambda=-\frac{\lambda^2}{\pi}\left[1-\frac{1-\beta}{ N}+\frac{\lambda}{2\pi N}\left(4-6(1-\beta)+3(1-\beta)^2\right)+ \frac{\lambda^2}{N}C_\lambda(\beta)\right]\\
	\mu\der{}{\mu}\left(1-\beta\right)=-\frac{\lambda}{\pi}\beta(1-\beta)\left[1+\frac{\lambda}{N\pi}\left(1+\beta\right)+\frac{\lambda^2}{N}C_\beta(\beta)\right].
\end{gather*}
The parameter $\beta$ along any given RG trajectory is implicitly a function of $\lambda$, which may be written as a power series,
\begin{align}
	1-\beta(\lambda)= \lambda\left(\gamma_0+\gamma_1 \lambda + \gamma_2 \lambda^2 + \order{\lambda^3}\right).
\end{align}
Since the limit $\lambda\rightarrow 0$ is the same as the UV limit, the coefficient $\gamma_0$ is consistent with our earlier definition of $\gamma_0$ as the UV limit of the parameter $\gamma$ defined in \eqref{Def gamma}. Now $\gamma_1$ and $\gamma_2$ may be solved in terms of $\gamma_0$ by enforcing consistency with the RG equations. A similar calculation for $N=2$ is done in \cite{BALOG2000367}. The results are
\begin{gather}
	\gamma_1 = -\gamma_0^2+\frac{1}{N}\gamma_0^2,\non
	\gamma_2 = \gamma_0^3 +\frac{C_{\beta}(1)-C_{\lambda}(1)}{2N}\gamma_0+\frac{\gamma_0^2}{2\pi N}\left[2-5\pi \gamma_0 - \frac{1}{N}\left(2-3\pi\gamma_0\right)\right].
\end{gather}
The three-loop results are already involved in the determination of $\gamma_2$, although in \cite{BALOG2000367} it is argued that the combination $C_\beta(1)-C_\lambda(1)$ should vanish. This result for $\beta(\lambda)$ may substituted back into the RG equations to get a one-parameter RG flow for $\lambda$ in terms of the invariant $\gamma_0$,
\begin{align}
\beta_\lambda=-\frac{\lambda^2}{\pi}\left[1+\frac{\lambda}{N}\left(\frac{2}{\pi}-\gamma_0\right)+\frac{\lambda^2}{N}\left(C_\lambda(1)+\left(1-1/N\right)\gamma_0^2-\frac{3\gamma_0}{\pi }\right)\right].\label{Eq.4 RG flow lambda one parameter}
\end{align}
This is actually already a three-loop expression. The $C_\lambda(1)$ term is still undetermined, but if we ignore scheme dependence for the moment it could simply be taken from existing three-loop results on the $CP^{N-1}$ model \cite{hikami1979renormalization,wegner1989four}.

Given a three-loop beta function of the form
\begin{align}
	M \derOrd{\lambda}{M}=-\lambda^2\beta_0\left(1 + \frac{1}{N}\left(\beta_1 \lambda + \beta_2 \lambda^2\right)\right),
\end{align}
it may be checked that the RG invariant $\Lambda$ takes the form
\begin{align}
	\Lambda = M e^{-\frac{1}{\beta_0\lambda}}\lambda^{-\frac{\beta_1}{\beta_0 N}}\left[1+\left(-\frac{\beta_2}{\beta_0 N}+\frac{\beta_1^2}{\beta_0N^2}\right)\lambda + \order{\lambda^2}\right]\non
	= M e^{-\frac{1}{\beta_0\lambda}} \left[1-\frac{1}{\beta_0 N}\left(\beta_1 \log \lambda+\beta_2 \lambda\right)+\order{\lambda^2,\,N^{-2}}\right].
\end{align}
So using \eqref{Eq.4 RG flow lambda one parameter} and the saddle point condition
\begin{align*}
	m^2 = M^2 e^{-\frac{2\pi}{\lambda}},\qquad \log \frac{M^2}{m^2}=\frac{2\pi}{\lambda},
\end{align*}
the $\order{\lambda, N^{-1}}$ correction to $\Lambda^2$ is
\begin{align}
	\delta\Lambda^2 =  -\frac{2\pi m^2}{N}\left[\left(\frac{2}{\pi}-\gamma_0\right) \log \lambda+\left(\gamma_0^2-\frac{3\gamma_0}{\pi }+C_\lambda(1)\right) \lambda\right].
\end{align}
Meanwhile the large $N$ correction $\delta m^2_z$ arising by expressing the logarithms of $M$ in \eqref{Eq.4 large N mass correction} in terms of $\lambda$ is
\begin{align}
	\delta m_z^2 = -\frac{2\pi m^2}{N}\left[\left(\frac{2}{\pi}-\gamma_0\right)\log \lambda+\left(\gamma_0^2-\frac{5\gamma_0}{2\pi}+\frac{3}{2\pi^2}\right)\lambda+\text{finite}\right],
\end{align}
where now ``finite'' means $\lambda$ independent and it will depend on the overall proportionality factor between $m_z^2$ and $\Lambda^2$.

We see clearly now that the large $N$ and weak coupling approaches only differ in the three-loop term depending on the coefficient $\beta_2$. On the weak coupling side if we use a different scheme with parameter $$\tilde{\lambda}^{-1} = \lambda^{-1} + a_0 + \frac{a_1}{N} \lambda,$$
then the three-loop beta coefficient is redefined as
$$\tilde{\beta_2}= \beta_2 +a_0 \beta_1 - a_1,$$
and it possible to agree with the large $N$ approach. If the scheme parameters $a_0$ and $a_1$ are independent of $\gamma_0$ then it is impossible to change the $\gamma_0^2 \lambda$ term and indeed we see that this term agrees between the two approaches. But given that the large $N$ results were found in a cutoff scheme and the weak coupling results relied on a minimal subtraction dimensional regularization scheme, there is no reason to think that the other terms should be the same. Indeed on the large $N$ side the change in scheme can be accomplished simply by a redefinition of cutoff $\tilde{M}= \left(b_0 + b_1 \lambda\right)M.$

So to summarize, we have argued earlier in Sec \ref{Sec 4.0} that in contrast to the interpretation in Azaria et al \cite{azaria1995massive}, the introduction of the auxiliary $A$ field is a completely innocent procedure that does not involve any new approximation beyond the large $N$ expansion itself. Instead we have shown here that the discrepancy depends on some three-loop results and the large $N$ approach agrees with the weak coupling approach as much as could possibly be expected given the difference in scheme.

\section{Squashed sphere model with a Wess-Zumino term}\label{Sec 4.3}

The idea that scheme dependence may appear already at two-loops in the weak coupling expansion for a sigma model involving more than one parameter was discussed from a different perspective by Metsaev and Tseytlin \cite{metsaev1987order}. That paper dealt with the equivalence between the equations of motion derived from tree-level string scattering and the vanishing of the beta functions of a generalized bosonic sigma model involving a dilaton and an antisymmetric $B$ field (see e.g. the review \cite{callan1988sigma}). The scheme dependence emphasized in that case was that arising from the inclusion of the $B$ field, which may be thought of as a generalized WZ term \cite{wess1971consequences,novikov1982hamiltonian,Witten:1983ar}. In this case the scheme dependence can be shown in a very concrete way by demonstrating that the ambiguity in extending the Levi-Civita symbol in dimensional regularization affects the results for the two-loop beta function.

After all of these caveats about scheme dependence we may be wary about trusting the weak coupling expansion beyond one loop, but in this section I will show that if the scheme is fixed by matching to a CFT then it appears that even some all-orders results for the weak coupling RG flow may be obtained.

\subsection{Introducing the model}
We will consider the $N=2$ case \eqref{Lagr squashed sphere N=2} with a WZ term constructed from the topological charge 3-form $\rho$ \eqref{Def top charge}.
\begin{align}
S= \int d^2 x\frac{1}{2g^2}\left[\left(J^i\right)^2-\beta\left(J^3\right)^2\right] + 2\pi i k \int d^3 x \epsilon^{\lambda\mu\nu}\rho_{\lambda\mu\nu},\label{Lagr squashed PCMWZ}
\end{align}
The $z$ field gives a map from spacetime $S^2$ to the target space $S^3$ and its domain may be extended to the interior of $S^2$ ---this is the meaning of the integral over three spacetime coordinates in a $d=2$ theory. The different ways of extending the map $z$ may lead to an integral that differs by an integer number of windings around $S^3$. But if the WZ term is multiplied by $2\pi i k$, where $k$ is the integer \emph{level} parameter, then since the action appears in the path integral in the form $e^{-S}$ there is no ambiguity.

Since $k$ is the only parameter multiplying the WZ term, it will be useful to define a new coupling constant $\eta$ where
\begin{align}
	\frac{1}{g^2}= \frac{k}{2\pi \eta}. \label{Def eta}
\end{align}
Now the action is proportional to $k$ and the weak coupling expansion may be thought of as a \emph{large level expansion}. The parameter $\eta = kg^2/2\pi$ is somewhat like a 't Hooft coupling that will be exact at each order of the $1/k$ expansion, as will $\beta$.

This model has been studied in detail in \cite{kawaguchi2011yangian, kawaguchi2014deformation} with respect to its integrability, and also the one-loop RG flow was found. More general Yang-Baxter models with a WZ term have also been considered \cite{demulder:2017zhz} and the stability of the locus of classical integrability under the one-loop RG flow was shown. In this section based on \cite{schubring2021sigma} we take the modest goal of extending the results to two-loops and note the potential appearance of new fixed points.

First let us briefly review the Wess-Zumino-Witten fixed point at $\beta=0$ \cite{Witten:1983ar}. In the unsquashed $\beta=0$ limit the model has $SU(2)_L\times SU(2)_R$ symmetry, but the corresponding currents, $J_L$ and $J_R$ respectively, are modified by the WZ term,
	\begin{align}
	J_{L,\mu}= i \partial_\mu U U^\dagger-\eta\epsilon_{\mu\nu}\partial_\nu U U^\dagger,\qquad J_{R,\mu}=-i U^\dagger \partial_\mu U-\eta\epsilon_{\mu\nu}U^\dagger \partial_\nu U.\label{Eq.4 noetherWZW}
\end{align}
At the special point $\eta=1$ these currents reduce to one independent component in holomorphic coordinates, $z=x^0+ix^1$,
\begin{align}
	J_{L,\bar{z}}= 2i \partial_{\bar{z}} U U^\dagger,\qquad J_{R,z}=-2i U^\dagger \partial_z U,\label{Eq.4 noetherWZWz}
\end{align}
and the Noether current conservation law implies that $J_{R,z}$ only depends on $z$, and $J_{L,\bar{z}}$ only depends on $\bar{z}$. These currents form a Kac-Moody algebra of level $k$ and this is of course just the WZW conformal fixed point.

Since we will be modifying the Lagrangian by terms of the form $J^\mu J_\mu \propto J_z J_{\bar{z}}$ we will need to consider not only $J_z$, which is another name for the non-trivial component of the conserved current $J_{R,z}$, but also $J_{\bar{z}}$ which is no longer conserved and thus can have an anomalous dimension. The scaling dimension of this operator was calculated by Knizhnik and Zamolodchikov \cite{knizhnik1984current}. It takes the form $1+2\Delta_1$, and in the case of the group $SU(2)$ the anomalous part is
\begin{align}
	2\Delta_1 = \frac{4}{2+k}= \frac{4}{k} -\frac{8}{k^2}+\order{k^{-3}}.\label{Eq.4 Idimension}
\end{align}

\subsection{RG flow to two-loops}
	The problem of renormalizing sigma models with a WZ term has been considered by a number of authors, (see \cite{braaten1985torsion,bos1987dimensional,ketov1987two,ketov1990three,metsaev1987order} and references therein). The WZ term may be written as a two-dimensional integral on the same footing as the ordinary metric term \eqref{Lagr generic sigma model} and then the same background field method which works to find the renormalization of the pure metric sigma model may be applied.

In order to bring our conventions closer to \cite{ketov1990three}, the WZ term in the action \eqref{Lagr squashed PCMWZ} may be rewritten as
\begin{align}
	S_{\text{WZ}}&=\frac{i}{3}\int dx^3 S_{\alpha\beta\gamma}\epsilon^{\lambda\mu\nu}\partial_\lambda\varphi^\alpha\partial_\mu\varphi^\beta\partial_\nu\varphi^\gamma\label{Eq.4 WZWterm2}
\end{align}
where $\varphi^\alpha$ are the coordinate fields in the sigma model, and $S_{\alpha\beta\gamma}\equiv 6\pi k \,\rho_{\alpha\beta\gamma}$. These tensors are also proportional to the volume form on the target space $\sqrt{g}\epsilon_{\alpha\beta\gamma}$, \begin{align}
	S_{\alpha\beta\gamma}=\frac{k}{2\pi}\frac{\lambda^3}{\sqrt{1-k}}\sqrt{g}\epsilon_{\alpha\beta\gamma}.
\end{align}

Of course since $S$ is a three-form on a three-dimensional manifold, it is a closed form. This means that locally we may write $S$ as the exterior derivative of a 2-form gauge field $B$,
$$S_{\alpha\beta\gamma}=\partial_{[\alpha}B_{\beta\gamma]},$$
and apply Stokes' theorem to rewrite the action in a form similar to the metric term \eqref{Lagr generic sigma model},
\begin{align}
	S_{\text{WZ}}&=\frac{i}{3}\int dx^2 \epsilon^{\mu\nu}B_{\alpha\beta}\partial_\mu\varphi^\alpha\partial_\nu\varphi^\beta.
\end{align}

The problem of finding the renormalization of a sigma model with general $G$ and $B$ has been solved up to three loops \cite{ketov1990three}, although there are still some ambiguities left to be cleared up, as we will discuss later. In our case, $S$ is proportional to the volume form, so its covariant derivative vanishes, and it also satisfies the identity
$$S_{[ab}^{\quad h}S_{c]gh}=0.$$
This will simplify some of the formulas for the beta functions slightly. To two loops the beta function is given as \cite{ketov1987two}
\begin{align}
	\mu\der{}{\mu}G_{ab}=\frac{1}{2\pi}\hat{R}^{c}_{\,\,abc}+\frac{1}{8\pi^2}\hat{R}_{acdf}\hat{R}_{b}^{\,\,cdf}-\frac{1}{(2\pi)^2}\hat{R}_{adfb}S^{d}_{\,\,gh}S^{fgh},\label{betaFunctionGeneral}
\end{align}
where $\hat{R}_{abcd}$ is the Riemann curvature tensor $R$ modified by $S$ in such a way that it has an interpretation as a Riemann curvature for a target space metric with torsion \cite{braaten1985torsion},
\begin{align}
	\hat{R}_{abcd}\equiv R_{abcd}-S_{fab}S^f_{\,\,dc}.
\end{align}
The last term in \eqref{betaFunctionGeneral} involved some of the scheme ambiguities in continuing the Levi-Civita tensor in dimensional regularization mentioned above. These ambiguities were fixed by matching the beta function to the dimension of the operator perturbing the conformal fixed point \cite{bos1987dimensional}.

%For the ordinary ($\kappa=0$) $SU(2)$ WZNW model,
%$$R_{abcd}=\lambda^2\left(g_{ad}g_{bc}-g_{ac}g_{bd}\right),$$
%and given that $S$ is proportional to the Levi-Civita tensor
%the contraction $S_{fab}S^f_{\,\,dc}$ produces an identical structure. So the full Riemann curvature with torsion $\hat{R}$ is
%$$\hat{R}_{abcd}=\lambda^2(1-\eta^2)\left(g_{ad}g_{bc}-g_{ac}g_{bd}\right),$$
%where we have defined the useful parameter\,\footnote{The parameter $\eta$ in (\ref{peta}) and below (without indices) is not to be confused
%	with the spacetime metric $\eta^{\mu\nu}$.}
%\begin{align}
%	\eta \equiv \frac{k\lambda^2}{2\pi}=\frac{\lambda^2}{\lambda_k^2}.
%	\label{peta}
%\end{align}
%When we are at the point $\eta=1$, the modified Riemann curvature $\hat{R}_{abcd}=0$ and thus the $\beta$ function vanishes. And this is of course just the point where $\lambda^2=\lambda_k^2$ which was introduced above in the context of the current algebra.
%
%\subsubsection{RG flow of the squashed sphere with WZNW term}

Now given this general beta function it merely remains to use the expressions for $G$ \eqref{Eq.2 metric} and $R$ \eqref{Eq.2 riemannAppendix} derived in chapter \ref{Sec 2}.\footnote{Recall that the antisymmetric $S$ tensor is proportional to the volume form so its contractions will be related to the metric $G$.} In particular note that the Riemann tensor $R_{ijkl}=R_{[ij][kl]}$ is
\begin{align}
	R_{1212}=-\frac{1}{g^2}(1+3\beta),\qquad R_{1313}=R_{2323}=-\frac{1}{g^2}(1-\beta)^2.
\end{align}

The beta functions are found to be,
\begin{gather}
	\mu \derOrd{}{\mu}\left(\frac{1}{g^2}\right)=\frac{1}{\pi}\left[(1+\beta)-\frac{1}{1-\beta}\eta^2\right]+\frac{\eta}{\pi k}\left[1+2\beta+5\beta^2-4\frac{1+\beta}{1-\beta}\eta^2+3\frac{1}{(1-\beta)^2}\eta^4\right],\non\non
	\mu \derOrd{}{\mu}\left(\frac{1-\beta}{g^2}\right)=\frac{1}{\pi}\left[(1-\beta)^2-\eta^2\right]+\frac{\eta}{\pi k}\left[(1-\beta)^3-4(1-\beta)\eta^2+3\frac{1}{(1-\beta)}\eta^4\right].
	\label{Eq.4 WZ RGequations}
\end{gather}
When $\beta=0$ this agrees with the usual two-loop RG equation for the PCM with a WZ term \cite{bos1987dimensional,ketov1987two}. When $\beta\neq 0$ but $k=0$ these agree with \eqref{Eq.2 RG 1} and \eqref{Eq.2 RG 2} for $\mathcal{N}=N$. Note that $\beta=0, \eta=1$ is the WZW fixed point.

\begin{figure}
	\begin{center}
		\includegraphics[width=7cm]{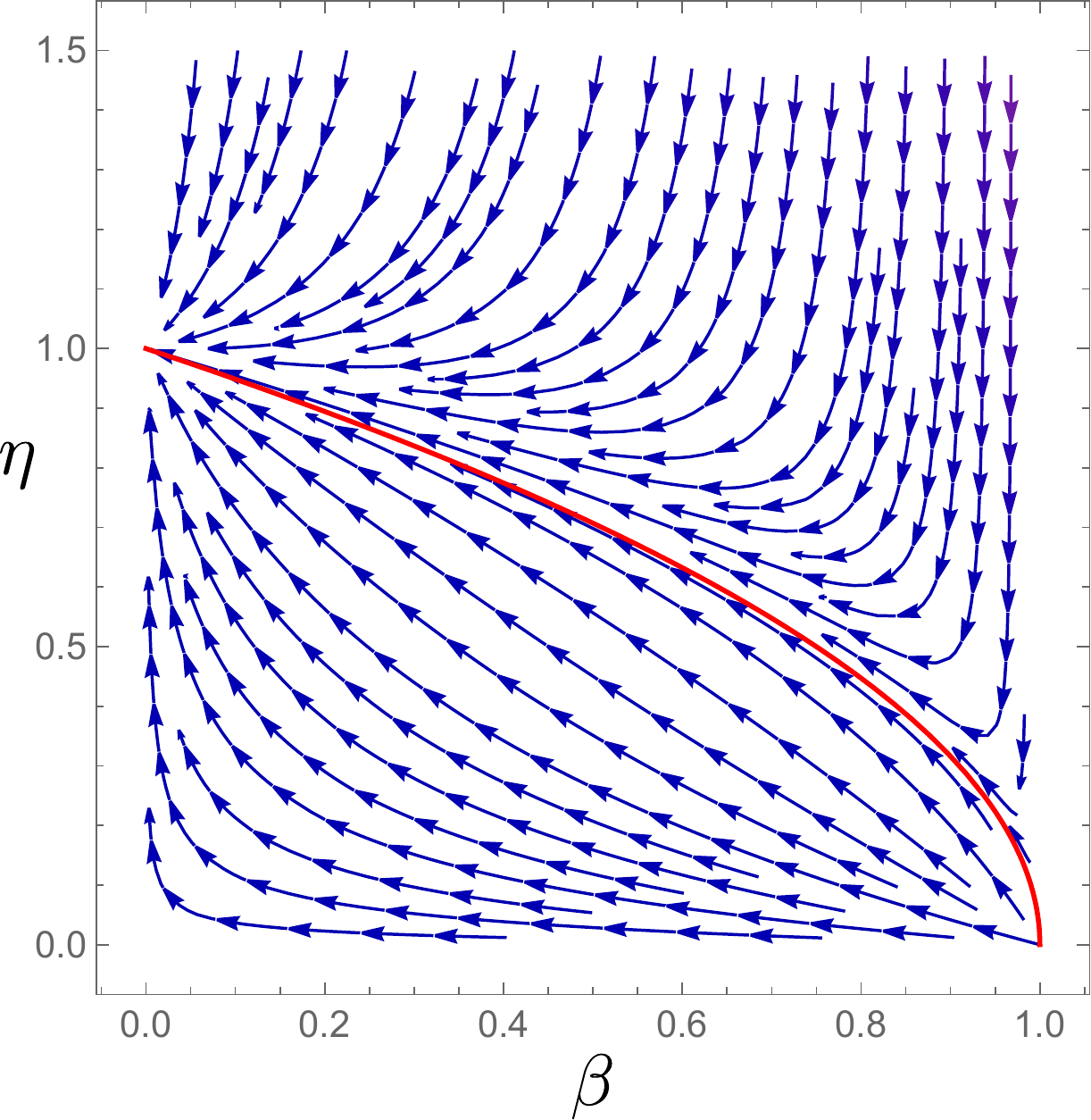}
	\end{center}
	\caption{\small RG flow for the squashed sphere sigma model with a WZ term to one loop. The flow is pointing towards the IR. The separatrix $\eta=\sqrt{1-\beta}$ is plotted in red.}\label{Fig1loop}
\end{figure}

The RG flow to one loop is plotted in Fig. \ref{Fig1loop}. Everything flows to the WZW fixed point in the IR. The curve $\eta=\sqrt{1-\beta}$ solves the RG equations, and is in fact a separatrix determining two classes of UV behavior.\footnote{This separatrix also appears in \cite{kawaguchi2011yangian} as the condition for the $SU(2)$ currents at non-zero $\beta$ to satisfy the flatness condition of \cite{brezin1979remarks} without modification by a topological current. This separatrix is not stable at higher loops.}  Below the separatrix every trajectory flows in the UV to the asymptotically free fixed point at $\beta=1$. And above the separatrix the coupling constant $\eta$ and thus $g^2$ appears to diverge in the UV. This divergence is quite possibly an artifact of taking only a finite order in perturbation theory, and we will have more to say on this later.

Far below the separatrix, for $\eta\ll\sqrt{1-\beta}$, the RG equations reduce to the RG equations for the squashed sphere sigma model without a WZ term, and the RG invariant $\gamma_0$ is valid in this case as well. The $CP^1$ decoupling limit of $\gamma_0\rightarrow 0$ may be identified as the separatrix trajectory.

Note that the UV fixed point at $\beta=1$ involves not just the $CP^1$ model but also the additional degree of freedom that Higgses the $U(1)$ gauge field. This can be seen by considering the Zamolodchikov c-theorem \cite{zamolodchikov1986irreversibility}. The central charge at the WZW fixed point is
$$c=\frac{3k}{k+2}$$
and since the central charge monotonically decreases along the RG flow and we are free to take $k$ as large as we want, there must be a central charge of at least $3$ in the UV, rather than the 2 that might be expected based on the degrees of freedom of the $CP^1$ model alone.

\subsection{A consistency check and an exact RG trajectory}\label{sectionTesting}

This weak coupling or large level expansion may be compared with a perturbative expansion about the WZW fixed point action $S_0$, in a calculation similar to the $\beta=0$ case \cite{bos1987dimensional,ketov1987two} which was used to fix the scheme dependence noted above. If the CFT action $S_0$ is perturbed by operators $\mathcal{O}^i$ each of which has a well-defined scaling dimension $\Delta_i$,
$$S=S_0 + \lambda_i \mathcal{O}^i,$$
then under a RG coarse graining from $\mu$ to $\mu'$, to lowest order in the small parameter $\lambda_i$ the action will transform to
$$S_0+\lambda_i \left(\frac{\mu'}{\mu}\right)^{\Delta_i} \mathcal{O}^i\approx S_0+\lambda_i \left(1+\Delta_i \log\frac{\mu'}{\mu}\right) \mathcal{O}^i.$$
So to the lowest order the $\beta$ function for $\lambda_i$ is just,
\begin{align}
	\beta_{\lambda_i}=\lambda_i \Delta_i + \mathcal{O}(\lambda^2).\label{betaFirstOrder}
\end{align}
In our case the total action is,
\begin{align*}
	S= S_0 + \frac{k}{2\pi}\left(\frac{1-\eta}{\eta}\right)\int dx^2\frac{1}{2}\left[\left(J^1\right)^2+\left(J^2\right)^2\right]+\frac{k}{2\pi}\left(\frac{1-\beta-\eta}{\eta}\right)\int dx^2\frac{1}{2}\left(J^3\right)^2.
\end{align*}
So the coefficients of the operators perturbing the fixed point action are
\begin{align}
	\lambda_\eta\equiv\frac{k}{2\pi}\left(\frac{1-\eta}{\eta}\right),\qquad \lambda_\beta \equiv \frac{k}{2\pi}\left(\frac{1-\beta-\eta}{\eta}\right).\label{gParameters}
\end{align}
Expanding the beta functions \eqref{Eq.4 WZ RGequations} in these parameters,
\begin{align}
	\mu \derOrd{}{\mu}\left(\frac{1}{g^2}\right)=\left(\frac{4}{k}-\frac{8}{k^2}\right)\lambda_\eta+\mathcal{O}(\lambda^2),\qquad \derOrd{}{\mu}\left(\frac{1-\beta}{g^2}\right)=\left(\frac{4}{k}-\frac{8}{k^2}\right)\lambda_\beta+\mathcal{O}(\lambda^2)\label{RGdimension}
\end{align}
This agrees with the anomalous dimension $2\Delta_1$ noted earlier \eqref{Eq.4 Idimension}. So the weak coupling expansion to two loops agrees with conformal perturbation theory to first order.

The condition that the weak coupling expansion agrees to all loops may be taken to be a condition partially fixing the scheme. We may use this to give an argument about the linear $\eta=1-\beta$ trajectory which satisfies the RG equations \eqref{Eq.4 WZ RGequations} to two loops. Incidentally this trajectory appears in the so-called $T^{1,q}$ model \cite{levine2021integrability} as the intersection of two parameter constraints where the model is classically integrable. In terms of the parameters perturbing the CFT, the stability of this trajectory means that if $\lambda_\beta$ is set to zero at one scale, it remains zero under the RG flow. No $\left(J^3\right)^2$ perturbation on the fixed point is generated through renormalization.

We conjecture that this will hold not just to two loops, but to all orders in perturbation theory. The reason that $\eta=1-\beta$ satisfies the RG equations is that the $\beta$ function for $(1-\beta)/g^2$  \eqref{Eq.4 WZ RGequations} takes the schematic form
\begin{align}
	\mu \dfrac{d}{d\mu}\left(\frac{1-\beta}{g^2}\right) &= \frac{1}{k^0}\left[a^{0}_{2,0}(1-\beta)^2+a^{0}_{0,2}\eta^2\right]
	+\frac{1}{ k^1}\left[a^{1}_{3,1}(1-\beta)^3+  a^{1}_{1,3}(1-\beta)\eta^3+a^{1}_{-1,5}(1-\beta)^{-1}\eta^5\right],\nonumber
\end{align}
where $a^n_{j,k}$ is just notation for the numerical coefficient of the term $(1-\beta)^j\eta^k$ in the $(n-1)$-loop correction to the beta function. Since $\beta=0,\,\, \eta=1$ is a fixed point, we must have the condition
\begin{align}
	{\sum_{i,j}a^n_{i,j}=0}.\label{condition1}
\end{align}
So, given this condition, the reason why $\eta=(1-\beta)$ solves the RG equations is because at each loop order $n$, the sum of the powers of $\eta$ and $(1-\beta)$ is the same for each term. In other words, for each $n$ all non-vanishing $a^n_{i,j}$ have the same value of $(i+j)$.

But consider now the general expansion of the beta function in the loop index $n$ and as a Legendre expansion in $(1-\beta)$ and $\eta$,
$$\mu \derOrd{}{\mu}\left(\frac{1-\beta}{g^2}\right)=\sum_{n,i,j}\frac{1}{k^n}a^{n}_{i,j}(1-\beta)^i\eta^j.$$
Expanding to first order in $\lambda_\eta,\lambda_\beta$ we have,
$$\mu \derOrd{}{\mu}\left(\frac{1-\beta}{g^2}\right)=\sum_{n}\frac{2\pi}{k^{n+1}}\left(\lambda_\beta\sum_{i,j}ia_{i,j}^n +\lambda_\eta\sum_{i,j}(i+j)a_{i,j}^n \right)+\mathcal{O}(\lambda^2).$$
But as in \eqref{betaFirstOrder} there should be no first order $\lambda_\eta$ term. So besides \eqref{condition1}, there is also the condition
\begin{align}
	{\sum_{i,j}(i+j)a_{i,j}^n=0}.\label{condition2}
\end{align}
This condition can also be derived by considering the $1/g^2$ beta function at $\beta=0$ and demanding that the dimension agrees with the $(1-\beta)/g^2$ beta function at $\lambda_\eta=0$, as in \eqref{RGdimension}.

The crux of the argument is that given \eqref{condition1} a simple way to satisfy this second condition \eqref{condition2} is if all non-vanishing $a^n_{i,j}$ have the same value of $(i+j)$. As mentioned above, this ensures that $\eta=1-\beta$ solves the RG equations. So while this is not a proof, it seems rather plausible that $\eta=1-\beta$, or equivalently $\lambda_\beta=0$, is preserved under the RG flow to all orders.

\subsection{Additional fixed points}\label{sectionNewFixedPoints}

\begin{figure}
	\begin{center}
		\includegraphics[width=7cm]{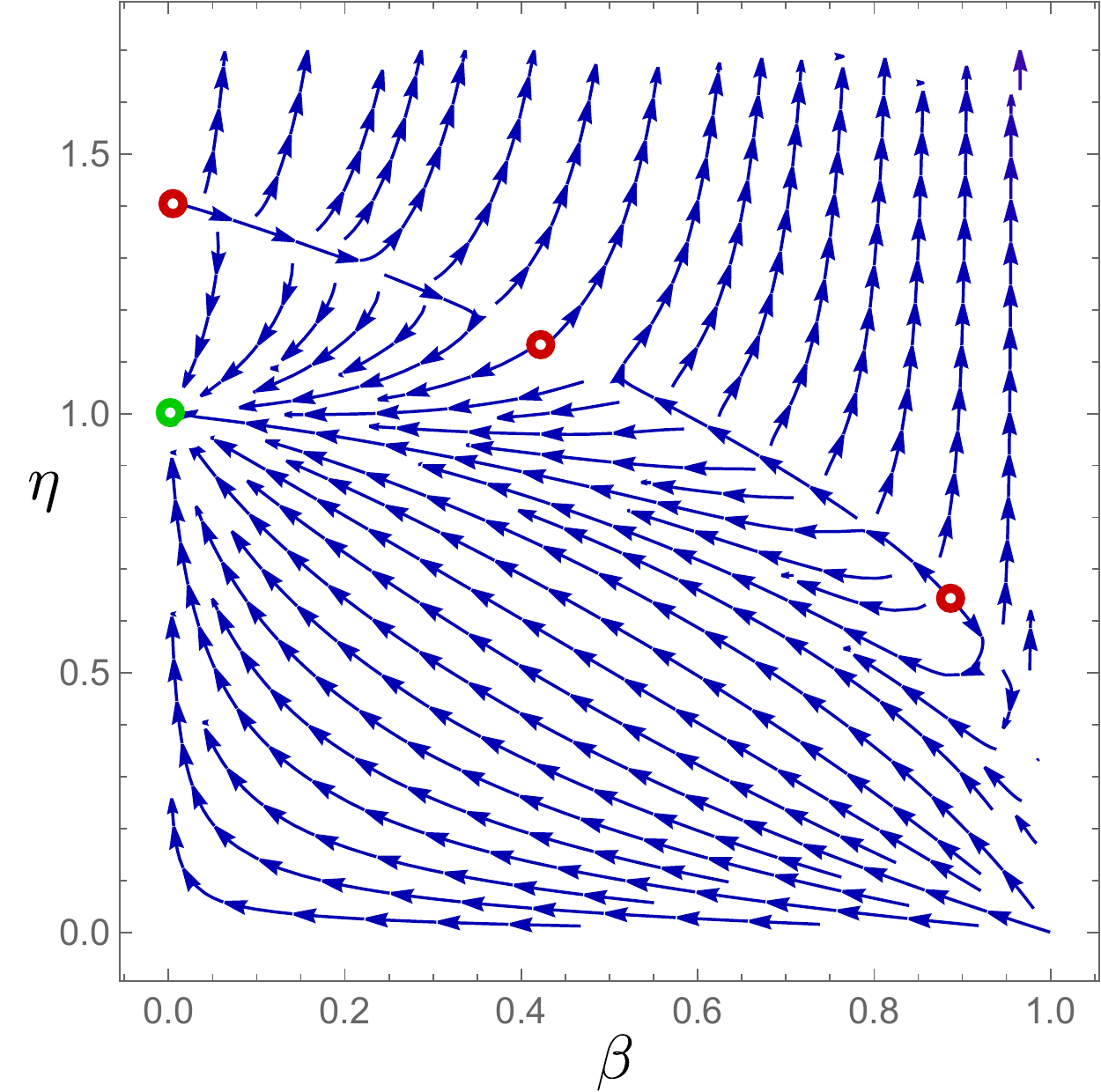}\qquad\includegraphics[width=7cm]{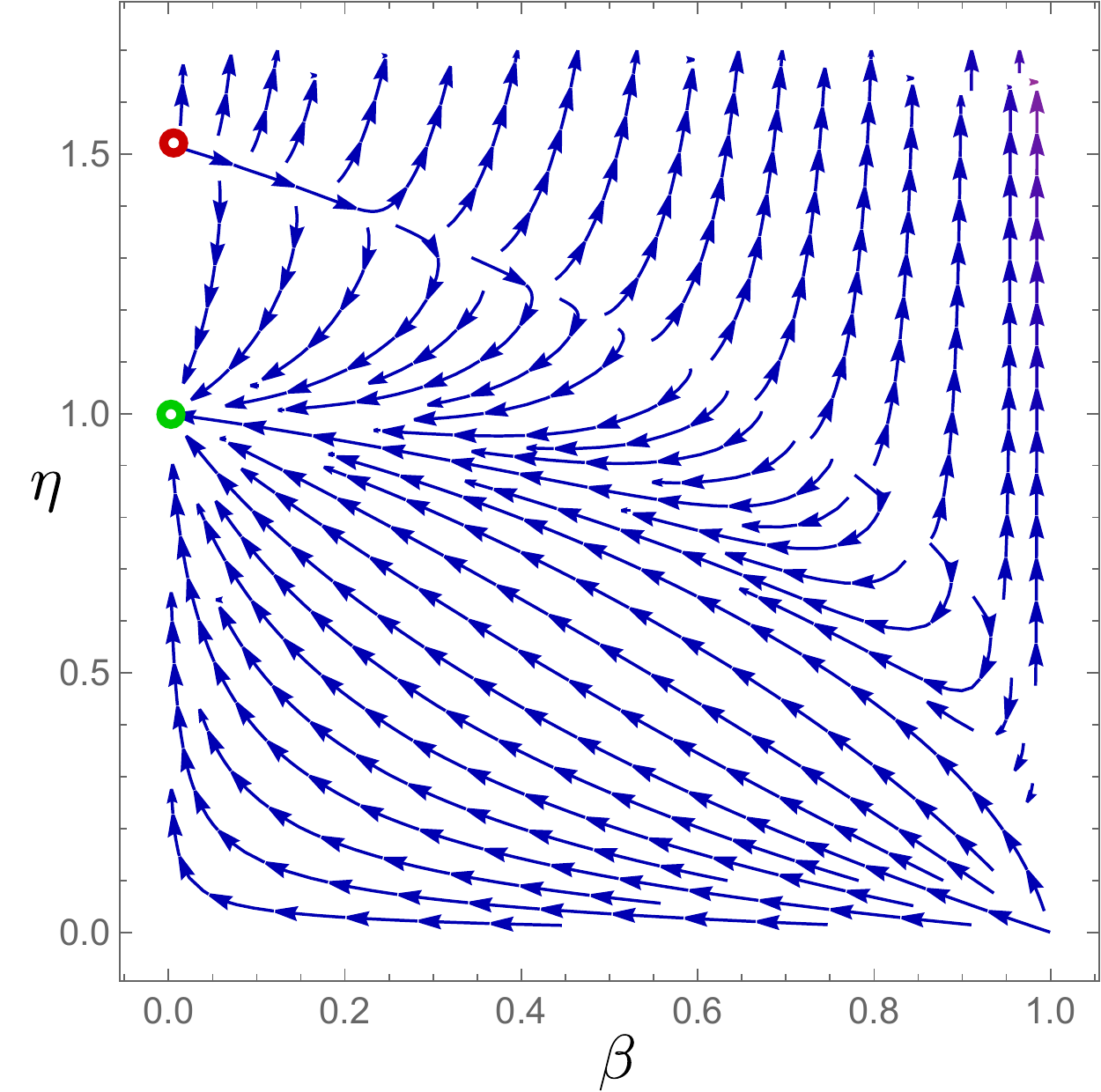}
	\end{center}
	\caption{\small RG flow for the squashed sphere sigma model with a WZ term to two loops. On the left is the level $k=7$ and on the right is $k=9$. The green dot is the ordinary WZW CFT, and the red dots are additional fixed points. The two fixed points with $\beta>0$ disappear for levels $k\geq 9$.}\label{Fig2loop}
\end{figure}

In Fig. \ref{Fig2loop}, the RG flow is plotted to two loops. The behavior below the first separatrix, which is now a slight deformation of the one loop result $\eta=\sqrt{1-\beta}$, is qualitatively the same as for one loop, but the behavior above the first separatrix is dramatically different. There is now a second separatrix, and between the first and second separatrices the trajectories flow to the WZW CFT in the IR, but flow to one or more new nontrivial fixed points in the UV, plotted in red in Fig. \ref{Fig2loop}.

Where the second separatrix meets $\beta=0$ there is an IR unstable fixed point. Setting $\beta=0$ in the RG equations \eqref{Eq.4 WZ RGequations}, and dividing out the factor of $(1-\eta^2)$, we see that the unstable fixed point is at the value $\eta=\eta_0$ which is the real root of the cubic equation
\begin{align}
	3\eta_0^3-\eta_0-k=0.\label{cubicFixedPoint}
\end{align} 

When $k$ is in the range $2<k\leq 8$, there are also two additional fixed points for $\beta>0$, as shown on the left of Fig. \ref{Fig2loop}. The coordinates of these fixed points can be found by first recalling that $\eta=1-\beta$ solves the RG equations. This means $(1-\beta)-\eta$ should be a factor in the RG equation for $((1-\beta)/g^2),$ and in fact, we can factor out $(1-\beta)^2-\eta^2,$
\begin{align*}
	\mu \derOrd{}{\mu}\left(\frac{1-\beta}{g^2}\right)&=\frac{(1-\beta)^2-\eta^2}{\pi k(1-\beta)}\left[k(1-\beta)+\eta(1-\beta)^2-3\eta^3\right].
\end{align*}
And we can also solve for the RG equation of $\beta$ alone,
\begin{align*}
	\mu \derOrd{\beta}{\mu}&=-\frac{4\eta}{k}\beta\left[(1-\beta) +\frac{2\eta}{k}\left((1-\beta^2) -2 \eta^2\right)\right].
\end{align*}
The fixed points $(\beta_0,\eta_0)$ are roots of the bracketed polynomials on the RHS of these two equations, and these can be further manipulated so that $\beta_0$ is given by the two real roots of a single quartic equation, and then $\eta_0$ can be easily found from the value of $\beta_0$,
\begin{align}
	4(1-\beta_0)&(1+3\beta_0)(1+5\beta_0)^2-k^2=0,\\
	\eta_0&=\frac{k}{10\beta_0+2}.
\end{align}
This quartic equation also indicates an upper bound on $k$ for which these two extra fixed points exist. At $k\geq 9$, these two fixed points disappear as in the right side of Fig. \ref{Fig2loop}, and since the loop expansion is really a large $k$ expansion, it is not clear whether these two fixed points are an artifact of the loop expansion or not.

Perhaps more interesting is the unstable fixed point at $\beta=0$ which solves \eqref{cubicFixedPoint}. Of course at $\beta=0$ this is just the ordinary well-studied PCM with a WZ term, so it would be surprising if there were an additional fixed point besides the WZW fixed point.

Note that the loop expansion is a large level expansion and the $\eta$ coordinate of the fixed point scales as $\eta_0\sim k^{1/3}$ at large $k$. So in the strict large $k$ limit it would seem to go off to infinity. The $\eta$ parameter is appropriate for large level expansion, but the $g^2$ coordinate of the fixed point which is associated with the target space geometry scales as $k^{-2/3}$. So in some sense for large $k$ the fixed point is at `weak coupling' (although much less weak than the WZW fixed point).

There are actually three-loop results available that can illuminate the issue \cite{ketov1990three}. The beta function at $\beta=0$  is
\begin{align*}
	\mu\frac{d}{d\mu}\left(\frac{1}{g^2}\right)=\frac{1}{\pi}\left(1-\eta^2\right)+\frac{\eta}{\pi k}\left(1-\eta^2\right)\left(1-3\eta^2\right)+\frac{\eta^2}{2\pi k^2}(1-\eta^2)\left[8+(1-\eta^2)q_2+(1-\eta^4)q_4\right].
\end{align*}
By dividing out the $1-\eta^2$ factor associated with the WZW fixed point, and keeping only the highest order terms in $\eta$ at each loop, we find the fixed points are solutions to the equation
\begin{align}-\frac{q_4}{2k}\eta^6-3\eta^3+k=0.\label{Eq.4 wz model 3 loop fixed points}\end{align}
Now the	numerical parameters $q_2$ and $q_4$ above are scheme-dependent and unlike the two-loop case they are not determined uniquely by matching to first-order perturbation theory about the WZW CFT. The authors of \cite{ketov1990three} use a scheme where
\begin{align}
	q_2=-\frac{10}{3},\qquad q_4=-\frac{5}{3}.
\end{align}
Something like the existence of a fixed point should not be scheme dependent, and we may tentatively consider these values for $q_2, q_4$ as at least one valid choice. 
From \eqref{Eq.4 wz model 3 loop fixed points} we see the unstable fixed point survives for large $k$ as long as $0>q_4>-9/2$, which is a condition that is satisfied for this scheme choice. In terms of the large level expansion, the third loop correction to the position of the fixed point $\eta_0$ is at the same order as the second loop, and only the coefficient of $k^{1/3}$ is modified.

Besides this unstable fixed point, there is also now a stable fixed point solution to \eqref{Eq.4 wz model 3 loop fixed points} also scaling as $k^{1/3}$. Although these three-loop results are only for $\beta=0$, let us consider what this might imply about the $\beta>0$ behavior. As seen in Fig. \ref{Fig2loop}, at the order of two loops, all RG trajectories above the second separatrix flow to strong coupling in the IR. Now given there is an IR stable fixed point that appears at three loops, it is very tempting to conjecture that the new stable fixed point at three loops lies at the end of a new third separatrix, above which all trajectories flow to strong coupling in the UV. If the pattern continues, at each loop order there could be a new separatrix and a new fixed point at $\beta=0$, alternating between IR stable and unstable. 

Of course as is clear from the discussion above, even the non-perturbative existence of the first unstable fixed point is inconclusive in the large $k$ expansion. Given the integrability of the PCM with a WZ term, further study of the RG trajectory for $\eta>1$ could be an interesting direction to pursue. There is also a natural large $N$ limit that may be taken, although the large $N$ limit for sigma models on Lie groups is more complicated than the squashed sphere models we are considering here (the two series happen to coincide at $N=2$). We may also go further in the conformal perturbation theory direction, although this is complicated by the fact that we are perturbing by an irrelevant operator and this leads to UV divergences. From this perspective the question of the existence of a UV fixed point or not seems very similar to the asymptotic safety scenario for gravity, and it may be fruitful to explore the connection further.

%	\subsection{[Literature notes]}
%\begin{itemize}
%	\item Balog, Forgacs, Horvath, Palla 1996. Also find 2 loop RG. Scheme dependence when comparing to T-dual. Solved by considering RG invariants.
%\end{itemize}

	\chapter{The operator product expansion and renormalons}\label{Sec 5}
	In the last chapter we have considered the squashed sphere model both in terms of the large $N$ expansion and the weak coupling expansion in $g$ and demonstrated that they are consistent as far as the RG flow is concerned. But there is a major difference in that the large $N$ expansion maintains manifest $SU(N)$ invariance at all stages and involves the non-perturbative scale $m$ from the very outset, whereas the weak coupling expansion involves a definite choice of coordinates on the target space and the non-perturbative scale only appears indirectly through the RG invariant $\Lambda$. Inasmuch as the squashed sphere model and related non-linear sigma models may be understood as toy models for 4D non-Abelian gauge theories it is important to better understand the weak coupling expansion, since this has a counterpart in the 4D gauge theories whereas the vector-like large $N$ limit does not.
	
	There are three related difficulties that occur in the weak coupling expansion. First of all, all non-perturbative corrections involving powers of $\exp(-1/g^2)$ are completely missed in a direct weak coupling expansion. In our approach these non-perturbative corrections may be understood to arise from non-trivial condensates of operators, much as $m^2$ may be understood to arise from a VEV of the $\alpha$ field.
	
	Another difficulty of the weak coupling expansion is that since the fields in the perturbative expansion are massless it generically leads to IR divergences. Both of these difficulties may in principle be dealt with by splitting up the fields in the theory into distinct UV and IR fields which are separated by an arbitrary factorization scale $\mu$. The VEVs of the IR fields will lead to the non-perturbative corrections and the perturbation theory in the UV fields is well-defined due to the explicit IR cutoff $\mu$. 
	
	 This is the background field method approach to the operator product expansion \cite{shifman1979qcd,novikov1984two} and it will be developed for the squashed sphere model in sections \ref{Sec 5 back field} and \ref{Sec 5 subleading}. Schematically a correlation function $f(p)$ as a function of external momentum $p$ may be written as
	 \begin{align}
	 	f(p)= \sum_{j=0}^{\infty}C_j(p;\,g)\langle O_j\rangle. \label{Eq.5 schematic ope}
	 \end{align}
	 Here the $O_j$ are operators constructed out of the IR fields, and $j$ indexes the engineering dimension, so $\langle O_j\rangle\sim m^{2j}$ gives powers of the non-perturbative scale. $C_j(p; g)$ are known as \emph{coefficient functions}, and they may be found from perturbation theory in the UV fields. On the other hand, the values $\langle O_j\rangle$ may not be calculated directly in perturbation theory, but this is still a useful asymptotic expansion for $p^2\gg m^2$ since by dimensional analysis the higher order terms must be surpressed by powers of $m^2/p^2$.
	 
	 In this approach there is a rigid factorization scale $\mu$ serving as an IR cutoff, but often the coefficient functions $C_j$ may be more easily calculated in another scheme such as dimensional regularization.\footnote{For suitable correlation functions it is possible to show that the IR divergences cancel \cite{Elitzur:1978ww,david1981cancellations,becchi1988renormalizability}.} In this case there is a third difficulty encountered in the weak coupling expansion. As is typical for the asymptotic expansions encountered in physics, the power series in $g$ for the coefficient functions $C_j$ diverge. We may attempt Borel summation to define $C_j$ as a non-perturbative function of $g$, but this is obstructed by poles in the Borel plane known as a renormalons (see the reviews \cite{beneke1999renormalons,Shifman:2013uka}). We may shift the integration contour involved in Borel summation to avoid the poles, but this leads to an ambiguity depending on which side of the poles we shift the contour.
	 
	 It might be hoped that these ambiguities cancel between $C_j$ for different values of $j$ in the expansion \eqref{Eq.5 schematic ope}, and this is partially correct. But as shown by David \cite{dave1984237,david1986operator} there are also ambiguities arising from the operator VEVs $\langle O_j\rangle$ which must be taken into account. These VEVs may be calculated directly in the squashed sphere model in the large $N$ approach, and their ambiguities are calculated in section \ref{Sec. 5 operator part}.
	 
	 For a complete understanding of the expansion in \eqref{Eq.5 schematic ope} it would be desirable to show that the ambiguities fully cancel to all orders, and this will be done in section \ref{Sec 5 mellin} following the approach of \cite{beneke1998308}. Essentially what we will do in this chapter is extend the ordinary weak coupling expansion in $g$ to a \emph{transseries} expansion which includes non-perturbative corrections and may be Borel summed in an unambiguous way. While transseries expansions are more commonly done for quantities such as the free energy, it is the goal of this chapter to provide an example of a transseries expansion for a correlation function involving some external momentum $p$. Since the large $N$ squashed sphere model is very closely connected to the $CP^{N-1}$ model \cite{dunne:2012ae} the $O(2N)$ model \cite{dunne:2015ywa} and the $N=2$ squashed sphere model \cite{demulder2016resurgence}, and involves the same operator VEVs $\langle F^{\mu\nu}F_{\mu\nu}\rangle$ that have begun to be considered in the SUSY $CP^{N-1}$ model \cite{ishikawa2020infrared}, it is hoped that the calculation in this chapter may shed some light on resurgence in these compactified models as well.

	\section{Introduction to the background field method}\label{Sec 5 back field}
		In this section we will begin to consider the OPE of the spinon two-point function in the squashed sphere model, and in particular show how at lowest order in the large $N$ expansion it reduces to the standard $(p^2+m^2)^{-1}$ propagator. The OPE for the spinon propagator in the $O(2N)$ model was first considered in \cite{David:1982qv} using an entirely different approach, and it was also considered in \cite{novikov1984two} using a background field approach similar in principle to that considered here. However the approach described here will differ in that we will use a background field method adapted from Polyakov \cite{polyakov1975interaction} that maintains manifest $SU(N)$ invariance in the background field at the cost of introducing $SU(N-1)$ gauge redundancy. The advantage compared to previous background field methods \cite{novikov1984two} or a direct asymptotic expansion of the large $N$ amplitude as in section \ref{Sec 5 mellin} is that the operator content of the VEVs will be exceptionally clear.
		
	\subsection{Polyakov-style background field method and one-loop renormalization}
	The idea of a Wilsonian background field method is to separate the $z$ field in the sigma model into a high frequency ``quantum" field $q$ and a low frequency ``classical" field $z_0$. If we integrate out the high frequency field $q$ we find how the action for $z_0$ is renormalized. Perhaps less often understood is that if we take the expectation value of some correlation function of $z$ and integrate  out $q$ then we can find the OPE of that correlation function.
	
	We will split up $z$ into the fields $z_0$ and $q$ in the manner of Polyakov \cite{polyakov1975interaction}, rather than the more straightforward linear manner $z= z_0+q$. Recall that the constrained action (with no $g$ field or $m^2$ term) is
	\begin{align*}
		\Lagr = \frac{1}{2g^2}\left[\partial \bar{z}\cdot\partial z+\beta\left(\bar{z}\cdot \partial z\right)^2\right],\qquad |z|^2=1.
	\end{align*}
	Now we will expand $z^i$ in terms of a spacetime dependent basis $e^i_\alpha(x)$, $$\bar{e}^i_{\alpha}(x)e^i_{\beta}(x)=\delta_{\alpha\beta}\qquad z^i(x)=e^i_\alpha(x) z^{\alpha}(x),$$ where both the original basis indexed by $i$ and the space dependent basis indexed by $\alpha$ range from $1$ to $N$.
	
	The transformation from $z^i$ to $z^{\alpha}$ may be thought of as an $SU(N)$ gauge transformation, and making this substitution in the action effectively promotes all derivatives to covariant derivatives with a flat gauge connection,
	\begin{align}	\Lagr = \frac{1}{2g^2}\left[\left| \mathcal{D}_{\alpha\beta}z^\beta\right|^2+\beta\left(\bar{z}^\alpha\mathcal{D}_{\alpha\beta} z^\beta\right)^2\right],\qquad \mathcal{D}_{\alpha\beta}\equiv\delta_{\alpha\beta}\partial +\bar{e}_\alpha\cdot \partial e_\beta.\label{Lagr SU(N) invariant}
	\end{align}
	
	Now at this point $e_\alpha$ is completely arbitrary, but we will connect this to Wilsonian renormalization by taking the $\alpha=N$ component to be the low frequency coarse grained $z$ field, which we call $e_N\equiv z_0$. We will use Latin indices for the remaining $N-1$ components $e_a$.
	
The $N$th component of $z$ in spacetime dependent basis will be referred to as $\sigma$, and the remaining $N-1$ components as $q^a$. To be clear, the original $z$ field is expanded in this notation as
\begin{align}z^i= \sigma z_0^i +q^a e^i_a ,\qquad |\sigma|^2+|q|^2=1,\label{Eq.5 renormalized fields}\end{align}
and the background fields satisfy the constraints,
\begin{align}
	|z_0|^2=1,\quad \bar{e}_a\cdot e_b = \delta_{ab},\quad \bar{e}_a\cdot z_0=0.
\end{align}
In the presence of the infrared background field the $SU(N)$ symmetry is explicitly broken and $\sigma$ will pick up a non-zero vacuum expectation value which is directly connected to the field renormalization of the $z$ field. This justifies the usual perturbative treatment where we single out the $N$th component and expand it in terms of the unconstrained $q$ fields and an unconstrained phase field $\psi$,
\begin{align*}
	\sigma(x) = e^{i\psi(x)}\sqrt{1-|q(x)|^2}.
\end{align*}

At this point we may expand the $SU(N)$ invariant Lagrangian \eqref{Lagr SU(N) invariant} in this background field scheme. The covariant derivatives become,
\begin{align}
\mathcal{D}_{N\beta}z^\beta = \partial \sigma + \left(\bar{z}_0\cdot \partial z_0\right) \sigma + \left(\bar{z}_0\cdot \partial e_b\right) q^b\non
 \mathcal{D}_{a\beta}z^\beta = \partial q^a + \left(\bar{e}_a\cdot \partial e_b\right) q^b +\left(\bar{e}_a\cdot \partial z_0\right) \sigma\label{Bgd. covariant der expansion} .
\end{align}

Substituting these expressions into the Lagrangian \eqref{Lagr SU(N) invariant} leads to many terms, not all of which are relevant for our purposes. For instance, to find the one-loop renormalization of the action we only need terms that are quadratic in the $q,\psi$ fields. And as we will discuss below, in order to calculate the $\langle \bar{z}(x)\cdot z(0)\rangle$ correlation function at sub-leading order in large $N$, we will be able to ignore all terms involving unfavorable contractions of the $SU(N-1)$ gauge indices.
	
	But first let us first verify that this scheme does give the correct one-loop renormalization with no large $N$ approximations made. We may also use this to calculate the anomalous dimension of the $z$ field and relate it to the invariant $\gamma_0$ discussed in the previous chapter \cite{schubring2020sigma}.
	
	As mentioned above, only terms quadratic in the fluctuating fields will matter at one-loop, and moreover terms such as $q\psi$ without a derivative may be disregarded since they will only lead to irrelevant corrections when contracted in a two-vertex bubble diagram. The terms which lead to relevant one-loop corrections are as follows,
	\begin{align}
		\Lagr=\frac{1}{2g^2} &\left[\left|\bar{e}_a\cdot\partial z_0\right|^2 +(1-\beta)\left|\bar{z}_0\cdot\partial z_0\right|^2 +|\partial q|^2+ (1-\beta)(\partial\psi)^2\right.\non
		&-\left(\left|\partial z_0\right|^2-2\beta\left|\bar{z}_0\cdot\partial z_0\right|^2\right)|q|^2\non
		&+(1-2\beta)|\bar{q}^a \bar{e}_a\cdot\partial z_0|^2+2(1-\beta)i\partial\psi\left(\bar{q}^a \bar{e}_a\cdot\partial z_0-\bar{z}_0\cdot\partial e_a{q}^a\right)\non
		&\left.+\left(\frac{1}{(N-1)^2}+\frac{2\beta}{N-1}\right)|\bar{z}_0\cdot\partial z_0|^2|q|^2+\left(\frac{1}{N-1}+\beta\right)\bar{z}_0\cdot\partial z_0\left(\bar{q}\partial q-q\partial\bar{q}\right)\right].\label{Lagr Bgd. renormalization}
	\end{align}
	Note that the factors of $(N-1)^{-1}$ arose from the structure $\bar{e}_a\cdot \partial e_b$, which can only lead to relevant terms through its scalar part $$(N-1)^{-1}\delta_{ab}\bar{e}_c\cdot \partial e_c = -(N-1)^{-1}\delta_{ab}\bar{z}_0\cdot\partial z_0.$$

	The first two terms in the first line of \eqref{Lagr Bgd. renormalization} give the Lagrangian $\Lagr_0$ of the background fields, which have a UV cutoff at the scale $\mu$. The second two terms on that line give the propagators for  $q^a$ and $\psi$, which have a UV cutoff at the scale $M$ and an IR cutoff at $\mu$. Integrating out $q$ and $\psi$ leads to,
	\begin{align*}
		\Lagr &=\frac{1}{2g^2} \left(\left|\bar{e}_a\cdot\partial z_0\right|^2 +(1-\beta)\left|\bar{z}_0\cdot\partial z_0\right|^2\right)-\frac{1}{2\pi}\log\frac{M}{\mu}\left((N-1+\beta)|\bar{e}_a\cdot\partial z_0|^2+(N-1)(1-\beta)^2|\bar{z}_0\cdot\partial z_0|^2\right).
	\end{align*}
	
	We can directly read off the beta functions,
	\begin{align}
		\mu\der{}{\mu}\left(\frac{1}{g^2}\right)=\frac{1}{\pi}\left(N-1-\beta\right),\qquad \mu\der{}{\mu}\left(\frac{1-\beta}{g^2}\right)=\frac{1}{\pi}\left(N-1\right)(1-\beta)^2,
	\end{align}
	which are identical to the one loop truncation of the beta functions \eqref{Eq.2 RG 1} and \eqref{Eq.2 RG 2}.
	
	This Wilsonian renormalization scheme also makes it quite easy to calculate the one-loop anomalous dimension of the $z$ field. Using \eqref{Eq.5 renormalized fields} a correlation function involving the bare field $z$ may be expressed in terms of the renormalized field $z_0$,
	\begin{align}
		\langle z\dots\rangle = \langle \sigma z_0 \dots \rangle &=\left\langle \left(1+i\psi -\frac{1}{2}\psi^2+\dots\right)\left(1-\frac{1}{2}|q|^2+\dots\right)z_0\dots\right\rangle\non&=\left[1-\frac{g^2}{4\pi}\left(\frac{1}{1-\beta}+2(N-1)\right)\log \frac{M}{\mu}\right]\langle z_0\dots\rangle.\label{Eq.5 fieldRenorm}
	\end{align}
	We are integrating out the high energy fields as though this were a one-point function, although of course non-perturbatively the one-point function $\langle z\rangle$ vanishes and so the ellipsis indicates the presence of other operators at distant spacetime points. From this expression the dimension $\Delta$ of the $z$ field may be read off directly,
	\begin{align}
		\Delta= \frac{g^2}{4\pi}\left(\frac{1}{1-\beta}+2(N-1)\right)\rightarrow \frac{1}{4\pi N \gamma_0},\label{Eq.5 anomalous dimension}
	\end{align}
where the arrow gives the non-vanishing dimension in the UV limit in which $g^2\rightarrow 0, \,\beta\rightarrow 1.$

This dimension may be understood if we go back to the gauge invariant form of the squashed sphere model involving an extra Stueckelberg field $\phi$ \eqref{Lagr squashed sphere A psi}. The gauge invariant spinon operator when $\phi$ is not fixed to zero must actually be understood as $z e^{-i\phi}$. Given the form of the kinetic term for $\phi$, the dimension of the vertex operator $e^{-i\phi}$ is exactly $\Delta$ above. So we again see that this somewhat hidden degree of freedom is very important to the squashed sphere model, even as $\beta\rightarrow 1$.
	
%	\ds{Verify this or take it out!}
%	
%	In the large $N$ limit this anomalous dimension should show up in the behavior of the spinon two-point function in the limit of $p^2\gg m^2$,
%	\begin{align}
%		\langle \bar{z}(p)\cdot z(-p)\rangle \sim \frac{1}{p^2}\left[1+\frac{1}{4\pi N\gamma_0}\log \frac{p^2}{m^2}+\order{\frac{m^2}{p^2}}\right]+\order{N^{-2}}.
%	\end{align}
%We will verify this as a consistency check later in Sec \ref{}.

	%The first term is just the action for the background field. The second set of terms is quadratic in $\varphi$. The final set of terms involves corrections higher order in $g^2$. Note that the equations of motion $e\cdot\partial^2 n=0$ were used to throw away a term linear in $\varphi$. Of course since $n_0$ is integrated over in the path integral it will not always satisfy the equations of motion, but for our purposes this would only matter if there would be operators in the OPE involving factors of $e\cdot\partial^2 n$, and that is certainly not possible for low-dimension operators and at low orders of the large-$N$ expansion. 
	
	\subsection{Operator product expansion in the large \boldmath{$N$} limit}\label{Sec 5 Large N}
	
	Ultimately our goal with this background field scheme is to discuss correlation functions at arbitrary loop order in the weak coupling expansion, and in principle this would involve many more terms in the Lagrangian beyond the relevant quadratic ones which appear in \eqref{Lagr Bgd. renormalization}. In particular there are terms which correspond to the usual higher order corrections in the weak coupling expansion, which arise from terms like $\left(\partial\sigma\right)^2$ in the Lagrangian. For this section it will be convenient to absorb a factor of $1/g$ into the definition of the $q$ and $\psi$ fields so that $g$ appears associated with these higher order vertices rather than the propagators. With this convention  $\sigma$ becomes
	\begin{align}
		\sigma = e^{ig\psi}\sqrt{1-2g^2|q|^2}.\label{Def sigma}
	\end{align}
	
	The background field expansion is now dramatically simplified by considering the large-$N$ limit, in which $g^2\sim N^{-1}$, so each factor of $g^2$ can only appear if it is compensated by a summation of indices producing a factor of $N$. It seems that this compensation might be possible if the factor of $|q|^2$ is contracted, but actually since the interaction terms will involve derivatives the net zero external momentum of a $|q|^2$ tadpole will always cause the diagram to vanish at lowest order in the large $N$ expansion.\footnote{If this point is not yet clear, it will be clarified further when we discuss the corrections to the coefficient functions in Sec. \ref{Sec 5 coefficient function}.} The main point is that if we are only considering the lowest order of the large $N$ expansion ---or if we are considering the sub-leading corrections to the operator VEVs but not the coefficient functions--- then we may ignore the higher-vertex corrections.
	
	Also as usual we may disregard any terms linear in the high energy fields. In general since the background fields are integrated over in the process of taking VEVs, they need not satisfy the equations of motion, and thus there may be non-vanishing linear terms. However these linear terms will imply a non-vanishing saddle point value for the high energy fields, and this saddle point value might more properly be grouped with the `classical' background fields as an external source.
	
	Rather than considering such a complicated scheme we will simply follow conventional wisdom and consider the background fields to always satisfy the equations of motion, with the justification that up to possible contact terms and niceties involving regularization of composite operators (see e.g. \cite{schubring2021lessons}) the background fields do obey the equations of motion within expectation values.

	Now let us consider the correlation function for the full field $z$. So as not to clutter our notation too much, the angled brackets will indicate integrating over the high energy fields first, and then afterwards integration over the background field $z_0$ is implied.
	\begin{align}
		\langle \bar{z}(x)\cdot z(0)\rangle&= 2g^2\langle\bar{q}^a(x)q^b(0)\rangle \, \bar{e}^a(x)\cdot e^b(0)+\langle\sigma(x)\sigma(0)\rangle \, \bar{z}_0(x)\cdot z_0(0) +\text{cross terms}.\label{Eq.5 corrTotal}
	\end{align}
	The cross terms are those terms involving expectations of $q\,\sigma$, and these vanish at lowest order for the reason mentioned above, since they require the use of the odd interaction terms arising from the expansion of $\sigma$.
	
	Now let us consider the first term $$\langle\bar{q}^a(x)q^b(0)\rangle \, \bar{e}^a(x)\cdot e^b(0)=\langle\bar{q}^a(x)q^b(0)\rangle \,\left(\,\delta^{ab}+x^\mu\partial_\mu \bar{e}^a(0)\cdot e^b(0)+\dots\right)$$
	The higher order $SU(N-1)$ gauge dependent terms in the expansion of $\bar{e}^a(x)\cdot e^b(0)$ should in principle cancel the gauge dependence of $\bar{q}^a(x)q^b(0)$. But at lowest order in the large $N$ expansion we will need a factor of $N$ to compensate the overall factor of $g^2$, and that may only be provided by the Kronecker delta term contracting with a corresponding Kronecker delta in the $\bar{q}^a(x)q^b(0)$ correlation function.
	
	So to produce this factor of $N$, we need only consider $\langle\bar{q}^a(x)q^b(0)\rangle\,\delta^{ab}$ evaluated according to a truncated version of the quadratic Lagrangian \eqref{Lagr Bgd. renormalization}, keeping only those terms where all $q$ are contracted with each other rather than factors of $e$, and of course dropping those terms which involve an explicit factor of $1/N$,
		\begin{align*}
		\Lagr_{large\,N}=\Lagr_0+|\partial q|^2+ \frac{1-\beta}{2}(\partial\psi)^2-\left(\left|\partial z_0\right|^2-2\beta\left|\bar{z}_0\cdot\partial z_0\right|^2\right)|q|^2+\beta\bar{z}_0\cdot\partial z_0\left(\bar{q}\partial q-q\partial\bar{q}\right).
	\end{align*}
This can be put in a more enlightening form by introducing Lagrange multiplier fields in the background field sector, as in \eqref{Lagr large N}. The equations of motion for $z_0$ and $A_0$ (in the normalization $|z_0|^2=1$) lead to
$$A_0 = -i\beta \bar{z_0}\cdot\partial z_0,\qquad \alpha_0 +m^2=  - 2iA_0\bar{z_0}\cdot\partial z_0 -|\partial z_0|^2,$$
So the relevant terms in the Lagrangian may be written as,
\begin{align}
	\Lagr_{large\,N}=\Lagr_0+|\partial q|^2+ \frac{1-\beta}{2}(\partial\psi)^2+\left(m^2+\alpha_0\right)|q|^2+iA_0\left(\bar{q}\partial q-q\partial\bar{q}\right).\label{Lagr background field large N}
\end{align}

Let us pause to remark on what we have done so far. We have factored the original field $z$ into high energy and low energy parts in such a way as to ensure the low energy fields $z_0$ maintain manifest $SU(N)$ symmetry. The expansion of the Lagrangian in this scheme leads to a great many terms, but we gave various arguments that most of these terms will not contribute to the correlation function $\langle \bar{z}\cdot z\rangle$ at lowest order in large $N$. The remaining terms look very much like the ordinary large $N$ expansion of the action \eqref{Lagr large N}, but the $z$ field is replaced by the unconstrained $q, \psi$ fields, and the Lagrange multiplier fields are associated to the background fields in $\Lagr_0$. A similar result was seen in the expansion of the SUSY $O(N)$ model \cite{schubring2021treating}, although the process was even more involved since the $e$ fields needed to be extended to superfields to maintain manifest SUSY throughout.

In keeping with the philosophy of the OPE, the quantity $m^2$ here should be understood as a background field `operator.' It may just be considered to be the zeroth order VEV of the operator $2\beta\left|\bar{z}_0\cdot\partial z_0\right|^2 -|\partial z_0|^2$, and this value can not be calculated within the OPE approach itself.

Neither $A_0$ nor $\alpha_0$ (given the separation of the $m^2$ part) has a VEV at lowest order, and since the VEVs of composite operators factor at lowest order in the large $N$ expansion, this means that effective only the $m^2$ term will matter. So the two-point function becomes simply
$$2g^2\langle\bar{q}^a(p)q^a(-p)\rangle = \theta\left(p^2-\mu^2\right)\frac{2Ng^2}{p^2+m^2}.$$
The $\theta$ factor is a step function to ensure that the high energy $q$ field propagator only involves momenta $p^2 > \mu^2$. This is already almost our full OPE at lowest order in large $N$, but we also must consider the effect of the second term of Eq. \eqref{Eq.5 corrTotal}, $\langle\sigma(x)\sigma(0)\rangle \, \bar{z}_0(x)\cdot z_0(0)$ which will matter for $p^2<\mu^2$.
	
	To lowest order $\langle\sigma(x)\sigma(0)\rangle$ factors as $\langle \sigma\rangle^2$, and this is exactly the field renormalization needed to convert back to the original fields $z_0\langle \sigma\rangle = z$. Explicitly the $\bar{z}_0(-p)\cdot z_0(p)$ part may be written as a propagator with a factor of $g_\mu$ which is renormalized to the UV cutoff $\mu$, and the field renormalization part \eqref{Eq.5 fieldRenorm} cancels with the renormalization of $g_\mu$ \eqref{Eq.4 RG large N 2},
	\begin{align*}
\theta\left(\mu^2-p^2\right)\langle \sigma\rangle^2 \frac{2Ng_\mu^2}{p^2+m^2}&=\theta\left(\mu^2-p^2\right)\left(1-\frac{Ng^2}{2\pi}\log \frac{M}{\mu}\right)^2\frac{2Ng^2\left(1+\frac{Ng^2}{\pi}\log \frac{M}{\mu}\right)}{p^2+m^2}\non&=\theta\left(\mu^2-p^2\right)\frac{2Ng^2}{p^2+m^2}.
	\end{align*}
	So for both $p^2$ above and below the factorization scale we have a result that agrees with the known large $N$ limit of the spinon 2-point function. And this may be expanded as a somewhat trivial OPE,
	\begin{align}
		\langle \bar{z}(-p)\cdot z (p)\rangle^{(0)}=\frac{2Ng^2}{p^2}\sum_{k=0}\left(-\frac{m^2}{p^2}\right)^k,\label{Eq.5 spinon ope lowest order}
	\end{align}
where the values of $m^2$ are understood as operator VEVs and the $p$ dependence at each order of the OPE indexed by $k$ is understood as a coefficient function.\footnote{This OPE does not work so well as a distribution to be integrated over $p$, and the Mellin transform approach taken in Sec. \ref{Sec 5 mellin} is better in this respect. This may also be regulated according to some scheme such as Hadamard's finite part \cite{estrada1989regularization}, which was used by David in \cite{David:1982qv}.}

\section{Background field method at subleading order}\label{Sec 5 subleading}

A somewhat less trivial application of the background field method is to go to the next order in the large $N$ expansion. The OPE for the correlation function of $z$ at subleading order  is schematically
\begin{align}
\langle \bar{z}(-p)\cdot z(p) \rangle^{(1)}=\sum_{j=0}C_j^{(0)}(p,g)\langle O_j\rangle^{(1)} +\sum_{j=0}C_j^{(1)}(p,g)m^{2j}.\label{Eq.5 schematic ope 2}
\end{align}
The first term is the `operator part' where the $1/N$ corrections are coming from the IR operator VEVs, and the second term is the `coefficient function part' where only $m^2$ appears as an operator, but there are non-trivial $1/N$ corrections arising from the UV fields.

\subsection{Operator part of the OPE}\label{Sec. 5 operator part}

To find the $1/N$ corrections due to the operator VEVs we may use the same lowest order action \eqref{Lagr background field large N} derived above, but also consider the VEVs arising from the $\alpha_0, A_0$ fields. Given the form of the action it will perhaps not be surprising that this expansion will end up being equivalent to the expansion of the large $N$ diagrams in Fig. \ref{Fig 4 self energy} in powers of $m^2$ and loop momentum $k^2$.

For instance there are diagrams contributing to $\langle \bar{q}^a(p)q^a(-p) \rangle$ at order $1/N$ that involve two $A_0$ insertions and arbitrary number of mass insertions. We may separate $A_0$ into transverse and longitudinal components, and let us restrict our attention to the transverse part $A_{\perp,0}$ here. The correction to the self energy of $\langle \bar{q}^a(p)q^a(-p) \rangle$ is
\begin{align}
\Pi_\perp(p)&=-\int\frac{d^2k}{(2\pi)^2} \left\langle A_{\perp,0}^\mu(k) A_{\perp,0}^\nu(-k)\right\rangle_{IR} \frac{\left(2p_\mu+ k_\mu\right)\left(2p_\nu+ k_\nu\right)}{(p+k)^2+m^2}\non &=-\frac{4p^2}{p^2+m^2}\sum_{j=0}\left(-\frac{1}{p^2+m^2}\right)^{j}\int\frac{d^2k}{(2\pi)^2} \frac{1}{N}\Delta(k)\left(1-\frac{\left(p\cdot k\right)^2}{p^2k^2}\right)\left(k^2+2p\cdot k\right)^j.\label{Eq.5 operator part ope 1}
\end{align}
This is essentially identical to the arc diagram self-energy \eqref{Eq.4 Pi perp arc} in the large $N$ approach, but we are working as if we are ignorant of the infrared physics. We do know that the two operators $A_\perp$ must contract with each other so there is a single loop momentum $k$. We also know that the general form of the IR two point function must be
$$\left\langle A_{\perp,0}^\mu(k) A_{\perp,0}^\nu(-k)\right\rangle_{IR}= 
\frac{1}{N}\Delta(k)\left(g^{\mu\nu}-\frac{k^\mu k^\nu}{k^2}\right),$$
but here we do not presume to know what $\Delta(k)$ is, nor do we presume to be able to calculate $m^2$ from bare parameters.

This integral can now be expanded in powers of the loop momentum $k^2$, by making use of a formula for the average over angle of $\left(p\cdot k\right)$, which may be calculated similarly to the $A_j$ coefficients in Section 5 of \cite{schubring2021lessons},
\begin{align}\frac{1}{p^{2n} k^{2n}
	\,\Omega }\int d\Omega \,\,(p\cdot k)^{2n} =\frac{1\times 3\times 5\times \dots}{d(d+2)(d+4)\dots}= \frac{(2n-1)!}{2^{2n-1}(n-1)!\left(d/2\right)_n}.\label{Eq.5 angular integral}
\end{align}
The terms in \eqref{Eq.5 operator part ope 1} involving one power of $k^2$ lead to
\begin{align}
	\frac{2p^2m^2}{\left(p^2+m^2\right)^3}\int\frac{d^2k}{(2\pi)^2} \frac{k^2}{N}\Delta(k)= -\frac{1}{p^2}\sum_{j=1} \frac{j(j+1)}{2}\left(-\frac{m^2}{p^2}\right)^j\left\langle \left(F_{\mu\nu}\right)^2\right\rangle,\label{Eq.5 operator part F2}
\end{align}
where the integral over $k^2\Delta(k)$ has been rewritten as a local VEV of $\left(\partial_\mu A_\nu - \partial_\nu A_\mu\right)^2$.

This is now clearly in the form of an operator product expansion, and the higher powers of loop momentum $k^{2(n+1)}$ in \eqref{Eq.5 operator part ope 1} will lead to higher derivatives $ \left\langle F^{\mu\nu}\partial^{2n}F_{\mu\nu}\right\rangle$. A general formula for the expansion in terms of these higher derivative operators as well as the gauge invariant operator $\langle A_\perp^2\rangle$ will be derived in Sec. \ref{Sec 5 mellin}, in which we make use of explicit knowledge of $\Delta(k)$. But that formula may in principle be derived only using the methods here involving 
\eqref{Eq.5 operator part ope 1}, \eqref{Eq.5 angular integral} and some combinatorics.\footnote{I tested this explicitly for the case with arbitrary non-zero powers of loop momentum but zero mass insertions (in which case it can eventually be shown to vanish) and for the $A_\perp^2$, $F^2$, and $F\partial^2 F$ operators with arbitrary mass insertions.}

	\subsection{Coefficient function part of the OPE}\label{Sec 5 coefficient function}
	
Now for the coefficient function part of the OPE we can ignore the effect of the background fields in the action except for the presence of the $m^2$ term. But the arguments we have given for dropping the higher order vertices in the background field action are no longer valid. In effect the problem of finding the coefficient function part of the spinon 2-point function ends up being much the same as the calculation of the 2-point function in the ordinary naive perturbation theory in $g$.\footnote{See \cite{bardeenLeeShrock1976} and \cite{Elitzur:1978ww} for related calculations in the $O(N)$ model, and \cite{HABER1980458} in the $CP^{N-1}$ model.} 

The difference is that there are three mass scales involved, $m<\mu<M$, where $m$ is the physical IR scale coming from non-perturbative effects, $\mu$ is an arbitrary factorization scale cutting off the propagators of the $q$ fields, and $M$ is the UV cutoff. In ordinary perturbation theory often a mass $\mu$ is introduced to regulate infrared divergences, but this is considered artificial and the hope is to remove it at the end of the calculation. From the perspective of the background field method, $\mu$ is an arbitrary dividing line between the IR and UV fields that should be there to make sense of the perturbation series. Even if $\mu$ is formally sent to zero, the physical mass $m^2$ from the VEV of the background fields is present to regulate IR divergences. From this perspective, the possibility of removing all IR regulators in the calculation of coordinate invariant quantities shown by Elitzur and others  \cite{Elitzur:1978ww,david1981cancellations,becchi1988renormalizability} is equivalent to a calculation of the coefficient function of the identity operator in the OPE, and a further power series expansion in $m^2$ gives the higher order coefficient functions.

So that the calculations are more tractable, in this section we will treat $\mu$ as a mass for the $q$ field rather than a hard IR cutoff on the propagators. This scheme is not entirely justified, but in the end we will take the limit $\mu\rightarrow m$ which would just be equivalent to setting a hard cutoff $\mu\rightarrow 0$ while keeping the physical mass $m$ from the background fields fixed. This limit will have much to with correlation functions calculated in the ordinary large $N$ approach.

Going back to the original Lagrangian \eqref{Lagr SU(N) invariant} and keeping the relevant terms, we have
\begin{align*}
|\partial q|^2 + \mu^2|q|^2 + \frac{1}{2}\left(1-2g^2|q|^2\right)\left(\partial \psi\right) ^2 &+ \frac{1}{2g^2}\left(\partial|\sigma|\right)^2+ \frac{\beta}{2}\left(i\left(1-2g^2|q|^2\right)\partial\psi+g\left(\bar{q}\partial q- q\partial \bar{q}\right)^2\right)^2 .
\end{align*}
Now we may decouple the final term with a Hubbard-Stratonovich vector field $a_\mu$ leading to
\begin{align}
\Lagr_{pert}=	\left(\partial+iga\right)\bar{q}\cdot\left(\partial-iga\right)q  + \mu^2|q|^2 + \frac{1}{2}\left(1-2g^2|q|^2\right)\left(\partial \psi-a\right) ^2 +\frac{1-\beta}{2\beta}a^2- \frac{1}{2g^2}|\sigma|\partial^2|\sigma|.\label{Lagr pert}
\end{align}

The coupling to the single $\psi$ degree of freedom should not matter for 2-point functions of $q$ at this order of the large $N$ expansion. Diagrams involving the $a$ field will involve arbitrary chains of $q$ bubbles as in Fig \ref{Fig.4 bubble chains} and so in the limit $\mu\rightarrow m$ they will straightforwardly correspond to diagrams in the large $N$ approach involving the $A$ field. So in what follows we will focus on the more subtle $|\sigma|\partial^2|\sigma|$ term.

Recall that $|\sigma|$ is defined as \eqref{Def sigma}
\begin{align*}
|\sigma|=\sqrt{1-2g^2|q|^2}.
\end{align*}

%\ds{Show action only including $q$ fields and an $\mu^2$ term, arguing that the $\psi$ fields don't matter at large $N$. Argue that the $(qdq)^2$ vertices simply reproduce the $A$ field. Repeat calculation for bubbles as in paper. Talk a little about apparent double counting}

%The second term in the correlation function involving $	\sigma_\varphi$ is necessary to cancel the IR divergences in the first term involving the $N-1$ components $\varphi^a$ \cite{Elitzur:1978ww,David:1980rr}. Strictly speaking $\varphi$ may still be thought of as a UV field defined up to some arbitrary IR cutoff $\mu>m$, beyond which it is more appropriate to consider the IR field $n_0$ via some non-perturbative method. The cancellation of IR divergences simply means $\mu$ can be taken arbitrarily small in a way which can be compared to asymptotic methods such as those in Section 6. Since it will be taken arbitrarily small anyway, for simplicity we will modify the IR cutoff to be a soft cutoff given by an ad-hoc mass term $\mu^2\varphi^2$ as is usual in perturbative treatments of the $O(N)$ model.

Once $|\sigma|\partial^2|\sigma|$ is expanded as a power series in $q$, there may be arbitrary powers of $g^2|q|^2$ on either side of the Laplacian $\partial^2$.  It is convenient to represent this Laplacian in Feynman diagram notation as a dotted line corresponding to a factor $-p^2/(2g^2)$, where $p$ is the net momentum of the $|\sigma|$ factors. In a slight abuse of language we will refer to this as a \emph{$\sigma$ propagator}. A similar notation is used in \cite{Elitzur:1978ww,david1981cancellations}.\footnote{In \cite{Elitzur:1978ww} the pairing of each factor of $\bar{q}^aq^a$ is also indicated explicitly in diagrams. Here this will not be necessary since large $N$ considerations drastically restrict the relevant diagrams.}

\begin{figure}[t]
	\centering
	\includegraphics[width=0.6\textwidth]{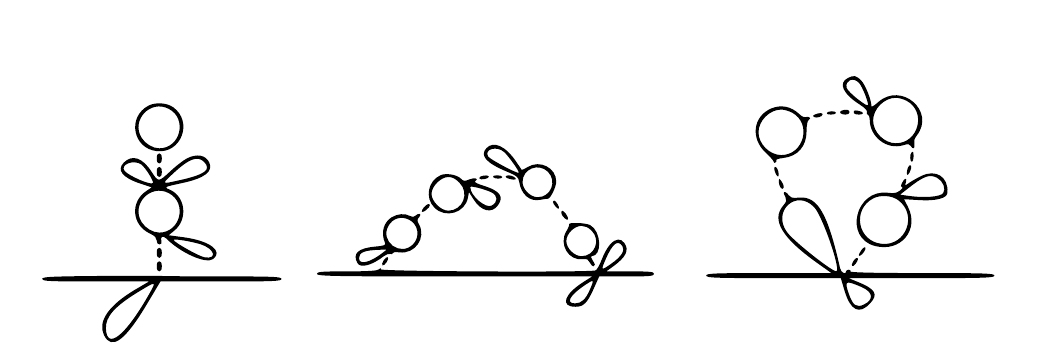}
	\caption{Some bubble chains involved in the correction of the $q$ two-point function. A solid black line indicates a $q$ propagator, and a dotted line indicates a $\sigma$ propagator as discussed in the text. The left diagram would be of $\order{N^{0}}$ but vanishes due to zero momentum through the dotted line. the right two diagrams contribute at order $\sim 1/N$.}\label{Fig BubbleChain}
\end{figure}

As in the discussion of the large $N$ limit of the OPE in Section \ref{Sec 5 Large N}, since each power of $g^2|q|^2$ in the expansion of $|\sigma|$ comes with a factor of $g^2\sim N^{-1}$ the $|q|^2$ factor must be contracted with itself as a tadpole to provide a compensating factor of $N$. If instead two separate factors of $g^2 |q|^2$ are contracted with each other as a bubble then there is only one compensating factor of $N$ so this is unfavorable. However, the $g^{-2}$ factor from the $\sigma$ propagator may compensate a single unfavorable bubble contraction, so we may form bubble chains in the large $N$ limit as in Fig \ref{Fig BubbleChain}.

Purely from considering factors of $N$ in this manner, the class of diagrams illustrated on the left of Fig \ref{Fig BubbleChain} would be expected to contribute at leading order in the large $N$ limit. However there is no net momentum flowing through the dotted lines of the bubble chain, so in fact these diagrams vanish. A bubble chain may be routed in a loop in order to introduce a net momentum through the chain, but this gives up the factor of $N$ associated to the contracted $|q|^2$ at the end of the chain, so these diagrams first appear at subleading order in large $N$. There are two distinct classes of diagrams forming a bubble chain loop at this order.

To calculate the class of \emph{arc diagrams} illustrated in the center of Fig \ref{Fig BubbleChain}, we reserve one factor of $\nor{|q|^2}$ in each $|\sigma|$ in the interaction term of \eqref{Lagr pert} to either form a bubble or connect to the external legs. The remaining $|q|^2$ are contracted into tadpoles which effectively renormalizes $g^2$ to $g_\mu^2$,
$$g^2_\mu\equiv \frac{g^2}{1-\frac{Ng^2}{2\pi}\ln \frac{M^2}{\mu^2}}.$$
So this manner of contraction leads to an effective interaction term of the form,
$$-|\sigma|\left(-\frac{p^2}{2g^2}\right)|\sigma| \rightarrow \frac{g^2_\mu p^2}{2}\nor{|q|^2}^2.$$ 
Using this it is now easy to sum over the bubble chain arc diagrams. The correction to the self energy of a $q$ propagator is
\begin{align}
	\Pi_{arc}(p)= \int \frac{d^2 k}{\left(2\pi\right)^2}\frac{g_\mu^2 k^2}{1+Ng_\mu^2 k^2 J(k;\mu)}\frac{1}{(k+p)^2+\mu^2}
\end{align}
where $J(k;\mu)$ is the $d=2$ bubble integral \eqref{Def J}.

The class of diagrams on the right of Fig \ref{Fig BubbleChain} will be referred to as \emph{tadpole diagrams} for reasons that will become apparent. The effective bubble chain interaction involving $\nor{|q|^2}^2$ presented above may be used for most of the diagram, but the part of the diagram that connects to the external $q$ legs requires one more factor of $\nor{|q|^2}$. The end result is
\begin{align}
	\Pi_{tadpole}= -g_\mu^2\int \frac{d^2 k}{\left(2\pi\right)^2}\left(\frac{1}{1+Ng_\mu^2 k^2 J(k;\mu)}-1\right).
\end{align}
The power law divergent $-1$ term may be shown to cancel exactly with the correction coming from the Jacobian in the path integral measure and so it will be disregarded.\footnote{Of course in dimensional regularization this simply vanishes directly. In \cite{schubring2021treating} the Jacobian correction was calculated but it rather confusingly was taken to cancel with part of the arc diagrams.}

Now notice that when we set $\mu\rightarrow m$, the renormalized $g_\mu$ diverges.\footnote{This relies on the knowledge that the VEV $m^2$ from the background fields is the same as the RG invariant scale $\Lambda^2$, which strictly speaking is information outside the purview of OPE approach.} In this limit $\Pi_{arc}$ which we calculated from naive perturbation theory goes exactly to the quantity $\Pi_{\alpha,arc}$ which we calculated from the $\alpha$ propagator in the large $N$ approach \eqref{Eq.4 alpha arc}. The tadpole correction goes to
$$-\int \frac{d^2 k}{\left(2\pi\right)^2}\frac{1}{ J(k;m)}\frac{1}{k^2},$$ 
which is equivalent to the $k^2\gg m^2$ part of the first term in the $\alpha$ tadpole \eqref{Eq.4 alpha tadpole}, and it serves the same role of cutting off the power law divergence of the arc diagrams.

Presumably the behavior of the tadpole integral for $k^2\sim m^2$ would be resolved by a more careful treatment of the $\mu^2$ regularization which keeps the background field VEV $m^2$ present throughout. There is also a correction to the spinon two-point function coming from the $\langle\sigma(x)\sigma(0) \rangle$ term in \eqref{Eq.5 corrTotal}, and it also involves non-trivial bubble chains at order $1/N$. We calculated this correction in \cite{schubring2021treating} and although it does not appear to have an interesting $\mu\rightarrow m$ limit, presumably it would become important in a more careful treatment.

However the point of this discussion is that in the formal limit $\mu\rightarrow m$ the $p$ dependence arising from the arc diagrams is exactly the same as in the large $N$ approach, and an expansion of $\Pi_{arc}(p)+\Pi_{tadpole}$ in powers of $m^{2j}$ will lead to coefficient functions $C_j^{(1)}(p)$ which correspond to those calculated in Sec. \ref{Sec 5 mellin}.

At this point we should clarify the distinction between the operator and coefficient function parts of the OPE. Also in the operator part, corrections from the $\alpha_0$ background field in \eqref{Lagr background field large N} will lead to integrals corresponding to the arc and tadpole corrections in the large $N$ approach, and these will also be expanded in powers of $m^{2j}$, so it seems as though we might be double counting.

In the background field method the resolution comes from the strict use of $\mu$ to separate the IR fields from the UV fields. In the operator part of the OPE it is the propagator $\left((p+k)^2+m^2\right)^{-1}$ that is expanded while the $\alpha_0$ arc itself is lumped into a VEV with UV cutoff $\mu$. In the coefficient function part it is the arc that is first expanded in powers of $m^2$ and then integrated. This leads to neither double counting nor an IR divergence since $k^2$ has an explicit IR cutoff.

In the more analytic large $N$ approach used in Sec. \ref{Sec 5 mellin} there is no scale $\mu$ to separate the two parts of the OPE. But there still will be no double counting since the two parts of the OPE will be shown to arise from different poles in the Mellin transform. Those poles associated to the coefficient function part may be interpreted as an expansion of the $\alpha$ or $A$ arcs in powers of $m^2$, so there is a direct analogy to background field approach to the OPE.

	\section{Ambiguities in operator condensates}\label{Sec.5 ambiguity operator}
	
	If the factorization scale $\mu$ in the background field method is not strictly enforced then the bubble chains of Fig \ref{Fig BubbleChain} will lead to coefficient functions which are factorially divergent when expressed as power series in $g$. As we will review further below, this divergence corresponds to IR renormalon poles in the Borel plane, and an attempt at Borel summation will lead to imaginary ambiguities which depend on whether we integrate on a contour slightly above or below these poles. But we would expect that exact expressions for the correlation functions do not have such ambiguities. So the IR renormalon ambiguities from the coefficient functions must somehow cancel in the full OPE, and in fact they will cancel with ambiguities arising in the precise definition of the operator VEVs. This was first pointed out in the context of the bosonic $O(N)$ model by 
	David \cite{dave1984237,david1986operator}.
	
	So far we have been using a hard cutoff $\mu > m$ separating UV fields from IR background fields as in \cite{novikov1984two}. If this IR cutoff on the UV fields is taken seriously in calculating the coefficient functions the factorial divergence is suppressed and no IR renormalons are seen, so the problem of ambiguities never arises (see the discussion in \cite{Shifman:2013uka}). Instead it is the arbitrary cutoff $\mu$ that must cancel in the full OPE. This lack of dependence on $\mu$ in the spinon propagator was seen explicitly in the calculation leading to \eqref{Eq.5 spinon ope lowest order}, although no IR renormalons would be seen in this simple example at lowest order in large $N$.
	
	However if we formally take the limit $\mu\rightarrow 0$, or use some form of analytic regularization, then there are indeed IR renormalons. And so the corresponding ambiguities in the operator VEVs must be taken seriously as well. We will calculate these ambiguities via two complementary ways. One method is somewhat closer in spirit to David's original calculation in dimensional regularization, although we will use a simpler scheme introduced by Beneke, Braun and Kivel \cite{beneke1998308}. This method will depend on the specific form of the gauge field propagator $\Delta(k)$ so it is not easily extendable beyond the integrable case $\gamma_0=1/\pi$. However it will be very powerful and we will also be able to use it in Sec \ref{Sec 5 mellin} to calculate the ambiguities in the coefficient functions and demonstrate a full cancellation of ambiguities in the OPE.
	
	The other method involves a hard cutoff $\mu$ and so it is perhaps easier to understand although it is somewhat heuristic. This was also first proposed by David \cite{dave1984237} in terms of lattice regularization. We will apply it here to calculate the ambiguities of the operators $F^{\mu\nu}\partial^{2n}F_{\mu\nu}$ for any value of $\gamma_0$. This will require us to uncover some subtleties about proper use of this heuristic method which have been overlooked in the literature.

	\subsection{Ambiguity in a cutoff scheme}
	To introduce the cutoff method and the idea of renormalons, consider the following integral,
	\begin{align}
		I_n\equiv\int_0^{\mu^2}dk^2 \frac{k^{2(n-1)}}{\log \frac{k^2}{m^2}}.
	\end{align}
	Given that the $k^2\gg m^2$ limit of the bubble integral $J(k)$ is proportional to a $\log$ of this form \eqref{Eq.4 J asymptotic limit}, these are exactly the kind of integrals that will arise in calculating the operator VEVs, although the subtlety alluded to above is that the higher order terms in the $m^2/k^2$ expansion of $J(k)$ will end up mattering. We will worry about this later and for now let us just discuss the integral itself.
	
	$I_n$ can actually easily be put in the form of the exponential integral special function $\text{Ei}$,
	\begin{align*}
		I_n= 
		m^{2n}\,\text{Ei}\left(n\log \frac{\mu^2}{m^2}\right),\qquad \text{Ei}(x)\equiv \int_{-\infty}^{x}dy\frac{e^y}{y},
	\end{align*}
and so in that sense it is well-defined function of the cutoff $\mu$ with no ambiguity. But a heuristic picture of how the ambiguity arises is as follows. If we calculate the power law UV divergences of $I_n$ as $\mu$ goes to infinity we will in fact find a factorially divergent series in $g$ which does have a Borel ambiguity. Since the full integral $I_n$ is unambiguous, the remaining non-power law divergent terms in $I_n$ should be considered to include an ambiguity of opposite sign to cancel that arising from the divergent series. Now if we recalculate the quantity corresponding to $I_n$ in dimensional regularization, the part that is power law divergent in $\mu$ never appears, however the ambiguity of opposite sign remains.
	
To represent $I_n$ as a factorially divergent series we may define a rescaled coupling constant $\hat{g}$, and change the variable of integration,
\begin{align*}
	\hat{g}\equiv \frac{Ng^2}{4\pi}=\left(\ln\frac{\mu^2}{m^2}\right)^{-1}, \qquad y\equiv \ln \frac{\mu^2}{k^2},
\end{align*}
\begin{align*}
I_n&=\mu^{2n}\int_0^{\infty} dy\frac{\hat{g}e^{-ny}}{ 1-\hat{g}y}=\mu^{2n}\sum_{j=0}^\infty j!\left(\frac{\hat{g}}{n}\right)^{j+1}.
\end{align*}
This series is clearly divergent, but it can be easily treated by taking the Borel transform, defined for a power series in $\hat{g}$ with arbitrary coefficients $r_j$ as
\begin{align*}
	\mathcal{B}\left(\sum_{j=0}^\infty r_j\hat{g}^{j+1}\right)\![t]\equiv \sum_{j=0}^\infty \frac{r_j}{j!}t^j,
\end{align*}
and the corresponding inverse Borel transform of a function $f(t)$ is defined as
\begin{align*}
	\mathcal{B}^{-1}\left(f \right)[\hat{g}]=\int_{0}^{\infty}dt f(t)e^{-\frac{t}{\hat{g}}}.
\end{align*}
So then,
\begin{align*}
	\mathcal{B}\left(I_n \right)[t]&=\frac{\mu^{2n}}{n-t},
\end{align*}
which has a pole on the positive $t$ axis so the inverse Borel transformation can not be applied without further consideration. Poles such as this are what we are referring to as renormalons. The $t$ integral may be regularized by deforming the contour of integration either above or below the pole, and the difference between the two will lead to an imaginary ambiguity which is clearly equivalent to taking a contour integral encircling the pole.\footnote{We will not need the full machinery of resurgence here, but this operation of calculating the ambiguity is closely related to a simple case of the {alien derivative} and the positive real axis where the renormalons are situated is referred to as a {Stokes line} (see e.g. \cite{dorigoni2019introduction}).} Using the definition of $\hat{g}$, the residue is calculated as
\begin{align*}
	\text{Res} \left[\frac{\mu^{2n}}{n-t}e^{-\frac{t}{\hat{g}}}\right]=- m^{2n}.
\end{align*}
This (multiplied by $2\pi i$) gives the ambiguity in the Borel summation of the divergent series, but again the heuristic idea is that there is a corresponding ambiguity of opposite sign in the non-perturbative part of $I_n$ and that is what remains when dimensional regularization is considered. To summarize, using the notation of curly braces to indicate the ambiguity,
\begin{align}
	\{I_n\}=+ m^{2n}.
\end{align}
\subsection{Ambiguity in an analytic scheme}

Let us now speak more about the physical VEVs themselves, which lead to integrals that are closely related to the $I_n$ above. The operators associated to the transverse gauge field\footnote{The VEV for $n=1$ is associated to the operator $-2\left\langle A_\perp^2\right\rangle$. } are
\begin{align}
	\left\langle F^{\mu\nu}\partial^{2(n-2)}F_{\mu\nu}\right\rangle &= 2\frac{(-1)^n}{N}\int \frac{d^2 k}{(2\pi)^2}k^{2(n-1)}\Delta(k)\non
	&=\frac{(-1)^n}{N}\int_0^{\infty}dk^2 \frac{k^{2(n-1)}}{\xi\log\left(\frac{\xi+1}{\xi-1}\right)+2\pi\gamma-2}.\label{Eq.5 F operators}
\end{align}
Recall that $\xi\equiv \sqrt{1+\frac{4m^2}{k^2}}$. If we restrict to the integrable case $\gamma=1/\pi$ and keep only the leading order of the integrand for $k^2\gg m^2$ this will reduce to the integral $I_n$ above, but now we wish to consider a scheme in which there is no cutoff $\mu$ and the integral is regulated by analytically continuing in the parameter $n$. This is essentially the same as keeping $n$ as an integer and analytically continuing in dimension, which introduces a factor of $k^{2\left(d/2-1\right)}$ from the measure. The only difference is we are always keeping the bubble integral $J(k)$ in the $d=2$ form rather than using the hypergeometric function \eqref{Eq.4 hypergeometric} for general $d$.

Let us refer to this integral where $\gamma=1/\pi$ as $I'_n$,
\begin{align}
	I'_n \equiv \int_0^{\infty}dk^2 \frac{k^{2(n-1)}}{\xi\log\left(\frac{\xi+1}{\xi-1}\right)}.
\end{align}
The ambiguity may be calculated easily using a trick from Beneke, Braun and Kivel \cite{beneke1998308}. Rewrite the $\log$ using an auxiliary integration over a parameter $t$,
\begin{align*}
	I'_n = \int_0^\infty dt\int_0^{\infty}dk^2 k^{2(n-1)}\xi^{-1}\left(\frac{\xi-1}{\xi+1}\right)^t.
\end{align*}
Although it is not obvious now, this $t$ will end up being identified with the parameter $t$ in the Borel plane. The interior integral over loop momentum may be put into the standard integral form for Euler's beta function, and the result is
\begin{align}
	I'_n = m^{2n}\int_0^\infty dt \frac{\Gamma(2n+1)\Gamma{(t-n)}}{\Gamma(t+n+1)}.\label{Eq.5 In beta function}
\end{align}
The integrand has poles at $t=n-j$ where $j$ is an integer ranging from $0$ to $n$. It will become clear in the next section that the poles for $t>0$ are associated to renormalons and may cancel in the full OPE whereas the $t=0$ pole is associated to UV divergences which must be removed by regularization. Ignoring the pole at $t=0$, the integral may be regularized by integrating slightly above or below the positive $t$ axis, and once again this leads to an ambiguity. The ambiguity may be quantified by residues of the poles for $t>0$, denoted again by curly braces,
\begin{align}
	\{I'_n\}=m^{2n}\sum_{j=0}^{n-1}\frac{(-1)^{j}}{j!}\frac{(2n)!}{(2n-j)!}=m^{2n}\frac{(-1)^{n+1}(2n)!}{2\left(n!\right)^2}.\label{Eq.5 ambiguity formula}
\end{align}

This expression allows us to calculate the ambiguity of all local operators constructed from $A$ and $\alpha$ in the integrable limit of the squashed sphere model, $\gamma=1/\pi$. To calculate for $\gamma\neq 1/\pi$ the cutoff scheme will be considered further.

\subsection{Calculation of operator ambiguities in the non-integrable case}
Clearly $\{I'_n\}\neq\{I_n\}=m^{2n}$ so it might seem at first as though the heuristic cutoff method does not work. Actually I will show that the cutoff method does work if the higher order powers in $m^2/k^2$ are kept in the expansion of $J(k)$ in the VEVs.

It appears that this has been overlooked in the literature so far since the cutoff method has essentially only been compared to the analytic method for the Lagrange multiplier field operators $\alpha$ and $\alpha^2$ in the $O(N)$ model \cite{dave1984237,david1986operator}.\footnote{We have also tested this for some additional fields in the SUSY $O(N)$ model \cite{schubring2021treating}, but the integrals involved were the same as for $\alpha$ and $\alpha^2$.} The integral for $\langle\alpha\rangle$ is just the tadpole \eqref{Eq.4 alpha tadpole}, and the relevant term is identical to $I'_1$. And in the special case $n=1$ it is true that $\{I'_n\}=\{I_n\}$.

The ambiguity for $\langle \alpha^2 \rangle$ is
\begin{align*}
	\{\alpha^2\}  =-\frac{2}{N}\left\{\int \frac{d^2 k}{(2\pi)}\frac{1}{J(k)}\right\}=-\frac{1}{N}\left\{I'_2 + 4m^2 I'_1\right\}=-\frac{m^4}{N}.
\end{align*}
In this case it is a pure accident that the $I'_2$ and $4m^2I'_1$ terms cancel in just the right way to lead to the naive result $\{I_2\}=m^4$.

But now if we consider a nontrivial case $I'_3$, and write the ambiguity as a summation over residues indexed by $j$ in \eqref{Eq.5 ambiguity formula},
\begin{align*}
\{I'_3\}=m^6\left(1-6+15\right).
\end{align*}
On the other hand if consider $I'_3$ regulated by a UV cutoff $\mu$ and expand the integrand up to $\order{\frac{m^4}{k^4}}$ we find\footnote{The expansion of the integrand may be facilitated by use of \eqref{Eq.App bubble}.},
\begin{align*}
I'_3 &=  \int_0^{\mu^2}dk^2 \frac{k^4}{\xi\log\left(\frac{\xi+1}{\xi-1}\right)}\non&=\left(I_3-6 m^2 I_2+15m^4I_1\right) +2\hat{g}^2m^2\mu^4-9\hat{g}m^4\mu^2-2\hat{g}^2m^4\mu^2+\order{\log\frac{\mu^2}{m^2}}.
\end{align*}
The ambiguities only come from the $I_n$ terms in parenthesis and the coefficients are just right to agree with the analytic scheme.

In dealing with expansions in the cutoff method like this it is useful to consider a slight generalization of $I_n$,
\begin{align}
	I_n(C;p)\equiv \int_0^{\mu^2}dk^2 \frac{k^{2(n-1)}}{\left(\log \frac{k^2}{m^2}+C\right)^p}
\end{align}
The $C$ term is needed for dealing with the non-integrable case $\gamma\neq 1/\pi$, and we will consider this first for $p=1$. Making the same substitutions as were done for $I_n$ we find,
\begin{align*}
	I_n(C;1)=\mu^{2n}\frac{\hat{g}}{1+\hat{g}C}\int_0^{\infty} dy\frac{e^{-ny}}{ 1-\frac{\hat{g}}{1+\hat{g}C}y}.
\end{align*}
This evidently identical with the integral for $I_n$ with $\hat{g}$ replaced by $\hat{g}/(1+\hat{g}C)$. So we may Borel transform in this parameter instead, which leads to
\begin{align*}
	\{I_n(C;1)\}=-\text{Res} \left[\frac{\mu^{2n}}{n-t}e^{-\frac{t}{\hat{g}}(1+\hat{g}C)}\right]=e^{-C}m^{2n}.
\end{align*}

The integral with $p>1$ can be reduced to the integral for $p=1$ by the following recursive relation
\begin{align}
	I_n(C,p)=-\frac{\mu^{2n}}{p-1}\left(\frac{\hat{g}}{1+\hat{g}C}\right)^{p-1}+\frac{n}{p-1}I_n(C,p-1)
\end{align}
This relation is the source of all the additional power law divergent terms involving $\hat{g}$ in the example of $I'_3$ above. The ambiguous part may be calculated by reducing to $I_n(C,1)$,
\begin{align}
	\left\{I_n(C,p)\right\}=\frac{n^{p-1}}{(p-1)!}e^{-C}m^{2n}.\label{Eq.5 ambiguity more general}
\end{align}

This is now everything needed to calculate the ambiguities of high order operators for the $\gamma\neq 1/\pi$ case, which is not easily treated with the analytic scheme. To conclude with an example, the calculation of the ambiguity of the operator $ \left(F_{\mu\nu}\right)^2$ is
\begin{align*}
\left\langle \left(F_{\mu\nu}\right)^2\right\rangle &= \frac{1}{N}\int_0^{\mu^2}dk^2 \frac{k^2}{\xi\log\left(\frac{\xi+1}{\xi-1}\right)+2\pi\gamma-2}\non
&=\frac{1}{N}\int_0^{\mu^2}dk^2 \left[\frac{k^2}{\log\frac{k^2}{m^2}+2\pi\gamma-2}-2m^2\frac{1+\log\frac{k^2}{m^2}}{\left(\log\frac{k^2}{m^2}+2\pi\gamma-2\right)^2}+\order{\frac{m^4}{k^2}}\right],
\end{align*}
so applying \eqref{Eq.5 ambiguity more general},
\begin{align}
	\left\{\left(F_{\mu\nu}\right)^2\right\}=-\left(7-4\pi\gamma \right)e^{2-2\pi\gamma}\frac{m^4}{N}.
\end{align}
	\section{Transseries expansion and cancelation of renormalons}\label{Sec 5 mellin}
			Now we wish to show that all ambiguities cancel in the full OPE of the two-point function. We may refer to the OPE as a transseries to emphasize that it is an expansion in both $g$ and the non-perturbative factor $m^2\propto \exp\left(-\frac{2\pi}{Ng^2}\right)$, and that sectors of the expansion with different powers of $m^2$ are connected to each other through the mutual cancelation of ambiguities.
	
	Following the results of Sec. \ref{Sec. 5 operator part} the operator part of the OPE is already calculable to any order, and following the results of last section the ambiguities of the operators are calculable for any value of the squashing parameter $\gamma_0$. However the coefficient function part of the OPE is a little more subtle. Using the background field method with a strict factorization scale $\mu^2$ we may calculate the coefficient functions for any order of the `operator' $m^2$, but if the factorization scale $\mu^2>m^2$ is maintained there are no renormalons \cite{Shifman:2013uka}. But it was shown in Sec \ref{Sec 5 coefficient function} that if $\mu^2$ is formally taken to zero with fixed $m^2$ then these coefficient functions should be equivalent to expanding the auxiliary field propagators in the diagrams of the large $N$ expansion (Fig \ref{Fig 4 self energy}) in powers of $m^2/p^2$. So in this section we will simply begin with the large $N$ diagrams and perform just such an expansion.
	
	We will follow the method of Beneke, Braun and Kivel \cite{beneke1998308} partially introduced above in the context of calculating operator ambiguities. As mentioned above, this will differ from true dimensional regularization in that we will always keep the specific $d=2$ form of the bubble integral $J(k)$, which will allow us to introduce the Borel plane parameter $t$ at an early stage of the calculation. A disadvantage is that because the specific form of $J(k)$ is important, the following calculation will be restricted to the integrable case $\gamma_0=1/\pi$.
		
	The $\order{1/N}$ corrections to the spinon self-energy were already listed in Sec. \ref{Sec 4.2}. The $\alpha$ arc correction \eqref{Eq.4 alpha arc} is the same as the correction appearing in the $O(N)$ model, and it has already been analyzed in \cite{beneke1998308}. The $\alpha$ tadpole \eqref{Eq.4 alpha tadpole} was not directly considered in that paper but we will show later that it was effectively included as part of the $\alpha$ arc calculation.
	
	The correction due to the longitudinal part of the gauge field \eqref{Eq.4 Pi longitudinal} involves no bubble chain factor $J(k)^{-1}$ and it will not lead to renormalons, so we will not consider it in what follows. Furthermore the tadpole correction associated with the transverse part of the gauge field \eqref{Eq.4 Pi perp tadpole} simply vanishes when $\gamma_0=1/\pi$. So apart from the subtlety with the $\alpha$ tadpole, we will concern ourselves solely with the transverse $A$ arc correction to the self energy \eqref{Eq.4 Pi perp arc},
	\begin{align*}
		\Pi_{\perp,arc}(p)=-\frac{4}{N}\int \frac{d^2k}{(2\pi)^2}\frac{2\pi}{\xi \log\left(\frac{\xi+1}{\xi-1}\right)}\frac{k^2p^2-(k\cdot p)^2}{k^2\left((k+p)^2+m^2\right)}.
	\end{align*}
	
	\subsection{The Mellin transform and its relation to the OPE}
	
	As before the first step in this method will be to introduce an auxiliary integration over $t$, which will end up being interpreted as the Borel transform parameter.
	\begin{align}
		\Pi_{\perp,arc}(p)=-\frac{8\pi}{N}\int_0^\infty dt\int \frac{d^2k}{(2\pi)^2}\xi^{-1}\left(\frac{\xi-1}{\xi+1}\right)^{t}\frac{k^2p^2-(k\cdot p)^2}{k^2\left((k+p)^2+m^2\right)}.\label{Eq.5 piArc 1}
	\end{align}
	To find the coefficient function part of the OPE we must expand the factor from the arc which involves the quantity $\xi=\sqrt{1+\frac{4m^2}{k^2}}$ in powers of $m^2/k^2$, but a direct expansion runs into difficulty when we integrate over $k$.
	
	An indirect way to expand involves taking the Mellin transform (see e.g. \cite{bertrand_bertrand_ovarlez_2010}),
	\begin{align*}
		\mathcal{M}\left[\xi^{-1}\left(\frac{\xi-1}{\xi+1}\right)^{t}\right]\!(s)&\equiv\int_0^\infty d\left(\frac{m^2}{k^2}\right)\left(\frac{m^2}{k^2}\right)^{s-1}\xi^{-1}\left(\frac{\xi-1}{\xi+1}\right)^{t} \non &= B(-2s+1,s+t).
	\end{align*}
The result is expressed in terms of the Euler beta function,
\begin{align*}
	B(-2s+1,s+t)\equiv\frac{\Gamma(-2s+1)\Gamma(s+t)}{\Gamma(t-s+1)}.
\end{align*}
The integral involved in the Mellin transform only converges when the real part of $s$ satisfies $-t<\text{Re}(s)<\frac{1}{2}$. This is known as the \emph{fundamental strip}. The Mellin transform may be inverted by integrating $s$ along a contour parallel to the imaginary axis lying within the fundamental strip.
\begin{align}
\xi^{-1}\left(\frac{\xi-1}{\xi+1}\right)^{t}=\int \frac{ds}{2\pi i}\left(\frac{m^2}{k^2}\right)^{-s}B(-2s+1,s+t).\label{Eq.5 inverse Mellin identity}
\end{align}
This contour integral may be closed in the negative real half plane, which encloses poles at $s=-t-j$ for $j\in \mathbb{N}_0,$ and so the residue theorem leads to,
\begin{align*}
\xi^{-1}\left(\frac{\xi-1}{\xi+1}\right)^{t}=\left(\frac{m^2}{k^2}\right)^{t}\sum_{j=0}\frac{(-1)^j}{j!}\frac{\Gamma(2t+2j+1)}{\Gamma(2t+j+1)}\left(\frac{m^2}{k^2}\right)^{j}.
\end{align*}
This is the sense in which the poles of the Mellin transform encode an expansion in terms of $m^2/k^2$. But the Mellin transform is more suitable than the series itself as a distribution to be integrated over.

Let us illustrate this with a simplified example which I think is also helpful to see how the distinct coefficient function and operator parts of the OPE will arise in this framework. Consider the integral of a propagator to a power $\lambda$ in dimensional regularization,
\begin{align}
	\int \frac{d^d k}{(2\pi)^d}\frac{1}{\left(k^2+m^2\right)^\lambda}&=\frac{1}{(4\pi)^{d/2}\Gamma(d/2)}\int_0^\infty dk^2 \frac{\left(k^2\right)^{d/2 - 1}}{\left(k^2+m^2\right)^{\lambda}}\non
	&=\frac{1}{(4\pi)^{d/2}\Gamma(d/2)}\left[\left(m^2\right)^{d/2-\lambda}B(d/2,\lambda-d/2)\right].\label{Eq.5 prop mellin transform}
\end{align}

Although it is never introduced in this manner, dimensional regularization is a Mellin transform from $k^2$ to $d/2$, and the poles in $d/2$ are directly related to an expansion of the integrand in powers of $k^2$.

Now consider using the inverse Mellin transform and exchanging order of integration, which is what we will do in the next subsection.
\begin{align}
	\int \frac{d^d k}{(2\pi)^d}\frac{1}{\left(k^2+m^2\right)^{\lambda+1}}&=\int \frac{d^d k}{(2\pi)^d}\left[\int \frac{ds}{2\pi i}\left(k^2\right)^{-s}\left(m^2\right)^{s-\lambda}B(s,\lambda-s)\right]\frac{1}{k^2+m^2}\non
	&=\int \frac{ds}{2\pi i}\left(m^2\right)^{s-\lambda}B(s,\lambda-s)\int \frac{d^d k}{(2\pi)^d}\frac{\left(k^2\right)^{-s}}{k^2+m^2}\non
	&=\frac{\left(m^2\right)^{d/2-\lambda-1}}{(4\pi)^{d/2}\Gamma(d/2)}\int\frac{ds}{2\pi i}B(s,\lambda-s)B(s-\lambda+d/2,1+\lambda-d/2-s).\label{Eq.5 mellin transform test}
\end{align}
The fundamental strip is $0<\text{Re}(s)<\lambda$, and the interchange of order of integration is justified if also $d/2-1<\text{Re}(s)<d/2$.

So if we close the contour of integration in the negative real half plane we enclose poles from $B(s,\lambda-s)$ which would be equivalent to expanding $(k^2+m^2)^{-\lambda}$ as a power series in $k^2$ to begin with and then integrating. These poles will be analogous to the coefficient function part of the OPE. But the second Euler beta function generated by the $k$ integration also introduces new poles, and these will be analogous to the operator part of the OPE.

And although it is maybe not so obvious, this expression \eqref{Eq.5 mellin transform test} is indeed correct, and the $s$ integral reduces to $B(d/2,\lambda+1-d/2)$, which is just \eqref{Eq.5 prop mellin transform} with $\lambda$ replaced by $\lambda+1$ (see formula (6.412) of \cite{gradshteyn2014table}).

	\subsection{A comment on the $\alpha$ corrections}
	Before discussing the $A_\perp$ arc \eqref{Eq.5 piArc 1} which is our main focus, let us briefly comment on the $\alpha$ arc which was studied in \cite{beneke1998308}. The $\alpha$ arc correction \eqref{Eq.4 alpha arc} may be expressed using \eqref{Eq.5 inverse Mellin identity},
	\begin{align*}
		\Pi_{\alpha,arc}(p)&=\frac{1}{N}\int\frac{d^2k}{(2\pi)^2}\frac{2\pi}{\xi \log\left(\frac{\xi+1}{\xi-1}\right)}\left(\frac{k^2+4m^2}{(k+p)^2+m^2}\right)\non
		&=\frac{2\pi}{N}\int_0^\infty dt \int \frac{ds}{2\pi i}B(-2s+1,s+t)\int\frac{d^2k}{(2\pi)^2}\left(\frac{k^2}{m^2}\right)^s\left(\frac{k^2+4m^2}{(k+p)^2+m^2}\right).
	\end{align*}
	But there is a problem with this interchange of order of integration because there is no value of $s$ for which the interior integral converges in both the UV and IR. However if the tadpole correction \eqref{Eq.4 alpha tadpole} is included the interior integral is modified to
	\begin{align*}
\int\frac{d^2k}{(2\pi)^2}\left(\frac{k^2}{m^2}\right)^s\left(\frac{k^2+4m^2}{(k+p)^2+m^2}-1\right).
	\end{align*}
Now this converges for $-1<\text{Re}(s)<0$, which is consistent with the fundamental strip $-t<\text{Re}(s)<1/2$ as long as $t>0$.\footnote{The swap of order of integration is still unjustified for $t=0$, and in what follows we will not concern ourselves with the behavior of the integral at $t=0$. This is presumably a remnant of the $\log \log$ divergence of the original unregularized integral. }

If we would blindly proceed with calculating the $\alpha$ arc correction without including the tadpole (as I did at first), there would be no practical difference in our calculation since the $k$ integral over a lone factor of $k^{2s}$ vanishes by the usual rules of analytic regularization. But in the end of our calculation for the OPE we would find that there was a contribution in the operator part of the OPE of the exact form of the tadpole integral $\langle \alpha\rangle$. Since Sec. \ref{Sec. 5 operator part} suggests that only operator VEVs of the form $\langle\alpha\partial^{2n}\alpha\rangle$ should appear in the OPE derived from the $\alpha$ arc it would be rather puzzling how the tadpole managed to sneak in until we consider the subtleties of the calculation mentioned above.

Note also that something equivalent to the tadpole is needed for the cancelation of ambiguities to all orders demonstrated in \cite{beneke1998308}. In that paper the authors explicitly regularized the self-energy by subtracting out the value at $p^2=0$. But that subtraction amounts to including the tadpole along with an additional piece with no ambiguities.

\subsection{Asymptotic expansion of the transverse gauge field correction}

Now we will apply these methods to the correction due to the gauge field \eqref{Eq.5 piArc 1},
\begin{align}
		\Pi_{\perp,arc}(p)&=-\frac{8\pi }{N}\int_0^\infty dt \int \frac{ds}{2\pi i}B\left(-2s+1,s+t\right)\left(m^2\right)^{-s}\int\frac{d^2k}{(2\pi)^2}\frac{k^2p^2-(k\cdot p)^2}{\left(k^2\right)^{1-s}\left((k-p)^2+m^2\right)}\non
		&=-\frac{p^2 }{N}\int_0^\infty dt \int \frac{ds}{2\pi i}B\left(-2s+1,s+t\right)B(-s,1+s)\,{}_2F_1\left(1-s,-s;\,2;\,-\frac{p^2}{m^2}\right).\label{Eq.5 piArc 2}
\end{align}
Once again the exchange of order of integration is valid provided $t>0$, and $-t<\text{Re}(s)<0$. The second equality may be derived by introducing Feynman parameters in the standard way.

The power series form of the hypergeometric function gives an expansion in terms of $p^2/m^2$, but the OPE should be expanded in terms of $m^2/p^2$. Fortunately there are formulas which may be used to invert the argument of the hypergeometric function. The results are rather unwieldy and are covered more fully in Appendix \ref{appendix asymptotic}. The most important terms for our purposes are
\begin{gather}
	\Pi_{\perp,A}(p)	=-\frac{p^2 }{N}\int_0^\infty dt \int \frac{ds}{2\pi i}B\left(-2s+1,s+t\right)\left(\frac{m^2}{p^2}\right)^{-s}\sum_{k=1}\frac{(-s)_{k-1}(-s+1)_{k-1}}{(k-1)!k!}A_k(s)\left(-\frac{m^2}{p^2}\right)^{k},\label{Eq.5 piArc 3}\\
	A_k(s)\equiv  \log \frac{p^2}{m^2} - \psi(1+s)-\psi(-s+k)-\psi(1-s+k)+\psi(-s)+2\psi(1+k).\label{Def Ak}
\end{gather}
This expression involves the Pochhammer symbol
\begin{align*}
	(x)_k\equiv \frac{\Gamma(x+k)}{\Gamma(x)},
\end{align*}
and the digamma function
\begin{align*}
	\psi(x)\equiv \frac{\Gamma'(x)}{\Gamma(x)}.
\end{align*}
The self-energy is now denoted $\Pi_{\perp,A}$ to emphasize that some terms dealt with in Appendix \ref{appendix asymptotic} are not presented here.

Given the integration contour is in the strip $-t<\text{Re}(s)<0$ there are two sets of poles enclosed in the negative real half plane. The first set of poles occurs at $s=-t-j$, with $j\in \mathbb{N}_0$. These poles are equivalent to expanding the gauge field propagator arc in powers of $m^2/k^2$ first and then integrating, and they lead to the \emph{coefficient function part} of the OPE
\begin{gather}
-\frac{p^2 }{N}\int_0^\infty dt \,e^{-\frac{t}{\hat{g}(p)}}\!\sum_{j=0, k=1}\frac{\Gamma(2t+2j+1)}{\Gamma(2t+j+1)}\frac{(t+j)_{k-1}(t+j+1)_{k-1}}{j!(k-1)!k!}A_k(-t-j)\left(-\frac{m^2}{p^2}\right)^{k+j}.\label{Eq.5 coefficient function}
\end{gather}
This is expressed as an inverse Borel transform from $t$ to the running coupling
\begin{align*}
	\hat{g}(p)\equiv \left(\log \frac{p^2}{m^2}\right)^{-1}.
\end{align*} 

The other set of poles arises from the term $-\psi(1+s)$ in the factor $A_k(s)$ defined by \eqref{Def Ak}. This has poles at $s= -n$, for $n$ in the positive integers,
\begin{align}
-\frac{p^2 }{N}\sum_{n=1}\int_0^\infty dt B\left(2n+1,t-n\right)\left(\frac{m^2}{p^2}\right)^{n}\sum_{k=1}\frac{(n)_{k-1}(n+1)_{k-1}}{(k-1)!k!}\left(-\frac{m^2}{p^2}\right)^{k}.\label{Eq.5 operator part 1}
\end{align}
This is the \emph{operator part} of the OPE, which may be written more clearly by using \eqref{Eq.5 In beta function} and \eqref{Eq.5 F operators},
\begin{align}
	\sum_{n=1} \left(-\frac{1}{p^2}\right)^{n-1}\sum_{k=1}\frac{(n)_{k-1}(n+1)_{k-1}}{(k-1)!k!}\left(-\frac{m^2}{p^2}\right)^{k}\left\langle F^{\mu\nu}\partial^{2(n-2)}F_{\mu\nu}\right\rangle.\label{Eq.5 operator part 2}
\end{align}
Clearly the $n=2$ case agrees with \eqref{Eq.5 operator part F2}, and this entire expression may be derived using the methods of Sec. \ref{Sec. 5 operator part}. 

Note that as mentioned earlier the $n=1$ case should be understood as
\begin{align*}
\sum_{k=1}\left(-\frac{m^2}{p^2}\right)^{k}\left\langle F^{\mu\nu}\partial^{-2}F_{\mu\nu}\right\rangle=\sum_{k=1}\left(-\frac{m^2}{p^2}\right)^{k}\left(-2\left\langle A_\perp^2\right\rangle\right).
\end{align*}
The summation for this special case may actually be extended to $k=0$ if an additional term appearing in Appendix \ref{appendix asymptotic} is included.

\subsection{Cancelation of ambiguities to all orders}

It is now easy to see the cancelation of ambiguities. The operator ambiguities have already been discussed and we may use \eqref{Eq.5 ambiguity formula} in \eqref{Eq.5 operator part 1} to find the total ambiguity of the operator part of the OPE
\begin{align*}
	-\frac{p^2 }{N}\sum^\infty_{n,k=1}\sum_{j=0}^{n-1}\frac{(-1)^{j+k}}{j!}\frac{(2n)!}{(2n-j)!}\frac{(n)_{k-1}(n+1)_{k-1}}{(k-1)!k!}\left(-\frac{m^2}{p^2}\right)^{n+k}.
\end{align*}

On the other hand, the coefficient function part \eqref{Eq.5 coefficient function} has poles with residue $-1$ arising from the term $-\psi(1-j-t)$ in $A_k(-t-j)$ for all $t=i\in \mathbb{N}$. So the total ambiguity of the coefficient function part is
\begin{gather*}
	+\frac{p^2 }{N}\sum^\infty_{i,k=1}\sum^\infty_{j=0}\frac{(-1)^{j+k}}{j!}\frac{(2(i+j))!}{(2(i+j)-j)!}\frac{(i+j)_{k-1}(i+j+1)_{k-1}}{(k-1)!k!}\left(-\frac{m^2}{p^2}\right)^{i+j+k}.
\end{gather*}
The terms may be regrouped by introducing a new index $n=i+j$, and summing $j$ only from $0$ to $n-1$, in which case the total ambiguity of both parts are clearly identical up to a sign, and so they cancel.\footnote{The ambiguity due to the lone $\langle A_\perp^2\rangle$ operator with no extra mass factors that was mentioned above will be shown to cancel with additional coefficient function terms in Appendix \ref{appendix asymptotic}.}

So to summarize what we have shown here, the spinon correlation function was explicitly put in the form of an OPE \eqref{Eq.5 schematic ope 2}.\footnote{Just to be clear, \eqref{Eq.5 coefficient function} and \eqref{Eq.5 operator part 2} are not the entire OPE. There are additional contributions in Appendix \ref{appendix asymptotic}, contributions due to the $\alpha$ field studied in \cite{beneke1998308}, and the longitudinal gauge field correction \eqref{Eq.4 Pi longitudinal} which has no ambiguities.} The Borel transform of the coefficient function part was given by \eqref{Eq.5 coefficient function} and the operator part was given by \eqref{Eq.5 operator part 2}, which is an expression that agrees fully with the background field method in Sec. \ref{Sec. 5 operator part}. All ambiguities were shown to cancel, which is a very non-trivial check on the validity of the OPE as a transseries. If there is any change in the ambiguity of VEVs after compactifying on a cylinder \cite{ishikawa2020infrared}, a corresponding change must occur in the coefficient functions, and so this perhaps opens another direction with which to explore the resurgence idea.

	\chapter{Discussion and future directions}\label{Sec 6}
	
	This dissertation has focused on a specific class of non-linear sigma models with a squashed sphere target space. This model is of direct relevance to condensed matter as an effective theory for frustrated magnets as in chapter \ref{Sec 3} and the beginning of chapter \ref{Sec 4}. It is also useful as a toy model for high-energy physics, as in the study of renormalons in chapter \ref{Sec 5}.
	
	As was discussed in the review in section \ref{Sec review}, this model has appeared in a number of different contexts in physics and especially since it has a number of dual representations in $d=2$ it seems that researchers are not always aware of previous work that has been done. Certainly that has happened to us in our papers \cite{schubring2020sigma, schubring2021sigma} as well as the early stages of work on \cite{schubring2021skyrmeHopf}, and with section \ref{Sec review} and some of the connections made later in this dissertation I hope to remedy that to some extent.
	
	Chapter \ref{Sec 2} was primarily about introducing the model and reviewing basic calculations on the RG flow and topological charge which are used in the later chapters. The method of calculating the RG flow using the structure coefficients of the Lie group in \eqref{Eq.2 riemannAppendix} is certainly equivalent to what has been done in \cite{azaria1995massive}, but the calculation is not entirely trivial and perhaps the independent presentation here and in appendices \ref{appendix degenerate} and \ref{appendix wavefunction} will be of use to others dealing with sigma models with homogeneous but non-symmetric target spaces.
	
	The somewhat minor but unambiguously new consideration in chapter \ref{Sec 2} and our paper \cite{schubring2020sigma} was the extension of the model to a target space that might be called a `Higgsed Grassmannian.' Such a model has been considered before in \cite{bratchikov19881} and related models involving massless fermions coupled to a Grassmannian have been considered in \cite{abdalla1985non,ABDALLA1988281,abdalla1986integrable}, but it is still rather esoteric. However note that it is rather interesting as a toy model. It involves $N$ flavors and $M$ colors which are gauged, and there is a natural `Seiberg duality' relating it to the model with $N-M$ colors. This is also true for the ordinary Grassmannian model which like the $CP^{N-1}$ model has a spectrum involving confined `mesons.' But given the Higgsing of the $U(1)$ part of the gauge group, there are also deconfined `baryons' which are $SU(M)$ singlets but have a $U(1)$ charge. Furthermore as in the squashed sphere model proper there should be an RG trajectory along which the model is integrable.
	
	Chapter \ref{Sec 3} and our paper \cite{schubring2021skyrmeHopf} should be of interest to researchers in skyrmionics \cite{Back2020} given both the increasing interest in 3D extensions of magnetic skyrmions (e.g. the recent papers \cite{seki2021direct,kent2021creation}) as well as magnetic skyrmions in inversion-symmetric magnets. One extension of this work which we are considering is to investigate translationally non-invariant ``Skyrmion crystal" ground states which arise beyond the Lifshitz point where the coefficient of the kinetic term in the effective action changes sign.
	
	Of course it would be very exciting if 3D Skyrmions in the sense described in this chapter were observed experimentally in frustrated magnets, and one direction to go would be to consider more realistic models involving $SO(3)$ Goldstone modes which nevertheless would expected to be qualitatively similar to the model described in section \ref{Sec 3 lattice}. One obvious choice to consider would be a non-coplanar phase of the antiferromagnetic pyrochlore lattice as in \cite{batistaEtAl2018} and Appendix \ref{appendix pyrochlore} here. Even though it was demonstrated in Appendix \ref{appendix pyrochlore} that there is a range of parameters in the pyrochlore lattice model with stable Skyrmions, it would be desirable to ensure that these parameters are consistent with real world materials and consistent with the constraint on the ground state, which was enforced by hand here.
	
	From my own perspective the most interesting aspect of chapter \ref{Sec 3} and our work \cite{schubring2021skyrmeHopf} is the way in which the topological defects change from polyhedral rational map-like solutions at $\kappa=0$ to linked rings and knots at $\kappa=1$, as shown in Fig \ref{fig Q10}. This is a very speculative point right now, but this chapter may have some relevance to the study of the Skyrme model and its extensions as an effective theory for QCD.
	
	The Faddeev-Niemi model \cite{faddeev1976some,faddeev1997stable} has already been introduced in a high-energy physics context as a model somewhat like the Skyrme model with Hopfion topological defects which are considered like loops of string. As discussed in section \ref{Sec 3 squashed skyrme}, the Faddeev-Niemi model is one limit of the ``squashed Skyrme model'' which was considered earlier in \cite{nasir2002effective,ward2004skyrmions}. This model makes it clear that the ordinary Skyrmions in the Skyrme model may also be considered like loops of string, albeit global strings much as in the XY model in three spatial dimensions. The stable Skyrmion solutions are not well described by thin strings with thickness much smaller than the radius of curvature, but this is also true for the Hopfions in the Faddeev-Niemi model. A couple of concrete directions along these lines would be to use the baryon string ansatz in section \ref{Sec 3 squashed skyrme} to find saddle point solutions in the Skyrme model corresponding to the unstable loops described in the Faddeev-Niemi model \cite{battye1999solitons}. It would also be interesting if the difference in BPS inequalities between the Faddeev-Niemi model and the squashed Skyrme model for $\beta<1$ could be related to the difference between local and global strings respectively.
	
	Chapter \ref{Sec 4} was organized in such a way that it centers on a rather technical point raised in \cite{azaria1995massive} concerning an apparent discrepancy between the RG flow found at two loops in the weak coupling expansion compared to the results of the large $N$ expansion. This discrepancy was resolved in section \ref{Sec 4.2} as a new result, but the real purpose of chapter \ref{Sec 4} was to tie together chapters \ref{Sec 3} and \ref{Sec 5} by connecting the squashed sphere model as seen in the context of condensed matter to the large $N$ approach used extensively in chapter \ref{Sec 5}. The chapter also gave an opportunity to present some of my work on QFT in $d=1$ \cite{schubring2021lessons,schubring2021yukawa} and the squashed sphere model with a WZ term \cite{schubring2021sigma}.
	
	The full spectrum of the squashed sphere model in $d=1$ was calculated in section \ref{Sec 4.1} and my paper \cite{schubring2021lessons}, demonstrating the validity of the large $N$ approach. The goal of that paper (as well as \cite{schubring2021yukawa}) was to present a different pedagogical approach to introducing QFT by applying Euclidean time-ordered correlation functions and Feynman diagrams to ordinary quantum mechanical systems. The simple homogeneous sigma models considered here are rather interesting in $d=1$ because the large $N$ methods are exact at sub-leading order. Much like the situation for scattering of integrable systems in $d=2$, in $d=1$ only two-body interactions end up mattering for the calculation of the spectrum. As discussed in \cite{schubring2021lessons}, this has consequences for the self-consistent screening approximation \cite{scsa} which aims to improve on a truncation of the large $N$ expansion.
	
	The squashed sphere with a WZ term was considered in section \ref{Sec 4.3}, and it demonstrates that the two-loop weak coupling expansion (which is essentially a \emph{large level expansion} in this model) is consistent with calculations at the WZW fixed point. Moreover if the renormalization scheme is always taken to be consistent with the WZW fixed point then there is a particular RG trajectory which is argued to be stable to all orders, which is a rather curious result when there is no obvious symmetry protecting the RG trajectory.
	
	Probably the most interesting implication of that section and \cite{schubring2021sigma} is that the ordinary PCM with a WZ term (without any squashing) may have additional fixed points at stronger coupling than the WZW fixed point. Since this was only shown at a finite order of the large level expansion, and since the PCM with a WZ term is apparently so well-studied (certainly the WZW fixed point itself is), any claim that these additional fixed points exist non-perturbatively should of course be met with skepticism. However this does raise the question, what exactly is the nature of the RG trajectory flowing to the WZW fixed point from the strong coupling side? Is this purely an effective field theory or is it UV complete? Given all the possible tools available in this model it seems this might be an issue that might be easily settled with some careful thought.
	
	Chapter \ref{Sec 5} is in the same vein as our paper \cite{schubring2021treating} on the SUSY $O(N)$ model, and it develops the OPE in the large $N$ expansion of the squashed sphere model and demonstrates a cancelation of renormalons to all orders in a particular amplitude. The background field method considered here and in that paper can lead to a particularly clear example of an OPE outside of conformal field theory, as evidenced by expressions like \eqref{Eq.5 operator part 2} for the part of the OPE involving the VEVs $\left\langle F^{\mu\nu}\partial^{2n}F_{\mu\nu}\right\rangle$ to all orders. Of course this is to some extent a manifestation of the fact that both the squashed sphere model and the SUSY $O(N)$ model have a nice vector-like large $N$ limit, and so it would be interesting to extend these methods to other asymptotically free theories without such a limit, in particular the principal chiral model.
	
	The main direction to proceed from chapter \ref{Sec 5} should be rather obvious given the discussion so far. It involves connecting the calculation more with the resurgence program, in which the ambiguities which cancel with the renormalons in perturbation theory are taken to arise from semiclassical objects which arise after compactification with twisted boundary conditions. Instead in chapter \ref{Sec 5} we are in the infinite volume limit, and the ambiguities are taken to arise from operator VEVs, which we are able to calculate via the large $N$ limit available to us in this model. The benefit of this large $N$ approach is that an all orders cancelation of ambiguities is a very restrictive check on the ambiguities of the individual operator VEVs, and indeed the discussion in section \ref{Sec.5 ambiguity operator} suggests that the calculation of the ambiguity of the ``gluon condensate" VEV in \cite{ishikawa2020infrared} should be reconsidered. Still that paper raises an interesting prospect of a shifting of the ambiguities in the VEVs upon compactification, and it would be interesting to calculate along the lines of \cite{beneke1998308}, \cite{schubring2021treating}, and chapter \ref{Sec 5} in the presence of twisted boundary conditions to explore more the extent to which the effective quantum mechanical models in the resurgence program are continuously connected to the infinite volume limit.

	\chapter*{Acknowledgements}
		\addcontentsline{toc}{chapter}{Acknowledgements}
		
	I would like to thank my advisor Misha Shifman for accepting me as a student and posing some interesting problems in a field of physics I am interested in. I would also like to thank my Master's advisor Vitaly Vanchurin and my undergraduate advisor Paul Ohmann for setting me in the right direction. Of course in addition to Misha, my other coauthors contributed to some of the papers reviewed in this thesis. In particular, Zhentao Wang is responsible for Fig \ref{fig profile lattice charge1}, \ref{fig high charge}, \ref{fig Q10} and Carlos Naya for Fig \ref{fig Hedgehog cross section}. The notes by Chao-Hsiang Cheu for our paper on the asymptotic expansion of the supersymmetric $O(N)$ model were very helpful for the closely related problem of the asymptotic expansion of the squashed sphere model which I present in Sec. \ref{Sec 5 mellin}.
	
	I would like to thank the members of my thesis committee: Aleksey Cherman, Vlad Pribiag, Sasha Voronov, Ke Wang. Besides some of the names already listed above there were a number of people with whom I discussed or collaborated to various extents on projects:  Yi-Zen Chu, Edwin Eireson, Alex Kamanev, Misha Katsnelson, Alyosha Tsvelik, Arkady Vainshtein. In particular, I learned a lot from Misha Voloshin, and I saved the notes from his class and a project on quarkonium we worked on. I also benefited from the classes of Keith Olive, Boris Shklovskii, and Jorge Vinals. Among others I had a good experience as a teaching assistant for Vuk Mandic, Elias Puchner, and Oriol Valls.
	
	I would like to thank Yusuf Buyukdag for co-organizing the high-energy theory journal club with me for several years until events in 2020 put it on hold. Some of the regular members included  Sergey Guts, Kunio Kaneta, Michael Kreshchuk, Evgeniy Kurianovych, Andrew Miller, Minh Nguyen, Alex Papageorgiou, Caner Unal. I also learned a lot from interaction with graduate students in condensed matter, especially the students of Andrey Chubukov with whom I shared an office. A few names include Dmitry Chichinadze, Saumit Kasturirangan, Dan Phan, Daniel Shaffer, and Mengxing Ye.
	
	Last but not least I would like to thank my family for being very supportive.
	
	\begin{appendices}
		\chapter{Notation guide}
				 \label{appendix notation}
		\begin{table}[h]
			\begin{center}
				\begin{tabular}{c||c|c|c|c}
					& \cite{CAMPOSTRINI1994680} & \cite{azaria1995massive} & \cite{BALOG2000367} & \cite{basso2013integrability} \\
					\hline &&&&\\
					$g^2$ &  & $T$ & $\lambda/4$ & $\frac{2\pi}{R^2}$ \\
					$\lambda$ & $f$ & $t$&  &  \\
					$\beta$ &  & $x$ & $-g$ & $\eta$ \\
					$\gamma$ & $\kappa$ & $M^2$ &  & $\frac{r^2}{2\pi N}$ \\
					$\gamma_0$ &  & $\left(NK_N\right)^{-1}$ & $p/\pi$ &  $p/\pi$ \\
					
				\end{tabular}
				\caption{The parameters in this paper in the left column compared to the parameters in Campostrini, Rossi (1993) \cite{CAMPOSTRINI1994680}; Azaria et al (1995) \cite{azaria1995massive}; Balog, Forg\'{a}cs (2000) \cite{BALOG2000367}; Basso, Rej (2012) \cite{basso2013integrability}.}
				\label{Table parameters}
			\end{center}
		\end{table}
		
		There is no consistent notation in the literature for the squashed sphere sigma model, and due to the many forms in which the action may be written even in the same paper there may be several more or less equivalent parameters which are useful to introduce. Here is a brief summary of the parameters used in this text and their relation to the notation in some other papers on this topic.
		
		The weak coupling parameter which determines the overall size of the target space manifold is referred to as $g^2$. In the large $N$ limit it may more be useful to use a parameter which does not scale with $N$, so we also introduce the 't Hooft coupling
		\begin{align}
			\lambda\equiv N g^2.
		\end{align}
		The squashing parameter multiplying $\left(\bar{z}\cdot\partial z\right)^2$ in the action \eqref{Lagr squashed sphere standard} is referred to as $\beta$. In a form of the action \eqref{Lagr squashed sphere A psi} involving an auxiliary gauge field $A$, this is replaced by the combination $\gamma$
		\begin{align}
			\gamma\equiv \frac{1-\beta}{\lambda\beta}.
		\end{align}
		This parameter is an RG invariant at the lowest order of the large $N$ expansion, and the limit of $\gamma$ as the RG scale goes to the UV limit is a true RG invariant which is referred to as $\gamma_0$.
		
		In the lattice model in chapter \ref{Sec 3} it is more convenient to parametrize the squashing by a parameter $\kappa$ which is more directly related to the colinearity of the spin vectors \eqref{def Kappa theta},
			 \begin{align}
			 	\kappa\equiv \frac{2\beta}{3-\beta}.
			 \end{align}

				 \chapter{Coordinates on the target space}
		\label{appendix coordinates}
		
		In the basic Lagrangian \eqref{Lagr squashed sphere standard} the squashed sphere is described in terms of $z^i$ which is the same as the homogeneous coordinates on $CP^{N-1}$. As usual, on a coordinate patch that excludes the points with $z^N=0$, we may define the complex coordinates $w^a$, where $a$ runs from $1$ to $N-1$.
		\begin{align}
			w^a=x^a+i y^a=\frac{z^a}{z^N},
		\end{align}
		
		To the $2(N-1)$ real coordinates $x^a, y^a$ we also add the single real coordinate $\phi$ parametrizing the overall phase of $z$. The coordinate $\phi$ is proportional to the coordinate $\theta$ along the integral curves associated to the \emph{phase} generator $\tau_0$ discussed in chapter \ref{Sec 2}. This may be seen either by directly considering the curves $U\exp\left(i\theta \tau_0\right)z_0$, or by finding $J^0$ in terms of $\partial \phi$ using \eqref{Def J explicit},
		\begin{align}
			\phi = -\sqrt{\frac{2(N-1)}{N}}\theta.\label{Eq.App phiCoord}
		\end{align}
		
		For convenience we will also define the quantity
		\begin{align}
			\chi\equiv 1+|w|^2.
		\end{align}
		Then by transforming the Lagrangian \eqref{Lagr squashed sphere standard} to these coordinates we can read off the components of the metric,
		\begin{gather}
			g_{\phi\phi}=1-\beta\,,\qquad g_{\phi x^a}=-(1-\beta)\chi^{-1}y^a\,,\qquad g_{\phi y^a}=+(1-\beta)\chi^{-1}x^a\,,\non
			g_{x^a x^b}=\chi^{-1}\delta_{ab}-\chi^{-2}\left(x^a x^b+\beta y^a y^b\right)\,,\qquad	g_{y^a y^b}=\chi^{-1}\delta_{ab}-\chi^{-2}\left(y^a y^b+\beta x^a x^b\right)\,,\non g_{x^a y^b}=-\chi^{-2}\left(x^a y^b - \beta y^a x^b\right)\,.
		\end{gather}
		It can be verified that the inverse metric is
		\begin{gather}
			g^{\phi\phi}=\frac{1+(1-\beta)|w|^2}{1-\beta}\,,\qquad	g^{\phi x^a}=\chi y^a\,,\qquad g^{\phi y^a}=-\chi x^a\,,\non
			g^{x^a y^b}=\chi (x^ay^b-y^ax^b)\,,\qquad g^{x^a x^b}=g^{y^ay^b}=\chi(\delta^{ab}+x^ax^b+y^ay^b)\,.
		\end{gather}
		
		Now it is straightforward to calculate the connection coefficients,
		\begin{align}
			\Gamma^\phi_{\phi\phi}&=\Gamma^{x^a}_{\phi\phi}=\Gamma^{y^a}_{\phi\phi}=\Gamma^{x^b}_{\phi x^a}=	\Gamma^{y^b}_{\phi y^a}=0\,,\non[2mm]
			\Gamma^\phi_{\phi x^a}&=-(1-\beta)\frac{x^a}{\chi},\qquad\Gamma^\phi_{\phi y^a}=-(1-\beta)\frac{y^a}{\chi}\,,\non[2mm]
			\Gamma^{y^b}_{\phi x^a}&=+(1-\beta)\delta_{ab},\qquad\Gamma^{x^b}_{\phi y^a}=-(1-\beta)\delta_{ab}\non\Gamma^\phi_{ x^a y^b}&=
			\frac{\beta}{\chi^2}(x^a x^b-y^a y^b)\,,\non[2mm]
			\Gamma^\phi_{ x^a x^b}&=-	\Gamma^\phi_{ y^a y^b}=-\frac{\beta}{\chi^2}(x^a y^b+y^a x^b)\,,\non[2mm]
			\Gamma^{x^c}_{ x^a x^b}&=-\chi^{-1}(\delta_{ac}x^b+\delta_{bc}x^a),\qquad\Gamma^{y^c}_{ x^a x^b}=\beta\chi^{-1}(\delta_{ac}y^b+\delta_{bc}y^a)\,,\non[2mm]
			\Gamma^{y^c}_{ y^a y^b}&=-\chi^{-1}(\delta_{ac}y^b+\delta_{bc}y^a),\qquad\Gamma^{x^c}_{y^a y^b}=\beta\chi^{-1}(\delta_{ac}x^b+\delta_{bc}x^a)\,,\non[2mm]
			\Gamma^{x^c}_{ x^a y^b}&=-\chi^{-1}(\delta_{ac}y^b+\beta\delta_{bc}y^a),\qquad\Gamma^{y^c}_{ x^a y^b}=-\chi^{-1}(\beta\delta_{ac}x^b+\delta_{bc}x^a)\,.	\end{align}
		
		From this point a short route to the one-loop RG equations is to calculate not the full Riemann tensor but only the tensor at the point $w=0$, and only calculate the $R_{\phi\phi}$ and $R_{x^1x^1}$ components. From the general argument that only the $g$ and $\beta$ parameters should renormalize, this shorter calculation gives us all the information about the full Ricci tensor.
		
		At $w=0$ the metric becomes diagonal in this coordinate system,
		\begin{align*}
G(0)=g^{-2}\text{diag}(1,1,\dots,1,1-\beta),
		\end{align*}
		This is almost the same as the metric in the non-coordinate basis associated to left-invariant vector fields \eqref{Eq.2 metric}. The difference for the phase coordinate is due to the proportionality factor between $\phi$ and $\theta$ \eqref{Eq.App phiCoord}.
		
		The Ricci tensor components are found to be
		\begin{align}
		R_{x^1 x^1}(0)&=2(N-1+\beta)\non
R_{\phi\phi}(0)&=2(N-1)(1-\beta)^2.
		\end{align}

These may be used to calculate the one-loop RG equations using \eqref{Eq.2 general RG} and the results are consistent with \eqref{Eq.2 RG 1} and \eqref{Eq.2 RG 2}.
			\chapter{Degenerate metrics on fiber bundles}\label{appendix degenerate}
		This section is a brief account of the mathematical justification for the method used in chapter \ref{Sec 2} to calculate the Riemann tensor on the squashed sphere and its Grassmannian generalization by considering the Riemann tensor on a Lie group.\footnote{More details may be found in Appendix A to the preprint version of \cite{schubring2020sigma}, \href{https://arxiv.org/abs/1809.08228}{arXiv:1809.08228}}
		
		In the spaces we are considering there is a total space $\mathcal{E}=SU(N)$ which projects to the base space $\mathcal{M}$ we are interested in.
		$$\pi:\mathcal{E}\rightarrow \mathcal{M}.$$
		The metric $G$ on $\mathcal{M}$ may be pulled back to $\mathcal{E}$,
		$$\bar{G} \equiv \pi^* G.$$
		This metric is degenerate since \emph{vertical} vectors $\eta$ that are in the kernel of the pushforward, $\pi_* \eta=0$, have vanishing norm.
		
		As discussed in the Sec. \ref{Sec 2.1}, the method we will use will be to introduce a section $\sigma:\mathcal{M}\rightarrow\mathcal{E}$ on a coordinate patch\footnote{It is not important for our purposes to make any distinction in notation between this coordinate patch and the global manifold.} and consider that as a local embedding of $\mathcal{M}$ as a submanifold in $\mathcal{E}$. Then we will calculate the Riemann tensor $\bar{R}$ in $\mathcal{E}$ and pull it back to $\mathcal{M}$ using $\sigma$. Throughout we will use barred notation to denote an object on $\mathcal{E}$ which was lifted from a corresponding object in $\mathcal{M}$. For instance given a vector field $X$ on $\mathcal{M}$, the (non-unique) extension of its pushforward $\sigma_* X$ will be denoted $\bar{X}$.
		
		To find the Riemann tensor we first need to introduce a Riemannian connection that is metric compatible and torsion free. Since $\bar{G}$ is degenerate this choice is actually not unique. However it can be shown that we can decompose any choice of Riemannian connection $\bar{\nabla}$ on $\mathcal{E}$ into a part tangent to the submanifold $\sigma(\mathcal{M})$ expressed in terms of the unique Riemannian connection $\nabla$ associated to $G$, and a vertical component $\eta$, 
		\begin{align}
			\bar{\nabla}_{{\bar X}}{\bar Y}=\sigma_\star{\nabla}_X Y+\eta.\label{Eq.AppA connectionPerp}
		\end{align}
	Using this relation between the covariant derivatives we can relate the Riemann tensors on $\mathcal{E}$ and $\mathcal{M}$. The result is very simple. The components of the Riemann tensor $\bar{R}$ on $\mathcal{E}$ in those directions tangent to the submanifold $\sigma(\mathcal{M})$ are equal to the components of the Riemann tensor $R$ on $\mathcal{M}$ itself.
	\begin{align}
		\hat{\bar W}(\bar{R}(\bar X,\bar Y)\bar Z)=\hat{W}({R}(X,Y)Z). \label{Eq.AppA riemannTensorEquality}
	\end{align}
	Here $\hat{W}$ and $\bar{\hat{W}}$ are dual vector fields related by the pullback $\pi^*$. This relation can be proven in much the same way as the ordinary Gauss equation for the curvature of a submanifold (see for instance \cite{do1992riemannian}).
	
	Note finally that we derived this relation by using a map $\sigma$, but the only appearance of $\sigma$ in \eqref{Eq.AppA riemannTensorEquality} is in the relation between $X,Y,Z$ and $\bar{X},\bar{Y},\bar{Z}$. This equality must be true for any possible $\sigma$, and in fact since it only involves vectors in the same local tangent space we may push forward vectors with a different map $\sigma_*$ at each point. In particular, we may choose $\sigma$ to be such that the tangent space $\sigma_*(T_p\mathcal{M})$ is just the space spanned by the \emph{horizontal} left-invariant vector fields in $\mathfrak{a}+\mathfrak{b}$ at $\sigma(p)$. So indeed in Sec. \ref{Sec 2.3} it is valid to conflate the Riemann tensor on $\mathcal{E}$ and $\mathcal{M}$ for components in the horizontal directions.
		\chapter{Wave functions on the squashed sphere}\label{appendix wavefunction}
			Here we will give a very brief sketch of the wavefunction approach to finding the spectrum of a sigma model in $d=1$ with a homogeneous target space such as the sphere $S^{2N-1}$ or complex projective space $CP^{N-1}$. A more general and rigorous account can found for instance in the appendices of Camporesi's review on harmonic analysis on homogeneous spaces \cite{camporesi1990harmonic}. Here the focus will be on deriving exact results for the sigma model on a squashed sphere which will agree with the two path integral methods considered in Sec \ref{Sec 4.1}.
		
		Given some Lie group $G$ and a subgroup $H\subset G$, consider the coset space $G/H=\{gH:g\in G\}$. There is a natural $G$ action on $G/H$ given by left multiplication, and we can consider a metric $g$ on $G/H$ which is invariant under this action.\footnote{The notation in this section will be somewhat different from most of the text where $G$ refers to the target space metric and $g$ is a parameter.}
		
		For example, the even dimensional sphere
		$S^{2N-1}$ may be considered to be the homogeneous space
		\linebreak$SO(2N)/ SO(2N-1)$, but a more general metric on $SU(N)/SU(N-1)$ that only has the required $SU(N)$ symmetry will instead describe a squashed sphere. If we mod out a larger subgroup $H=SU(N-1)\times U(1)$, then the homogeneous space $SU(N)/H$ will describe $CP^{N-1}$. 
		\section{Irreducible representations}
		We can define a Hilbert space $\mathcal{H}$ consisting of complex valued wavefunctions $\psi: G/H\rightarrow \mathbb{C}$, with the obvious inner product given by integration over $G/H$ using the measure induced by the $G$ invariant metric $g$. $G$ has a natural representation $\rho$ on the wavefunctions induced by its left action on $G/H$. Given $g\in G$, define the linear map $\rho(g):\mathcal{H}\rightarrow \mathcal{H}$ by
		$$\left[\rho(g)\psi\right](x)= \psi(g^{-1}x).$$
		Given the $G$ invariant measure involved in the inner product it is easy to see this is in fact a unitary representation of $G$. However, it is a very reducible representation. It will be useful to decompose the representation of $G$ into irreducible representations, much in the same way as rotations of general wavefunctions on $S^2$ may be decomposed into representations on spherical harmonics.
		
		Suppose we consider some arbitrary unitary irreducible representation $\lambda$ of $G$ acting on some vector space $V$. Given a $g\in G$ the action of the representation on $v\in V$  may be written as $U^{(\lambda)}(g)v$. Given a basis on $V$ we can write this in components as $U^{(\lambda)}(g)^i_j v^j$.
		
		Now suppose there is some unit vector $\xi$, that is invariant under the representation for all $h\in H$,
		$$U^{(\lambda)}(h)^i_j \xi^j= \xi^i.$$
		For convenience we may choose $\xi$ to be one of the basis vectors, say at the index $i=I$. Now consider the following functions defined on $G/H$,
		$$\psi^{(\xi)}_j(gH) \equiv U^{(\lambda)}(g^{-1})^I_j.$$
		Due to the invariance of $\xi$ under the representation restricted to $H$ this is in fact a well-defined function on $G/H$ that does not depend on choice of coset representative. Furthermore under action by $\rho$ these wavefunctions transform as
		$$\rho(g)\psi^{(\xi)}_j = U^{(\lambda)}(g)^k_j\psi^{(\xi)}_k,$$ 
		so this is in fact an irreducible subrepresentation of $\rho$ unitarily equivalent to $\lambda$. In fact conversely the number of times that the irreducible representation $\lambda$ appears in the decomposition of $\rho$ is equal to the number of linearly independent vectors $\xi$ which are invariant under $H$ in the irrep $\lambda$.
		
		In particular this means that if some irreducible representation of $G$ has no vector invariant under $H$, then that irreducible representation can not appear in the Hilbert space. To be more concrete, in the case of $CP^{N-1}$, the fundamental representation of $SU(N)$ given by complex vectors $z^a$ does have an invariant vector under the $SU(N-1)$ factor of $H$. Indeed every vector has a stabilizer isomorphic to $SU(N-1)$, and the factor of $SU(N-1)$ in $H$ is just the stabilizer of some reference vector. But that reference vector is not invariant under the $U(1)$ factor of $H$, which is specifically chosen to change the phase of that reference vector. The statement that the only representations that exist in the Hilbert space are those which have a vector invariant under the full subgroup $H$ amounts to being the same as the notion that only gauge invariant combinations of operators acting on the vacuum are true states in the Hilbert space.

		\section{Hamiltonian and eigenvalues}

		The Hamiltonian $\hat{H}$ will be given by the Laplace-Beltrami operator on $G/H$
		$$\hat{H} = -\frac{1}{2}g^{ab}\nabla_a\nabla_b,$$
		where $\nabla$ is the Levi-Civita connection associated to $g$. $\hat{H}$ commutes with the unitary representation $\rho$ of $G$, so by Schur's lemma the irreducible representations $\psi^{(\xi)}$ will be eigenvectors of $\hat{H}$. To find the eigenvalues, it will be helpful to rewrite $\hat{H}$ in a form that takes advantage of the properties of $G$ as a Lie group.
		
		$G$ may be considered as a fiber bundle over the base space $G/H$ with the natural projection map $\pi(g)=gH.$ $G$ has a family of right-invariant vector fields $R_a$, where $a$ is an index of the Lie algebra, and because they are right-invariant the pushforward $$K_a\equiv\pi_\star R_a$$ does not depend on the choice of coset representative. Since the metric on $G/H$ was chosen to be invariant under left action by $G$,  $K_a$ are in fact Killing vectors on $G/H$.
		
		$G$ also has left-invariant vectors $L_a$, but the pushforward of the left-invariant vectors does depend on the choice of coset representative. Even so, a section $\sigma:G/H\rightarrow G$ may be chosen and basis vectors $e_a$ on $G/H$ may be defined through the pushforward of $L_a$ at points of $G$ specified by the section
		$$e_a\equiv \pi_\star|_\sigma L_a.$$ 
		The metrics that are being considered here have the nice property that they take the same form in the basis $e_a$ regardless of choice of $\sigma$. In particular, the metric of the squashed sphere target space is \eqref{Eq.2 metric}
		$$g_{ab}=\frac{1}{g^2}C_a \delta_{ab},$$
		where $C_a$ is equal to $1$ for every index except that associated to the left-invariant vector $L_0$ that moves in the direction of overall change of phase of the $z$ fields, $C_0$
		\begin{align*}
			C_0 = (1-\beta)\frac{2(N-1)}{N}.
		\end{align*}
		
		To proceed with the evaluation of the Laplace-Beltrami operator for this metric, it is first useful to state a rather well-known result\footnote{This is stated without proof in \cite{camporesi1990harmonic}. To prove it, it is helpful to consider the properties of the adjoint map relating $L$ and $R$, which is directly related to the components $K_a = K^b_a e_b$. A useful lemma is to first show $K^b_a\nabla_b K^c_a = 0.$} for the slightly simpler metric of $g_{ab}=\delta_{ab}$, which is the metric of $CP^{N-1}$, or $S^{2N-1}$ considered as the homogeneous space $SO(2N)/SO(2N-1)$. In this special case, the Laplace-Beltrami operator may be written in terms of the Killing vectors considered as differential operators,
		\begin{align}
			\delta^{ab}\nabla_a\left(\nabla_b \psi\right)= K_a\left( K_a \psi\right).
		\end{align}
		Then the more general Laplace-Beltrami operator with $C_0\neq 1$ may be written in these terms as well 
		\begin{align}
			\hat{H}=-\frac{g^2}{2}\left(\left(K_a\right)^2 + \left(C_0^{-1}-1\right)\left(e_0\right)^2\right).
		\end{align}
		
		Now consider how the differential operators in $\bar{H}$ act on the eigenfunctions $\psi_j(gH)= U^{(\lambda)}(g^{-1})^I_j$ above,
	\begin{align*}
	K_a \psi_j&=-i\psi_k\,{\tau^{(\lambda)}_{a}}^k_j \qimplies \left(K_a \right)^2\psi_j = -C_2^{(\lambda)} \psi_j\\
	e_0 \psi_j &= -i{\tau^{(\lambda)}_{0}}^I_k{U^{(\lambda)}}^k_j\qimplies \left(e_0\right)^2 \psi_j = -\left(C_\xi\right)^{2}\psi_j.
\end{align*}
		Here $C_2$ is the quadratic Casimir $\left(\tau_a\right)^2=C_2I$. $C_\xi$ is the diagonal coefficient $C_\xi\equiv {\tau_{0}}^I_{\,\,I}$. Note that in the squashed sphere case $\tau_0$ commutes with all the generators of $H$, and takes a diagonal form, so $\psi_j$ is indeed an eigenvector of $e_0$. To summarize,
		\begin{align}
			\hat{H}\psi_j=\frac{g^2}{2}\left(C_2^{(\lambda)}+\left(C_0^{-1}-1\right)\left(C_\xi\right)^2\right)\psi_j.\label{Eq.App eigenvalue}
		\end{align}
		
		This formula then may be used to find the exact eigenvalue for any representation $\lambda$ with a singlet vector $\xi$ under $H$.
		
		As an important special case, consider the fundamental (i.e. \emph{spinon}) representation of $SU(N)$, which has eigenvalue $m_z$. The singlet vector $\xi$ may be taken as the unit vector with $1$ in the $N$th component, and the phase generator that commutes with the stabilizer $H$ of $\xi$ is
		$$\tau_0 = \sqrt{\frac{2}{({N}-1){N}}}\text{diag}\left(1,1,\dots,1,-({N}-1)\right).$$
		The quadratic Casimir for the fundamental representation, given our normalization convention $\text{Tr}\left(\tau_a\tau_b\right)=2\delta_{ab}$ is $$C_2=\frac{2}{N}\left(N^2-1\right).$$
		So now plugging in $C_2, C_0$ and $C_\xi$ into \eqref{Eq.App eigenvalue}, the eigenvalue is
		\begin{align}
			m_z = \frac{g^2}{2}\left(2N-2+\left(1-\beta\right)^{-1}\right)=m\left(1-\frac{1}{2N}\right)+\frac{1}{2N\gamma}.
		\end{align}
	Which agrees with the corrections found in section \ref{Sec 4.1} due to the $\alpha$ field \eqref{Eq.4 spinon alpha correction} and the $A$ field \eqref{Eq.4 deltam1}.
		\chapter{Skyrmions on the pyrochlore lattice}\label{appendix pyrochlore}		
		\section{Description of the lattice}
		
			A pyrochlore lattice involves an outer face-centered-cubic (FCC) lattice and an inner tetrahedral cell with four sites which are analogous to the three sites indexed by $r$ in the model of Sec \ref{Sec 3 lattice} (See e.g. Fig. 1 in \cite{moessner1998low}). The vector from a point of the FCC lattice with a conventional cell of length $b$ to the four points of the inner lattice is given by $\mathbf{p}_r$,
		\begin{align}
			\mathbf{p}_0 = (0,0,0)\frac{b}{4},\quad \mathbf{p}_1 = (0,1,1)\frac{b}{4},\quad \mathbf{p}_2 = (1,0,1)\frac{b}{4},\quad \mathbf{p}_3 = (1,1,0)\frac{b}{4}.
		\end{align}
		The lattice spacing $a$ between nearest neighbors is
		\begin{align}
			a=\frac{b}{2\sqrt{2}}.
			\end{align}
		For any lattice site of type $r$, the two unit vectors pointing to the nearest lattice site of type $s\neq r$ are $\pm\mathbf{\delta}_{rs}$ (the ordering of $rs$ is irrelevant due to the $\pm$)
		\begin{gather}
			\mathbf{\delta}_{10}=\frac{1}{\sqrt{2}}(0,1,1),\quad \mathbf{\delta}_{20}=\frac{1}{\sqrt{2}}(1,0,1),\quad \mathbf{\delta}_{30}=\frac{1}{\sqrt{2}}(1,1,0)\non
			\mathbf{\delta}_{32}=\frac{1}{\sqrt{2}}(0,1,-1),\quad \mathbf{\delta}_{31}=\frac{1}{\sqrt{2}}(1,0,-1),\quad \mathbf{\delta}_{21}=\frac{1}{\sqrt{2}}(1,-1,0).\label{Eq.app unit vector def}
		\end{gather}
	
	The second nearest neighbors are at a distance of $\sqrt{3}a$ and there are four distinct unit vectors from $r$ to $s$, denoted $\pm\mathbf{\varepsilon}^{\sigma}_{rs}$. Compared to the nearest neighbor case there is an extra choice of sign  $\sigma=\pm 1$.
	\begin{align}
		\mathbf{\varepsilon}^{\sigma}_{10}=\sqrt{\frac{2}{3}}\left(\sigma,\frac{1}{2},-\frac{1}{2}\right),\quad \mathbf{\varepsilon}^{\sigma}_{20}=\sqrt{\frac{2}{3}}\left(-\frac{1}{2},\sigma,\frac{1}{2}\right),\quad \mathbf{\varepsilon}^{\sigma}_{30}=\sqrt{\frac{2}{3}}\left(\frac{1}{2},-\frac{1}{2},\sigma\right)\non
		\mathbf{\varepsilon}^{\sigma}_{32}=\sqrt{\frac{2}{3}}\left(\sigma,\frac{1}{2},\frac{1}{2}\right),\quad \mathbf{\varepsilon}^{\sigma}_{31}=\sqrt{\frac{2}{3}}\left(\frac{1}{2},\sigma,\frac{1}{2}\right),\quad \mathbf{\varepsilon}^{\sigma}_{21}=\sqrt{\frac{2}{3}}\left(\frac{1}{2},\frac{1}{2},\sigma\right).\label{Eq.app epsilon def}
	\end{align}

We will also consider third nearest neighbors at a distance of $2a$. Third nearest neighbors always connect sites with same type of index $r$ much as in the model of Sec \ref{Sec 3 lattice}. There are two distinct types of third nearest neighbors which are not related by symmetries on the lattice. The first type of third nearest neighbor from $r$ is reached by the same unit vectors as the nearest neighbor case $\pm\mathbf{\delta}_{rs}$ with $s\neq r$, although just to be clear this vector is now connecting two sites both with index $r$. The second type of third nearest neighbor was not considered in \cite{batistaEtAl2018} and the unit vector from site $r$ is given by $\pm\mathbf{\delta}_{st}$ with both $s,t\neq r$.\footnote{The second type of third nearest neighbor is located on the opposite side of a hexagon in a kagome lattice slice of the pyrochlore lattice. See e.g. Fig. 1 in \cite{lapa2012ground}.}

\section{Continuum limit}
The classical Hamiltonian is
\begin{align}
	H= \frac{1}{2}\sum_{Ir,Js}K_{Ir,Js}\,\mathbf{S}_{Ir}\cdot \mathbf{S}_{Js},\label{eq.app Hamiltonian}
\end{align}
where an index like $I,J$ refers to the outer FCC lattice, $r,s$ to the four sites of the inner lattice, and the the couplings $K_{Ir,Js}$ are not necessarily nearest neighbor. The factor of $1/2$ is there because we are double-counting each link in our summation.

We may denote the sites $Ir$ by a spatial variable $x$, and write the spins in terms of a matrix $R(x)\in SO(3)$ and a basis $\mathbf{e}_r$,
\begin{gather}
	S^i_r(x)=R(x)^i_{\,j}e^{j}_r,\qquad
	e^j_{r}\equiv
	\left(\begin{array}{cccc}
		\frac{1}{\sqrt{2}}\sin\theta& -\frac{1}{\sqrt{2}}\sin\theta &\frac{1}{\sqrt{2}}\sin\theta&-\frac{1}{\sqrt{2}}\sin\theta\\
	\frac{1}{\sqrt{2}}\sin\theta& -\frac{1}{\sqrt{2}}\sin\theta &-\frac{1}{\sqrt{2}}\sin\theta&\frac{1}{\sqrt{2}}\sin\theta\\
		\cos\theta & 	\cos\theta & 	-\cos\theta & -\cos\theta
	\end{array}\right),
\end{gather}
and as before it may be more convenient to use the parameter $\kappa$ \eqref{def Kappa theta} rather than $\theta$
$$\kappa\equiv \frac{3}{2}\cos^2\theta-\frac{1}{2}.$$

If we keep the lattice points distinct, $\mathbf{e}_r$ is just a basis which defines the $R$ matrix. But now we will do some sleight-of-hand where we take $S_r^i(x)$ to be a continuous field taking values at all $x$. This is very similar to \cite{dombre1989nonlinear,batistaEtAl2018} but we will not introduce any fields like $L, b$ which describe departure of the four spins at a given point from a configuration $R\mathbf{e}_r$. So $\mathbf{e}_r$ is really a constraint describing some additional physics beyond the Hamiltonian \eqref{eq.app Hamiltonian} itself.\footnote{The classical Hamiltonian \eqref{eq.app Hamiltonian} with only nearest neighbor antiferromagnetic interactions actually has a macroscopically degenerate ground state \cite{moessner1998low}, so either a consideration of the coupling to higher order neighbors \cite{lapa2012ground}, quantum effects, or additional interactions such as in \cite{batistaEtAl2018} is necessary.} For instance, at $\kappa=0$ the spins form an ``all-in-all-out'' configuration, which was argued to be the ground state in \cite{batistaEtAl2018} based on additional biquadratic interactions. In this section we will simply take the constraint as given and focus on the $SO(3)$ modes, although this makes the model less realistic.

Now let us focus on the part of the Hamiltonian $H_1$ due to nearest neighbor interactions. The second and third nearest neighbors will follow similarly.
	\begin{align}
	H_1= K_1\rho\int d^3x\sum_{r\neq s}\mathbf{S}_r\cdot\left(\mathbf{S}_s+\frac{a^2}{2}\left(\mathbf{\delta}_{rs}\cdot \nabla\right)^2 \mathbf{S}_s + \frac{a^4}{4!}\left(\mathbf{\delta}_{rs}\cdot \nabla\right)^4 \mathbf{S}_s\right).\label{Eq.app H1}
\end{align}
Here $\rho$ is the number of FCC sites per unit volume,
	\begin{align}
	\rho = \frac{1}{4\sqrt{2}a^3}.
\end{align}
Given that $\sum_r \mathbf{S}_r=0$, up to a constant shift in energy this is equal to,
\begin{align}
	H_1= K_1\rho\int d^3x\sum_{r\neq s}\text{Tr}\left[R^{-1}\left(\frac{a^2}{2}\left(\delta_{rs}\cdot \nabla\right)^2+\frac{a^4}{4!}\left(\mathbf{\delta}_{rs}\cdot \nabla\right)^4\right)R\,\,E_{rs}\right], \label{Eq.app H1 2}
\end{align}
where $E_{rs}\equiv \mathbf{e}_{(r}\otimes \mathbf{e}_{s)}$.

It will be useful to define some additional matrices related to $E_{rs}$,
\begin{gather}
%I_\kappa\equiv\text{diag}\left(1-\kappa, 1-\kappa, 1+2\kappa\right),\non
\Delta_1\equiv \text{diag}\left(1-\kappa, 0, 0\right),\quad\Delta_2\equiv \text{diag}\left(0, 1-\kappa, 0\right),\quad\Delta_3\equiv \text{diag}\left(0, 0, 1+2\kappa\right),
\end{gather}
\begin{gather}
	\Sigma_{23}\equiv 	\frac{3}{\sqrt{2}}\sin\theta\cos\theta\left(\begin{array}{ccc}
			0& 0 & 0\\
			0 & 0 & 1\\
			0& 1 & 0
		\end{array}\right),\quad \Sigma_{31}\equiv 	\frac{3}{\sqrt{2}}\sin\theta\cos\theta\left(\begin{array}{ccc}
		0& 0 & 1\\
		0 & 0 & 0\\
		1& 0 & 0
	\end{array}\right),
\non
\Sigma_{23}\equiv 	\frac{3}{2}\sin^2\theta\left(\begin{array}{ccc}
	0& 1 & 0\\
	1 & 0 & 0\\
	0& 0 & 0
\end{array}\right),
\end{gather}
and also the matrix $I_\kappa\equiv\sum_r \Delta_r$ which equals the identity matrix in the $\kappa=0$ limit. Then we have the following relations with $r, s, t \neq 0$ all distinct,
\begin{align}
	E_{0r}+E_{st}&=-\frac{2}{3}I_\kappa +\frac{4}{3}\Delta_r\\
	E_{0r}-E_{st}&=-\frac{2}{3}\Sigma_{st}\\
	\sum_r E_{rr}&= \frac{4}{3}I_\kappa\\
	E_{00}+E_{rr}-E_{ss}-E_{tt}&=\frac{4}{3}\Sigma_{st}.
\end{align}

Now we may go back to \eqref{Eq.app H1 2} and using the explicit form of $\delta_{rs}$ \eqref{Eq.app unit vector def} group the terms by derivatives. The fourth order derivative terms will be presented next section. The second order derivative terms are
\begin{align}
	H_1=-\frac{2}{3}a^2K_1\rho\int d^3x\,\text{Tr}\left[\sum_r R^{-1}\partial_r^2R\,\Delta_{r}+\frac{1}{2}\sum_{r\neq s}R^{-1}\partial_r\partial_s R\,\Sigma_{rs} \right].
\end{align}
A similar calculation may be done for the second nearest neighbors with coupling $K_2$ and links $\varepsilon_{rs}^\sigma$ \eqref{Eq.app epsilon def},
\begin{align}
	H_2  =a^2K_2\rho\int d^3x\text{Tr}\left[-\frac{8}{3} R^{-1}\nabla^2R\,I_\kappa+4\sum_r R^{-1}\partial_r^2R\,\Delta_{r}+\frac{2}{3}\sum_{r\neq s}R^{-1}\partial_r\partial_s R\,\Sigma_{rs} \right],
	\label{Eq.app H2}
\end{align}
and also the third nearest neighbors which have distinct couplings $K_3, \tilde{K_3}$ depending on the type,
\begin{align}
	H_3= \frac{8}{3}a^2\rho\int d^3x\text{Tr}\left[(K_3+\tilde{K_3})R^{-1}\nabla^2R\,I_\kappa+(K_3-\tilde{K_3})\frac{1}{2}\sum_{r\neq s}R^{-1}\partial_r\partial_sR\,\Sigma_{rs}\right]. \label{Eq.app H3}
\end{align}

Adding all the contributions together we recover an expression that appears as Eq.(25) in \cite{batistaEtAl2018},\footnote{Compared to \cite{batistaEtAl2018} we have the additional parameters $\kappa$ and $\tilde{K_3}$, and lack the $\mathbf{L}, \mathbf{b}$ fields}
	\begin{gather}
		H=	-\int d^3x\text{Tr}\left[A\, R^{-1}\nabla^2R\,I_\kappa+B_1\sum_i R^{-1}\partial_i^2R \,\Delta_{r}+B_2\sum_{r\neq s}R^{-1}\partial_r\partial_s R\,\Sigma_{rs} \right],\label{A B definition}\\
		A\equiv \frac{8}{3}a^2\rho\left(K_2-K_3-\tilde{K_3}\right),\quad B_1\equiv \frac{2}{3}a^2\rho\left(K_1-6K_2\right),\quad B_2\equiv \frac{1}{3}a^2\rho\left(K_1-2K_2-4(K_3-\tilde{K_3})\right).\nonumber
	\end{gather}
Given the extra parameter $\tilde{K_3}$ this may be fine-tuned to a squashed sphere by setting $K_1=6{K_2}$, and $K_2=K_3-\tilde{K_3}$.

Finally, note that there appears to be an error in Eq.(27) in \cite{batistaEtAl2018}. Given \eqref{Eq.2 J R}, the continuum field theory in terms of $J$ at $\kappa=0$ is
\begin{align*}
	H=4\left[(2A+B_1)\sum_{r,s}(J^r_s)^2-B_1\sum_r(J^r_r)^2-B_2\sum_{r\neq s}\left(J^r_rJ^s_s+J^r_sJ^s_r\right)\right].
\end{align*}
\section{Fourth order derivatives and the hedgehog ansatz}
	Now using the same method as above we can find the part of the energy functional involving the fourth order derivative terms $H'$. The result is	
	\begin{align}
		H'&=\frac{a^4\rho}{4!}\left[K_1\left(-\frac{2}{3}C_1-2C_2+4C_3-\frac{4}{3}C_4\right)+\frac{36}{3}\left(K_3+\tilde{K_3}\right)\left(C_0+3C_2\right)+\frac{36}{3}\left(K_3-\tilde{K_3}\right)\left(2C_4\right)\right.\non
		&\qquad\qquad\left.+16K_2\left(-\frac{2}{3}C_0+\frac{5}{4}C_1-\frac{1}{4}C_2-\frac{3}{2}C_3+2C_4+\frac{1}{6}C_5\right)\right],
	\end{align}
	where there are multiple integrals over the $R$ field,
\begin{align*}
	C_0&\equiv\int d^3x \sum_i \text{Tr}\left[R^{-1}\partial_i^4 R \,I_\kappa\right]\non
	C_1&\equiv\int d^3x \sum_i \text{Tr}\left[R^{-1}\partial_i^4 R \,\Delta_i\right]\\
	C_2&\equiv\int d^3x \sum_{i<j} \text{Tr}\left[R^{-1}\partial_i^2\partial_j^2 R \,I_\kappa\right]\\
	C_3&\equiv\int d^3x \sum_{i<j, \,k\neq i,j} \text{Tr}\left[R^{-1}\partial_i^2\partial_j^2 R \,\Delta_k\right]\\
	C_4&\equiv\int d^3x \sum_{i<j} \text{Tr}\left[R^{-1}\left(\partial_i^3\partial_j+\partial_j^3\partial_i\right)R \,\Sigma_{ij}\right]\\
	C_5&\equiv\int d^3x \sum_{i<j,\,k\neq i,j} \text{Tr}\left[R^{-1}\partial^2_k\partial_i\partial_jR \,\Sigma_{ij}\right].
\end{align*}

These integrals may be evaluated for the hedgehog configuration \eqref{def Hedgehog} in terms of the $R$ matrix,
	\begin{align}
	R_{ij}= \cos 2f\,\delta_{ij} +\left(1-\cos 2f\right)\hat{r}^i\hat{r}^j + \sin 2f\,\epsilon_{ijk}\hat{r}^k.
\end{align}
As a rough test whether there is a range of parameters with stable Skyrmions at $\kappa=0$, we may simply input a trial profile function like $f(r)=\pi \exp\left[-\frac{r^2}{r_0^2}\right]$ and solve for the value of $r_0$ with minimum energy,
	\begin{align}
	r_0^2=a^2\frac{4.33\times 10^2 K_1-1.09\times 10^5 K_2+1.74\times 10^5 \left(K_3+\tilde{K_3}\right)-3.89\times 10^4 \left(K_3-\tilde{K_3}\right)}{1.55\times 10^3K_1+2.53\times 10^4K_2 -3.87\times 10^4\left(K_3+\tilde{K_3}\right)+4.00\times 10^3\left(K_3-\tilde{K}_3\right)}.
\end{align}
For this to be valid, the factor multiplying $a^2$ must be much greater than $1$, and there are clearly choices of parameters for which this is possible. For instance, consider the parameter restriction$$K_1=6{K_2},\quad K_2=K_3-\tilde{K_3},$$
in which the second derivative terms are spherically symmetric. The optimal Skyrmion radius reduces to
$$r_0^2=a^2\left(-.369\frac{K_2}{\tilde{K_3}}-4.50\right),$$
and so there are stable Skyrmions for $K_2\approx K_3 \gg -\tilde{K_3}.$ 

		\chapter{Details of the asymptotic expansion}\label{appendix asymptotic}

		This appendix deals with some additional technical details of the asymptotic expansion considered in Sec. \ref{Sec 5 mellin}.
		
		Beginning with the expression \eqref{Eq.5 piArc 2} for the self-energy correction which we reproduce here,
		\begin{gather}
			\Pi_{\perp,arc}(p)=-\frac{p^2 }{N}\int_0^\infty dt \int \frac{ds}{2\pi i}B\left(-2s+1,s+t\right)B(-s,1+s)\,{}_2F_1\left(1-s,-s;\,2;\,-\frac{p^2}{m^2}\right),\label{Eq.App eq1}
		\end{gather}
		we desire to find an expansion in powers of $m^2/p^2$. This can be accomplished by using following formula for the inversion of the argument of a hypergeometric function.
			\begin{gather}
				{}_2F_1(\alpha,\alpha;\gamma; z)=\frac{(-z)^{-\alpha}}{B(\alpha,\gamma-\alpha)}\sum_{j=0} \frac{(\alpha)_j(\alpha+1-\gamma)_j}{j!j!}\left(\frac{1}{z}\right)^{j}G_j(\alpha,\gamma;z),\non
				G_j(\alpha,\gamma;z)\equiv \log(-z)+\psi(1-\gamma+\alpha)-\psi(\gamma-\alpha)+2\psi(1+j)-\psi(\alpha+j)-\psi(1-\gamma+\alpha+j),\label{Eq.App main formula}
			\end{gather}
		This was derived for specific values of $\alpha$ and $\gamma$ by my coauthor in the second appendix of \cite{schubring2021treating}.\footnote{The derivation in \cite{schubring2021treating} involved deforming the arguments of the hypergeometric function by a parameter which happened to arise from dimensional regularization, but of course the resulting formula is simply a mathematical fact and the derivation of the OPE in \ref{Sec 5 mellin} did not involve analytic continuation in dimension.}
		
		As a side note, this formula can be applied to the general formula \eqref{Eq.4 hypergeometric} for the bubble integral in $d=2$,
		\begin{align}
J(p)&=\frac{1}{4\pi m^2}\,\,{}_2 F_1\left(1,1;\frac{3}{2};-\frac{p^2}{4m^2}\right)\non
&=\frac{1}{2\pi p^2}\sum_k \frac{(1/2)_k}{k!}\left[\log \frac{p^2}{4m^2}+\psi(1+k)-\psi(1/2+k)\right]\left(-\frac{4m^2}{p^2}\right)^k.\label{Eq.App bubble}
		\end{align}
	This gives a closed form expression for the large $p$ expansion of the bubble integral which may be useful in applications such as in the cutoff scheme ambiguity calculations in Sec \ref{Sec.5 ambiguity operator}.
	
	Returning to the problem of expanding \eqref{Eq.App eq1}, we can put the hypergeometric function into the form of the left hand side of \eqref{Eq.App main formula} by using a standard formula for shifting arguments,
	\begin{align*}
{}_2F_1\left(1-s,-s;\,2;\,-z\right)=\frac{(1+s)F(-s,-s;2 ;-z )+s(z+1)F(1-s,1-s;2 ;-z)}{2s+1},
	\end{align*}
where $z\equiv p^2/m^2$.

Now applying \eqref{Eq.App main formula} and simplifying,
		\begin{align}
			&B(-s,1+s)\,{}_2F_1\left(1-s,-s;\,2;\,-z\right)= z^s\sum_{k=1}\left(-z\right)^{-k}\frac{(-s)_{k-1}(-s+1)_{k-1}}{(k-1)!k!}A_k\non	
			&\quad+\frac{z^s}{2s+1}\left[B_0-A_0+\sum_{k=1}\left(-z\right)^{-k}\frac{(-s)_{k-1}(-s+1)_{k-1}}{k!k!}\left(s(s+1)\left[B_k-A_k\right]+ k^2\left[A_{k-1}-A_k\right]\right)\right],\label{Eq.App expansion}
		\end{align}
	where $A_k\equiv G_k(1-s,2;-z),\quad B_k\equiv G_k(-s,2;-z)$.
	
	The first set of terms in \eqref{Eq.App expansion} were dealt with in the main text. Let us now consider the remaining terms involving the large bracket on the second line.
	
	Recall that the operator part of the OPE \eqref{Eq.5 operator part 1} arose from the poles of the term $-\psi(1+s)$ in $A_k$. There is a corresponding term $-\psi(2+s)$ in $B_k$ which cancels all the poles in the bracketed expression of \eqref{Eq.App expansion} except the pole at $s=-1$ in the $B_0-A_0$ term. So let us focus on the self-energy $\Pi_{\perp,k=0}$ corresponding to this $B_0-A_0$ term alone,
	\begin{align}
\Pi_{\perp,k=0}=\frac{p^2 }{N}\int_0^\infty dt \int \frac{ds}{2\pi i}B\left(-2s+1,s+t\right)\left(\frac{p^2}{m^2}\right)^s\frac{1}{s(s+1)}.
	\end{align}
	
	The pole at $s=-1$ leads to the additional operator VEV mentioned below \eqref{Eq.5 operator part 2}
	\begin{align}
-\frac{m^2}{N}\int_0^\infty dt B(3,t-1)=-2\left\langle A_\perp^2\right\rangle,
	\end{align}
and this has ambiguity
	\begin{align*}
-2 \left\{A_\perp^2\right\}=-\frac{m^2}{N}.
\end{align*}

As in the main text, the coefficient function part of the OPE is found from the poles at $s=-t-j$,
\begin{align}
\frac{p^2}{N}\int_0^\infty dt\sum_{j=0}\frac{(-1)^j}{j!}\frac{\Gamma(2(t+j)+1)}{\Gamma(2t+j+1)} \left(\frac{m^2}{p^2}\right)^{t+j}\frac{1}{\left(t+j\right)(t+j-1)}.
\end{align}
This only has a pole for $t>0$ when $j=0$ and $t=1$, and the residue is $+m^2/N$, so the ambiguities cancel.

Finally, we still must consider the remaining terms in the second line of \eqref{Eq.App expansion} involving the sum over $k\geq 1$. It may be shown that
\begin{align}
\frac{	s(s+1)\left[B_k-A_k\right]+ k^2\left[A_{k-1}-A_k\right]}{2s+1}=\frac{s(s+1)+k^2 }{(k-s)(k-s-1)}
\end{align}
This has no poles in relevant region $\text{Re}(s)<0$ for any value of $k$ in the summation, so it corresponds to no operator VEVs. It will contribute to the coefficient part upon taking the residues at $s=-t-j$, but again there are no poles for $t>0$ that may contribute to ambiguities. So that ties up all the remaining loose ends in the demonstration of the cancelation of ambiguities in the OPE.
	\end{appendices}

\printbibliography
\end{document}